\documentclass[12pt]{article}
\usepackage{comment}
\usepackage{textcomp}
\usepackage{chetnew}
\usepackage{amssymb,amsmath,amsfonts,mathrsfs,braket}
\usepackage[utf8]{inputenc}
\usepackage{graphicx,longtable,afterpage}
\usepackage{xcolor}

\usepackage{floatrow}
\newfloatcommand{capbtabbox}{table}[][\FBwidth]

\def\sp{\;\;\;,\;\;\;}
\def\ar{{~~~\Rightarrow~~~}}
\renewcommand{\d}[1]{\ensuremath{\operatorname{d}\!{#1}}}

\def\one{{\,\hbox{1\kern-.8mm l}}}

\def\be{\begin{equation}}
\def\ee{\end{equation}}

\newcommand{\Tr}{\mathrm{Tr}}

\newcommand{\CC}{\mathcal{C}}

\allowdisplaybreaks

\def\makeatletter{\catcode`\@=11}% 11:letter
\makeatletter
\def\mathbox#1{\hbox{$\m@th#1$}}%
\def\math@ccstyles#1#2#3#4#5#6#7{{\leavevmode
      \setbox0\mathbox{#6#7}%
      \setbox2\mathbox{#4#5}%
      \dimen@ #3%
      \baselineskip\z@\lineskiplimit#1\lineskip\z@
      \vbox{\ialign{##\crcr
             \hfil \kern #2\box2 \hfil\crcr
             \noalign{\kern\dimen@}%
             \hfil\box0\hfil\crcr}}}}
\def\mathaccstyles{\math@ccstyles\maxdimen}
\def\maththroughstyles{\math@ccstyles{-\maxdimen}}
\def\unity%
 {\maththroughstyles{.45\ht0}\z@\displaystyle {\mathchar"006C}\displaystyle 1}

\def\AA{{\cal A}}

\def\CC{{\cal C}}
\def\DD{{\cal D}}

\def\KK{{\cal K}}

\def\MM{{\cal M}}

\def\OO{{\cal O}}
\def\PP{{\cal P}}

\def\SS{{\cal S}}
\def\TT{{\cal T}}

\def\ZZ{{\cal Z}}

\def\Tr{{\rm {Tr}}}

\def\d{{\partial}}

\def\im{{\rm Im}}

\def\r{{\mathfrak r}}

\def\beq{\begin{equation}}
\def\eeq{\end{equation}}
\newcommand{\bea}{\begin{eqnarray}}
\newcommand{\eea}{\end{eqnarray}}
\def\bal{\begin{align}}
\def\eal{\end{align}}

\newmuskip\pFqmuskip

\newcommand*\pFq[6][8]{%
  \begingroup % only local assignments
  \pFqmuskip=#1mu\relax
  \mathchardef\normalcomma=\mathcode`,
  \mathcode`\,=\string"8000
  \begingroup\lccode`\~=`\,
  \lowercase{\endgroup\let~}\pFqcomma
  {}_{#2}F_{#3}{\left(\genfrac..{0pt}{}{#4}{#5}\Big | #6\right)}%
  \endgroup
}
\newcommand{\pFqcomma}{{\normalcomma}\mskip\pFqmuskip}
\preprint{CCTP-2024-11 \\ ITCP-IPP-2024/11}

\title{\vspace{-1.cm}
Scattering, Absorption and Emission \\ \vspace{0.3cm}
of Highly Excited Strings}

\author{
M.~Firrotta\;$^{\flat}$, E.~Kiritsis\;$^{\flat\natural\dagger}$, V.~Niarchos\;$^{\flat}$}

\affiliation{
$^\flat$ \href{http://hep.physics.uoc.gr}{Crete Center for Theoretical Physics}, Institute for Theoretical and Computational Physics,
Department of Physics, Voutes University Campus,
GR-70013, Heraklion, Greece $ $ \\
~\\
$^\natural$  \href{http://www.apc.univ-paris7.fr}{Universit\'e de Paris, Cit\'e, CNRS, Astroparticule et Cosmologie,  F-75013 Paris, France}\\
~\\
$^\dagger$ Arnold Sommerfeld Center for Theoretical Physics,
Ludwig-Maximilians-Universit\"at M\"unchen, 80333 M\"unchen, Germany
\vspace{0.5cm} $ $

{\bf Abstract}
\vspace{-0.5cm}
}

\abstract{We study tree-level scattering processes of arbitrary string states using the DDF formalism and suitable coherent vertex operators. We obtain new exact compact formulae for heavy-heavy-light-light scattering amplitudes in open or closed bosonic string theories, and derive explicit exact expressions for the absorption cross-sections, and corresponding emission rates, of highly excited string states using the optical theorem and time reversal symmetry. We show that these expressions are independent of the microscopic structure of the excited string states without averaging. For the absorption of massless modes in open string theory, in particular, we find a constant, frequency-independent cross-section. In contrast, the corresponding cross-section for the absorption of massless modes by excited closed strings depends linearly on the frequency, implying a non-trivial grey-body factor. In both cases, at energies below the scale set by the mass of the highly excited strings, we find emission rates with a Boltzmann factor at Hagedorn temperature.
}

\date{\today}

\begin{document}

\maketitle

\hypersetup{pageanchor=true}

\setcounter{tocdepth}{2}

\toc

\section{Introduction}
\label{intro}

In this paper we revisit the computation of tree-level scattering amplitudes between light and heavy (highly-excited) string states. With the use of the DDF formalism \cite{DelGiudice:1971yjh}, we obtain new compact formulae for heavy-heavy-light-light (HHLL) scattering amplitudes and provide an exact derivation for the absorption cross-section of light modes from highly-excited string states with the use of the optical theorem. Corresponding emission rates follow from this result using microscopic reversibility, according to time-reversal symmetry.

Already at the classical level, excited string states are interesting examples of complex systems with high entropy. One would, therefore, like to understand how thermal and chaotic properties arise in these states and how they compare with other similarly complex systems, most notably black holes in gravity. Indeed, investigations of the correspondence between black holes and strings is an old subject that goes back to \cite{Susskind:1993ws,Halyo:1996vi,Halyo:1996xe,Horowitz:1996nw}. In the picture advocated in these works, a highly excited string state behaves as a black hole at strong coupling, but as the coupling is decreased its horizon shrinks and eventually becomes a long highly-excited string at weak coupling. At the correspondence point, where the curvature of the black hole geometry in the string frame reaches the string scale, the two descriptions are expected to match and there is suggestive evidence, \cite{Horowitz:1996nw}, through extrapolations and up to numbers of order one, that this is indeed happening for several quantities, e.g.\ the number of degenerate states. These observations have motivated a significant amount of work in the past, which has been focused on the comparison between the properties of black holes and strings. Some of the older works on the subject include \cite{Mitchell:1990cu} on string form factors, and \cite{Amati:1999fv,Manes:2001cs,Chialva:2003hg,Chialva:2004xm} on the decay rate of highly-excited fundamental string states. A lot of parallel related work, motivated by black hole microstate counting \cite{Strominger:1996sh} using D-branes, has also been performed using effective string descriptions.

In the present paper, we discuss exclusively fundamental strings, drawing lessons about their interactions in flat space through the study of tree-level scattering amplitudes. For convenience, we focus on bosonic strings on flat backgrounds. In order to implement efficiently the physical state conditions on the string states, we shall use the DDF formalism. This formalism, which has a long history, has been instrumental in several recent works on string scattering.

Since the era of the dual resonance model (DRM), a classification of physical excited states attracted a lot of attention, leading to the development of various techniques based on the light-cone quantization applied to the Heisenberg algebra of Fock spaces  \cite{DelGiudice:1971yjh,Brower:1971nd,Brower:1970tm,Brower:1972ev}. Three-point interactions of highly excited string states were first studied in the context of DRM \cite{Ademollo:1974kz}, using a quantization prescription based on a suitable isomorphism of the Heisenberg algebra, named Di Vecchia-Del Giudice-Fubini formalism. Highly excited string interactions, in particular three-string interactions, were first computed in \cite{Cremmer:1974jq}. The compatibility of the results in \cite{Cremmer:1974jq} with the results in \cite{Ademollo:1974kz}, provided the first evidence of how powerful the DDF formalism is in the context of highly excited string interactions.

A modern picture of the DDF formalism was proposed and intensively studied in \cite{Skliros:2009cs,Hindmarsh:2010if,Skliros:2011zz,Skliros:2011si,Skliros:2013pka}, where for the first time a fully BRST-covariant coherent vertex operator in bosonic string theory was formulated. The expansion of this coherent vertex operator reproduces level-by-level all the vertex operators associated to the infinite tower of string states. The supersymmetric extension of this construction was formulated in \cite{Aldi:2019osr}.

The covariant embedding of the DDF formalism was the bridge between the old fashioned DDF formalism and the Friedan-Martinec-Shenker formalism \cite{Friedan:1985ge} providing a powerful tool for the explorations of highly excited string scattering amplitudes. Relying on the technology developed in \cite{Hindmarsh:2010if}, the first computation of string scattering amplitudes involving coherent states (or arbitrarily excited string states after expansion), was done in \cite{Bianchi:2019ywd}. In particular, tree-level three- and four-point amplitudes with several coherent states and light states were computed in that work, providing a systematic way of including coherent states in the string $S$-matrix.

The properties of scattering amplitudes of highly-excited string states were also studied in \cite{Gross:2021gsj,Rosenhaus:2021xhm} within the covariant embedding of the DDF formalism, demonstrating how the erratic behavior of such interactions was related to chaotic features in $S$-matrix theory \cite{Rosenhaus:2020tmv,Polchinski:2015cea}.
This important observation opened a new window into the physical interpretation of highly-excited string interactions. Additional connections with chaos and random matrix theory were proposed in \cite{Firrotta:2022cku,Bianchi:2022mhs,Firrotta:2023wem,Bianchi:2023uby,Bianchi:2024fsi} and recently studied further in \cite{Savic:2024ock,Das:2023cdn}.

\subsection*{New results}

The present study employs the above powerful techniques to obtain a new systematic derivation of generic heavy-heavy-light-light scattering amplitudes and related absorption cross-sections following from the application of the optical theorem.

The mass  of the states considered, proportional to $\sqrt{N}$, cannot be independent of the string coupling $g_s$. In particular, such states  must have horizons that are smaller than the string length. When the horizon is at  the string length, this is  the correspondence limit of \cite{Horowitz:1996nw}
and gives
\be
N ~~\lesssim~~ {1\over g_s^4}
\label{9}\ee

In general, three- and four-point string amplitudes depend not only on the masses and momenta of particles, but also their spin configuration and other microscopic properties. In this paper, we use such amplitudes to calculate the absorption cross-section, without any averaging over initial states, and find that the result is independent of the spin configuration and other microscopic properties of both the absorber and the absorbed particle.

For the absorption of massless open string states (photons or massless scalars) from \underline{any} excited open string state, the cross-section in the lab frame is independent of the energy $\omega$ of the massless state
\beq
\label{introaa}
\sigma_{abs}^{(\rm open)} = \pi \ell_s^{d-2} g_o^2
~,
\eeq
where $\ell_s$ is the string length and $g_o$ is the open string coupling.\footnote{Here we assume $d$ non-compact dimensions.} This is an exact tree-level result in open bosonic string theory. It refers to a single absorbing excited string state and \underline{does not} involve any averaging. The corresponding emission rates of excited strings can be inferred using time-reversal symmetry. For Highly-Excited String (HES) states, with large mass $M\gg \ell_s^{-1}$ and emitted massless modes with energy $\omega$, one obtains a differential rate of emission of the form
\beq
\label{introaaExact}
\frac{d\Gamma^{(\rm open)}_{em}}{d\omega d\Omega_{d-2}} = 2^{1-d}\pi^{2-d} \ell_s^{d-1} g_o^2\, M\, {\omega^{d-2} \over 1 - {\omega \over M}}
{\rho\Big(\ell_s^2M(M - 2 \omega)\Big)\over \rho(\ell^2_sM^2)}
~,\eeq
where $\rho(N)$ denotes the degeneracy of string states at level $N$. To obtain this expression we have averaged over the initial decaying string state. At leading order in the regime $\omega \ll M$, \eqref{introaaExact} simplifies further to
\beq
\label{introaaa}
\frac{d\Gamma^{(\rm open)}_{em}}{d\omega d\Omega_{d-2}} = 2^{1-d}\pi^{2-d} \ell_s^{d-1} g_o^2\, M\, \omega^{d-2}
e^{-\frac{\omega}{T_H}}
~,
\eeq
where $T_H$ is the Hagedorn temperature (see Eq.\ \eqref{emiab}).

The corresponding absorption cross-section for massless states in closed string theory is found to be proportional to the energy of the absorbed state,
\beq
\label{introab}
\sigma_{abs}^{(\rm closed)} = 2\pi \ell_s^d g_{c}^{2} M \omega
~,
\eeq
implying that highly-excited closed string states are characterized by a non-trivial grey-body factor at tree-level that depends linearly on the energy $\omega$ of the massless mode. $M$, the mass of the excited closed string state, is the only feature of the excited state entering the final expression. Once again, this is an exact tree-level result in closed bosonic string theory. Using time-reversal symmetry one can also determine the corresponding emission rate. The exact result
\beq
\label{introaabex}
\frac{d\Gamma^{(\rm closed)}_{em}}{d\omega d\Omega_{d-2}} =
(2\pi)^{2-d} \ell_s^{d+1} g_{c}^{2} M^2\, {\omega^{d-1}\over 1-{\omega\over M}}
{\rho\Big(\ell_s^2M(M - 2 \omega)/4\Big)\over \rho(\ell^2_s M^2 /4)}\Bigg|_{L}\, {\rho\Big(\ell_s^2M(M - 2 \omega)/4\Big)\over \rho(\ell^2_s M^2 /4)}\Bigg|_{R}
\eeq
 is proportional to the product of state degeneracies from the left- and right-moving sectors of the string. In the regime $M\gg \ell_s^{-1}$, $\omega\ll M$ the result is similar to \eqref{introaaa}
\beq
\label{introaab}
\frac{d\Gamma^{(\rm closed)}_{em}}{d\omega d\Omega_{d-2}} =
2^{1-d}\pi^{2-d} \ell_s^{d+1} g_{c}^{2} M^2\, \omega^{d-1}
e^{-\frac{\omega}{T_H}}
~.
\eeq

These emission rates exhibit thermal features at low energies $\omega\ll M$, but not the full Bose factor, which is common in black body physics. The deviations from the Bose factor arise at energies $\omega \lesssim T_H$ and become more and more suppressed as the space-time dimension is formally increased. Some parallels with black hole physics are present, which would be worth understanding further. For example,
the linear dependence of the absorption cross-section in the closed string case \eqref{introaa}, matches the corresponding black hole result \cite{Maldacena:1996ix}, when $l_s^{-1} \ll \omega \ll M$. The main difference lies at super-low energies below the string mass scale, $\omega \ll l_s^{-1}$, where excited closed strings do not absorb or emit massless modes exactly as black holes (the grey-body factors are different). For instance, we find that the absorption cross-section of a highly-excited string vanishes as the energy of the absorbed massless particle tends to zero, which is unlike the behaviour of a black hole with a regular horizon. Of course, this is not in contradiction with the black hole-string correspondence in \cite{Susskind:1993ws,Halyo:1996vi,Halyo:1996xe,Horowitz:1996nw}, which requires an extrapolation to finite string coupling.

The differential rates of emission were also computed in the past in Ref.\ \cite{Amati:1999fv} from tree-level 3-point amplitudes in open and closed bosonic string theory using the light-cone gauge formalism (see also Ref.\ \cite{Chialva:2004xm} for a related analysis in superstring theory). Our results are consistent with these papers in the regime $l_s^{-1} \ll \omega \ll M$, but differ at energies below the string mass scale. Using the optical theorem we provide a new, independent computation of the absorption cross-section in this paper. Unlike \cite{Amati:1999fv}, the corresponding emission rates are deduced from the absorption cross section using a straightforward time-reversal symmetry argument that allows to extend the computation of emission rates to more general cases than previously considered. It would be interesting to clarify the tension with the results of Ref.\ \cite{Amati:1999fv} in the substringy regime. We comment further on this point in subsection \ref{comment}.

\subsection*{Outlook}

The recent successful applications of the DDF formalism to generic tree-level computations in string theory (including the applications and results in this paper) raise the exciting potential for several interesting future developments. One of the main ultimate goals is a better understanding of the complicated properties of string interactions and their intriguing relation to black hole physics.

Another perspective on our calculations is provided by Lloyd's theorems, \cite{Lloyd}. They imply that expectation values and correlators in generic pure states of high-energy subspaces in QFT are nearly thermal. In our example, the relevant subspace is that of string states with $N\gg 1$, which can clearly serve as  a microcanonical ensemble in weakly coupled string theory. Although some details differ, our findings are in accordance with \cite{Lloyd}. Moreover, our setup provides further possibilities for checks of the ``thermalization" properties of complex pure states, and this is an issue that will be studied in the near future.

Further interesting avenues for future research along the lines of the work in this paper include the study of the total emission and decay width of single string states, the tree-level scattering of heavy states and the structure of higher-point scattering amplitudes. We hope to return to some of these problems in future work.

\subsection*{Brief plan of paper}

In Section \ref{notation} we summarize the main components of the formalism that we employ throughout the paper and set our notation.

In Section \ref{4point} we discuss heavy-heavy-light-light 4-point amplitudes in open and closed string theory separately. We provide general formulae for the scattering of tachyons or photons with arbitrary highly-excited string states and present compact expressions of the final result in certain special cases, e.g.\ the special case of excited string states on the leading Regge trajectory. Analogous formulae are also derived in the case of scattering between tachyons or massless modes with highly-excited string states in closed string theory. Further explicit examples are reported in Appendix \ref{exemAmplLR}.

Using the results of Section \ref{4point} and the optical theorem, we proceed in Section \ref{absorb} to compute the absorption of a light state (a tachyonic scalar or a massless mode) from a highly-excited string state. We discuss separately the absorption of light states in open and closed string theory, obtaining compact exact expressions, like \eqref{introaa}, \eqref{introab} above. These results are also generalized to the most general case of absorption that involves arbitrary string states. In Section \ref{emission} we conclude with a derivation of the corresponding emission rates using the time-reversal symmetry of scattering amplitudes and compare with the existing results in the literature.

Further useful relations and technical results are relegated to eight Appendices at the end of the paper.

\section{Notation and other useful material}
\label{notation}

In this paper, we consider tree-level scattering amplitudes in string theory. For simplicity, and clarity of exposition, we examine in detail scattering in bosonic open and closed strings. It will be convenient to organise the presentation by demonstrating all the pertinent details first in the context of open string theory. The details of the analogous closed string computations will be left implicit, with the precise form of the corresponding final expressions reported separately in subsections \ref{closed1}, \ref{closed2}.

In this section, we summarize the main elements of the formalism that will be employed, and the relevant notation. To be concise, we focus on the case of open strings.

\subsection{String states}
\label{notation-states}

A generic open string state will be characterized by its mass $M$ and the total occupation number $J$ (defined in Eq.\ \eqref{notaf} below). The mass is related to the level $N$ through the standard relation
\beq
\label{notaa}
\alpha' M^2 = N-1
~,\eeq
where $\alpha'=\ell_s^2$ is the Regge slope and $\ell_s$ is the string length. The vertex operator of a BRST invariant state on the worldsheet will be constructed using the Del Giudice-Di Vecchia-Fubini (DDF) approach \cite{DelGiudice:1971yjh}. In this approach, the creation operators $\AA_{-n}^\mu$ (for the $n$-th harmonic in the $\mu$ target-space coordinagte $X^\mu$) are realized via the relation
\beq
\label{notab}
 \AA^{\mu}_{-n} := \oint \frac{dz}{2\pi i} i\partial X^{\mu}(z)\, e^{-inqX}
~.
\eeq
$z$ is the open string worldsheet coordinate and $q^\mu$ the DDF null momentum with the properties
\beq
\label{notac}
q \cdot \AA_{-n} = 0~,~~ q^2=0
~.
\eeq
The generic, normalized, BRST-invariant, arbitrarily excited, string state can be written as
\beq
\label{notad}
|\epsilon,\{n,g_{n}\}, p_N \rangle :=\epsilon_{(\mu_{1})_{g_{1}}(\mu_{2})_{g_{2}}\dots}\, \lim_{z\to 0} \left(\prod_{n}{1\over \sqrt{n^{g_{n}} g_{n}!}} \prod_{r=1}^{g_n} :{\cal A}_{-n}^{\mu_{n}^{r}}:\right)\, :e^{ip{\cdot}X}(z): |0 \rangle
~,
\eeq
with normalization condition
\be\label{NormCond}
\Big\langle\epsilon^{*},\{n,g_{n}\},p_{N}\, \Big|\,\epsilon,\{n,g_{n}\},p_{N}\Big\rangle=1 \quad \Rightarrow \quad \Tr(\epsilon^{T}\epsilon^{*})=1
~,
\ee
with
\be\label{DDFqp}
\alpha' p^2 = 1\,,\quad 2\alpha' p\cdot q = 1
\ee
and physical momentum $p_N=p-Nq$. The generic polarization tensor contains {$g_{1}$ symmetric  indices, $\mu_{1}$, $g_2$ symmetric indices $\mu_2$, etc}.
\be\label{genpoltens}
\epsilon_{(\mu_{1})_{g_{1}}(\mu_{2})_{g_{2}}\ldots} :=\epsilon_{(\mu^{1}_{1}...\mu^{g_{1}}_{1})(\mu^{1}_{2}...\mu^{g_{2}}_{2})\ldots}\,,\quad (\mu_{r})_{g_{r}}:=(\mu_{r}^{1}\mu_{r}^{2}....\mu_{r}^{g_{r}}) ~,
\ee
in accordance with the set of creation operators of the state.

After the action of the creation operators in \eqref{notad}, which is implemented by the operator product expansion between DDF operators \eqref{notab} and the tachyonic momentum operator $e^{ip{\cdot}X}$, each index of the generic polarization tensor \eqref{genpoltens} is acted upon by the matrix
\beq
\label{notaeR}
R_{(q)\nu^{r}_{k}}^{\mu^{r}_{k}}= \delta^{\mu^{r}_{k}}_{\quad\nu^{r}_{k}} - 2\alpha'   p^{\mu^{r}_{k}} q_{\nu^{r}_{k}}
~,
\eeq
leading to
\be\label{RotPol}
R_{(q)\nu_{1}^{1}}^{\mu_{1}^{1}} \ldots R_{(q)\nu_{1}^{g_{1}}}^{\mu_{1}^{g_{1}}} \ldots
\epsilon_{(\mu^{1}_{1}\ldots \mu^{g_{1}}_{1})\ldots }\equiv{\cal E}_{(\nu_{1})_{g_{1}} (\nu_{2})_{g_{2}}\ldots }
\ee
with
\be
R_{(q)\nu}^{\mu} p^{\nu} =0\;\;,\;\; R_{(q)\nu}^{\mu} q^{\nu} = q^{\mu}
\label{double}\ee
The first equation in (\ref{double}) follows from the condition $2\alpha' p\cdot q = 1$.
The second condition in (\ref{double}) implies that  such an action does not contribute on the state because
of the constraint $q{\cdot}{\cal A}_{-n}=0$, $i.e$ the orthogonality of $q$ to the transverse components.

 Note that the DDF rotations \eqref{notaeR} leave unchanged the normalization of the state
\be\label{NormRot}
\Tr(\epsilon^{T}\epsilon^{*})=\Tr({\cal E}^{T}{\cal E}^{*})
\ee
since $R_{q}^{T}R_{q}$ gives the identity matrix in the transverse space of physical components,
and $\cal E$ is defined in (\ref{RotPol}).

 The distribution of creation operators in the excited state is controlled by the set of non-negative integers $\{g_n\}$, which express the level $N$ and the total occupation number $J$ as
\beq
\label{notaf}
N = \sum_n n g_n ~, ~~ J = \sum_n g_n \leq N
~.
\eeq

Two relatively simple (and extreme) cases include:
\begin{itemize}
    \item The leading Regge trajectory states with $g_n=1$ for all $n=1,\ldots,N$ that have the maximum possible value of $J$, $J=N$.
    \item The states with $g_n=\delta_{n,N}$ that have a single excited harmonic at $n=N$. These states have $J=1$.
\end{itemize}

\subsection{Coherent states}
\label{notation-coherent}

The action of the DDF oscillators in \eqref{notad} can be computed explicitly using standard OPE relations in the free worldsheet CFT. The final result can be conveniently packaged into a coherent state of creation operators that provides a generating function of vertex operators \cite{Skliros:2009cs,Hindmarsh:2010if,Skliros:2011zz,Skliros:2011si,Skliros:2013pka}.
More specifically, for the general class of states in \eqref{notad} the relation
\beq
\label{cohaa}
:e^{\sum_{n}\lambda_{n}{\cdot}{\cal A}_{-n}}:\, e^{ipX}(z) =\exp{\left(\sum_{n,m}\frac{\zeta_{n}{\cdot}\zeta_{m}}{2}{\cal S}_{n,m}e^{-i(n{+}m)q{\cdot}X}{+}\sum_{n}\zeta_{n}{\cdot}{\cal P}_{n}e^{-inq{\cdot}X}{+}i p{\cdot}X\right)}(z)
\eeq
with
\beq
\label{notae}
\zeta_{n\,\nu} := \lambda_{n\,\mu}{R^{\mu}}_{\nu}=\lambda_{n\,\nu} - 2\alpha' (\lambda_{n} \cdot p) q_{\nu}
~,
\eeq
allows us to write the normalized state
\bea
\label{cohab}
|\epsilon,\{n,g_{n}\}, p_N \rangle =&&\epsilon_{(\mu_{1})_{g_{1}}(\mu_{2})_{g_{2}}\ldots}
\prod_{n}{1\over \sqrt{n^{g_n} g_n!}}\prod_{r=1}^{g_n}  \frac{\d}{\d \lambda_{n,\mu_{n}^{r}}}
\\
&&\exp{\left(\sum_{\ell,m} \frac{\zeta_{\ell}{\cdot}\zeta_{m}}{2}{\cal S}_{\ell,m}e^{-i(\ell{+}m)q{\cdot}X}{+}\sum_{m}\zeta_{m}{\cdot}{\cal P}_{m}e^{-imq{\cdot}X}{+}ip{\cdot}X\right)}\Bigg|_{\lambda_{n}=0 \atop z=0}
|0\rangle
\nonumber
\eea
with $\PP_m$, $\SS_{\ell,m}$ suitable polynomials of the worldsheet operators $\d^s X$ of the form\footnote{We are using the notation $\d := \d_z$.}
\beq
\label{cohac}
\zeta_{n}{\cdot}{\cal P}_{n}(z) := \sum_{k=1}^{n} \frac{\zeta_{n}{\cdot}i\partial^{k} X(z)}{(k{-}1)!}{\cal Z}_{n{-}k}(a_{s}^{(n)})~, ~~
a_s^{(n)} :=-in \frac{q{\cdot}\partial^{s}X}{(s{-}1)!}
~,
\eeq
\beq
\label{cohad}
{\cal S}_{n,m} := \sum_{r=1}^{m}r\,{\cal Z}_{n{+}r}(a_{s}^{(n)}){\cal Z}_{m{-}r}(a_{s}^{(m)})
\eeq
and $\ZZ_n$ the cycle index polynomial
\beq
\label{cohae}
 {\cal Z}_{n}(x)=\frac{1}{2\pi i}\oint \frac{dw}{w^{n{+}1}} e^{\sum_{s=1}^{\infty}\frac{x}{s}w^{s}}
~.
\eeq

\subsection{4-point amplitudes}
\label{notation-4point}

Before gauge-fixing, the generic 4-point amplitude of string states can be extracted from the generating amplitude \cite{Firrotta:2024qel}
\beq
\label{not4aa}
{\cal A}_{4}(s,t)=g_o^2\int_{\DD_2} \prod_{\ell=1}^{4}dz_{\ell}\,\Big\langle V_{{\cal C}}(p_{1},z_{1})\,V_{\CC}(p_{2},z_{2})\,V_{\CC}(p_{3},z_{3})\,V_{{\cal C}}(p_{4},z_{4}) \Big\rangle
~,
\eeq
which is expressed as an integral over the worldsheet disc $\DD_2$, in terms of the coherent vertex operators
\beq
\label{not4ab}
V_{{\cal C}}(p_{j},z_{j})=\exp{\left(\sum_{n,m} \frac{\zeta_{n}^{(j)}{\cdot}\zeta_{m}^{(j)}}{2}{\cal S}_{n,m}e^{-i(n{+}m)q_{j}{\cdot}X}{+}\sum_{n}\zeta_{n}^{(j)}{\cdot}{\cal P}_{n}e^{-inq_{j}{\cdot}X}{+}ip_{j}{\cdot}X\right)}(z_{j})
~.
\eeq
$g_o$ denotes the open string coupling.

\begin{figure}[t!]
\centering
\includegraphics[scale=0.35]{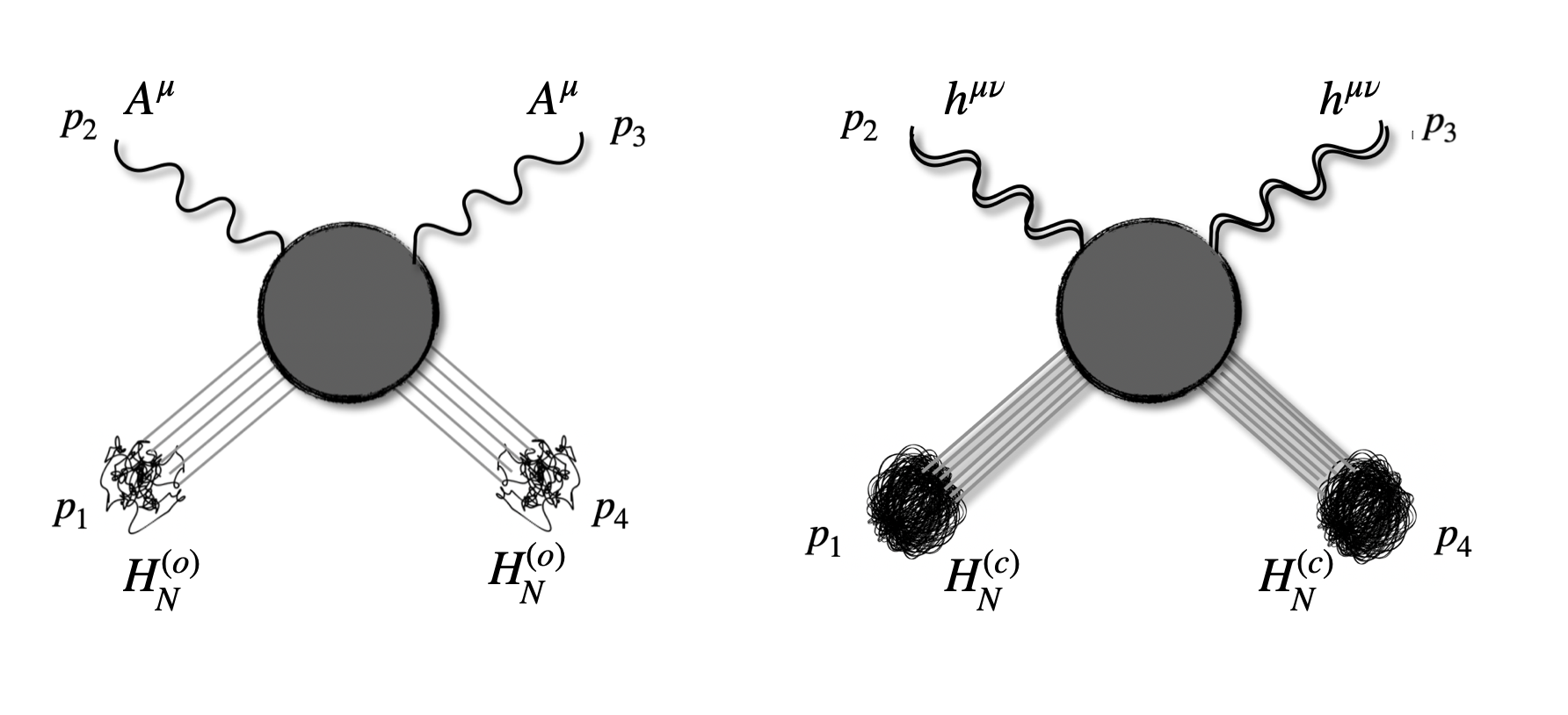}
\caption{Left figure: the amplitude of a photon scattering with an open HES state.
Right figure: the amplitude of a graviton/antisymmetric tensor/dilaton scattering with a closed HES state.}
\label{fig:4point}
\end{figure}

In what follows, we focus on 4-point amplitudes, Fig.\ \ref{fig:4point}, with 2 heavy (highly excited) string states (HES) and 2 light (tachyon or photon) states. We shall be referring to such amplitudes as Heavy-Heavy-Light-Light (HHLL).

\section{HHLL 4-point amplitudes}
\label{4point}

\subsection{Open strings}
\label{open1}

The 4-point amplitudes in this subsection involve two highly excited string states and two light states. All the states are open string states and we are keeping implicit the potential dependence on Chan-Paton indices. We are following the notation of \cite{Firrotta:2024qel}, where we refer the reader for additional details. Here we are reviewing the main elements of the construction for completeness.

On the worldsheet, the tachyon vertex operator is
\be
\label{open1aa}
V_{T}(p_{j},z_{j})=e^{ip_{j}{\cdot}X}(z_{j})
\ee
and the photon vertex operator
\be
\label{open1ab}
V_{\gamma}(p_j, z_j) = A\cdot i\partial X e^{i p_j \cdot X}(z_j)
~.
\ee
$A^\mu$ denotes the photon polarization.

For concreteness, we focus momentarily on the case of the tachyon. Then, the generating 4-point scattering amplitude \eqref{not4aa} becomes an integral over the disc of the form
\be
\label{open1ac}
{\cal A}_{gen}^{HHTT}(s,t)=g_o^2
\int_{{\DD}_2}
\prod_{\ell=1}^{4}dz_{\ell}\,\Big\langle V_{{\cal C}}(p_{1},z_{1})\,V_{T}(p_{2},z_{2})\,V_{T}(p_{3},z_{3})\,V_{{\cal C}}(p_{4},z_{4}) \Big\rangle
~.
\ee
The computation of this amplitude proceeds along the following lines.

First, we note that the world-sheet contractions
\beq
\label{open1ad}
\langle p_{j}{\cdot}X(z_{j})\,p_{\ell}{\cdot}X(z_{\ell})\rangle=-\,p_{j}{\cdot}p_{\ell}\log(z_{j\ell}) \,, \quad z_{j\ell}=z_{j}{-}z_{\ell}
\eeq
generate the standard Koba-Nielsen term
\beq
KN(\{z_{j}\})=z_{12}^{p_{1}{\cdot}p_{2}}z_{13}^{p_{1}{\cdot}p_{3}}z_{14}^{p_{1}{\cdot}p_4}z_{23}^{p_{2}{\cdot}p_{3}}z_{24}^{p_{2}{\cdot}p_{4}}z_{34}^{p_{3}{\cdot}p_{4}}
~.
\eeq
Assume for a moment that the amplitude in eq.\ \eqref{open1ac} involves excited string states at levels $N$ and $N'$ at the insertions $z_1$ and $z_4$, respectively, instead of the coherent vertex operators. Then, in terms of the Mandelstam variables $s,t,u$,
\beq
\label{open1ae}
(p_{3}{+}p_{4})^{2}=-s=(p_{1}{+}p_{2})^{2}\,,\,\,\,(p_{2}{+}p_{3})^{2}=-t=(p_{1}{+}p_{4})^{2}\,,\,\,\, (p_{1}{+}p_{3})^{2}=-u=(p_{2}{+}p_{4})^{2}\,,
\eeq
we would have the general kinematic relations $s+t+u=2M_{T}^{2}+M_{N}^{2}+M_{N'}^{2}$ with $\alpha' M_{N}^{2}=2(N{-}1)$
and $\alpha' M_{N'}^{2}=2(N'{-}1)$. The Koba-Nielsen term would then take the form
\beq
\label{open1af}
KN(\{z_{j}\})=\left(z_{12}z_{34}\over z_{13}z_{24} \right)^{-{s\over 2}{-}2}\left(z_{14}z_{23}\over z_{13}z_{24} \right)^{-{t\over 2}{-}2} (z_{13}z_{24})^{{-}2}\, \left({z_{34}z_{14}\over z_{13}} \right)^{N'}  \left({z_{12}z_{14}\over z_{24}} \right)^{N}
~.
\eeq
In the full amplitude (\ref{open1ac}), the last two factors in \eqref{open1af} will exponentiate giving rise to linear, $\OO(\zeta)$, and bilinear, $\OO(\zeta^2)$, contributions. Other contributions are related to the contractions
\beq
\label{open1ag}
\langle p_{j}{\cdot}X(z_{j})\,\zeta_{n}{\cdot}\partial X(z_{4}) \rangle={\zeta_{n}{\cdot}p_{j}\over z_{j4}}\,,\quad \langle \zeta_{n}^{(1)}{\cdot}X(z_{1})\,\zeta^{(4)}_{m}{\cdot}\partial X(z_{4}) \rangle={\zeta^{(1)}_{n}{\cdot}\zeta^{(4)}_{m}\over z_{14}^2}
~.
\eeq
The results of the Wick contractions are summarized in App.\ \ref{polynomials}.

At the end of the computation, the amplitude can be written in the form
\be
\label{open1ai}
{\cal A}_{gen}^{HHTT}(s,t)={\cal A}_{Ven}(s,t)\,e^{{\cal K}\left(\{\zeta^{(\ell)}_{n}\};\partial_{\beta_{s}},\partial_{\beta_{t}}\right)}\Phi_{\beta_{s},\beta_{t}}\left(s,t\right)\Big|_{\beta_{s,t}{=}0}
~,
\ee
where the exponential dressing factor involves the differential operator
\be
\label{Kfunca}
\begin{split}
{\cal K}\left(\{\zeta^{(\ell)}_{n}\}; \partial_{\beta_{s}},\partial_{\beta_{t}}\right)=&\sum_{\ell=1,4}\left(\sum_{n}\zeta_{n}^{(\ell)}{\cdot}V^{(\ell)}_{n}(\partial_{\beta_s},\d_{\beta_t}){+}\sum_{n,m} \zeta_{n}^{(\ell)}{\cdot}\zeta_{m}^{(m)}W_{n,m}^{(\ell)}(\partial_{\beta_s}, \d_{\beta_t})\right)\\
%%%
&\quad\quad+\sum_{n,m}\zeta^{(1)}_{n}{\cdot}\zeta^{(4)}_{m} I_{n,m}^{(1,4)}(\partial_{\beta_s}, \d_{\beta t})
\end{split}
\ee
acting on
\be
\label{ffunca}
\Phi_{\beta_{s},\beta_{t}}(s,t)=\sum_{r=0}^{\infty}\sum_{v=0}^{\infty}{\beta_{s}^{r}\over r!}{\beta_{t}^{v}\over v!}{(-\alpha's{-}1)_{r}(-\alpha't{-}1)_{v}\over (-\alpha's{-}\alpha't{-}2)_{r+v}}
~.
\ee
The specific form of the polynomials $V^{(\ell)}_n$, $W^{(\ell)}_{n,m}$, $I_{n,m}^{(1,4)}$ can be found in Eqs.\ \eqref{exprze1t}-\eqref{finI14t} in Appendix \ref{polynomials}. $(a)_n = \frac{\Gamma(a+n)}{\Gamma(a)}$ denotes the standard Pochhammer symbol.

The result in \eqref{open1ai} has a manifest Veneziano factor
\be
\label{open1aj}
{\cal A}_{Ven}(s,t)=g_{o}^{2}{\Gamma({-}\alpha's{-}1)\Gamma({-}\alpha't{-}1)\over \Gamma({-}\alpha's{-}\alpha't{-}2)}
\ee
that contains all the poles of the amplitude.
The dressing operator \eqref{Kfunca} encodes the non-trivial structure of the external HES states.

Specific HHTT amplitudes of excited string states can now be deduced from Eq.\ \eqref{open1ai} with suitable derivatives. For two states of the form \eqref{cohab} at levels $N$ and $N'$, respectively, one finds the expression
\be
\label{open1ak}
\begin{split}
&\AA_{H_N+T \rightarrow H_{N'}+T}(s,t) =
\\
&\epsilon^{(1)}_{(\mu_{1})_{g_{1}}(\mu_{2})_{g_{2}}\ldots} \prod_{n=1}^{N}{1\over \sqrt{n^{g_n}g_n!}}\prod_{r=1}^{g_n}{\partial \over \partial \lambda^{(1)}_{n \,\mu_{n}^{r}}}
\epsilon^{(4)}_{(\nu_{1})_{g'_{1}}(\nu_{2})_{g'_{2}}\ldots}\prod_{m=1}^{N'}{1\over \sqrt{m^{g'_m}g'_m!}}\prod_{v=1}^{g'_m}{\partial \over \partial \lambda^{(4)}_{m\,\nu_{n}^{v}}}{\cal A}_{gen}^{HHTT}(s,t)\Bigg|_{\lambda^{(1,4)}_n=0}
.
\end{split}
\ee
Below we present several examples where we calculate the systematics of four-point amplitudes.
We simplify the calculations by using a factorized ansatz  for the polarization tensors defined in \eqref{notad} and \eqref{RotPol}
\be
\label{polansatz1}
\epsilon_{(\mu_{1})_{g_{1}}(\mu_{2})_{g_{2}}\ldots} \rightarrow \lambda_{1 (\mu_{1}^{1}}...\lambda_{1\mu_{1}^{g_{1}})}
\lambda_{2 (\mu_{2}^{1}}...\lambda_{2 \mu_{2}^{g_{2}})}\cdots=\prod_{n}\prod_{r=1}^{g_{n}}\lambda_{n \mu_{n}^{g_{n}}}
\ee
\be
\label{polansatz2}
{\cal E}_{(\mu_{1})_{g_{1}}(\mu_{2})_{g_{2}}\ldots} \rightarrow \zeta_{1 (\mu_{1}^{1}}...\zeta_{1\mu_{1}^{g_{1}})}
\zeta_{2 (\mu_{2}^{1}}...\zeta_{2 \mu_{2}^{g_{2}})}\cdots=\prod_{n}\prod_{r=1}^{g_{n}}\zeta_{n \mu_{n}^{g_{n}}}
~.
\ee
We shall use this ansatz in the ensuing discussion of examples, but we note that the contractions with general polarization tensors, can be reintroduced at the end of every computation.

\subsubsection{Example: scattering with leading Regge states}

As an illustration, we apply \eqref{open1ak} to leading Regge states, specifically to equal mass states with $N=N'$ and $N$ single oscillators $\AA_{-1}$. This translates to the parameters $n_{1,4}{=}1$, $g_{n_{1,4}}=N$. In the polarization ansatz \eqref{polansatz2} we obtain the amplitude
\be
\label{ampReggeint}
{\cal A}_{H_N{+}T\rightarrow H_N{+}T}(s,t)={1\over \sqrt{N!}} \prod_{r=1}^N \left(\zeta_1^{(1,r)}{\cdot}{\d \over \d\zeta^{(1)}_{1}}\right)
{1\over \sqrt{N!}} \prod_{v=1}^N
\left(\zeta_1^{(4,v)}{\cdot}{\d\over \d\zeta^{(4)}_{1}}\right)
{\cal A}_{gen}^{HHTT}(s,t)\Bigg|_{\zeta_1^{(1,4)}=0}
~.
\ee

For instance, consider the more specific, simple case of $N=2$. Then, \eqref{ampReggeint} gives
\be\label{DApolb}
{1\over 2}\left(\zeta_1^{(1,1)}{\cdot}{\d \over\d\zeta^{(1)}_{1}}\right)\left(\zeta_1^{(1,2)}{\cdot}{\d \over\d\zeta^{(1)}_{1}}\right)
\left(\zeta_1^{(4,1)}{\cdot}{\d\over \d\zeta^{(4)}_{1}}\right)\left(\zeta_1^{(4,2)}{\cdot}{\d\over \d\zeta^{(4)}_{1}}\right)
{\cal A}_{gen}^{HHTT}(s,t)\Bigg|_{\zeta_1^{(1,4)}=0}
~.
\ee
Since leading Regge states are transverse, symmetric and traceless, one can simply consider polarizations of the form
\be
\label{DApola}
\zeta_{1\mu}^{(1,1)} = \zeta_{1\mu}^{(1,2)} := \zeta_{1\mu}^{(1)}~, ~~
\zeta_{1\mu}^{(4,1)} = \zeta_{1\mu}^{(4,2)} := \zeta_{1\mu}^{(4)}
\ee
and write the derivative action (\ref{DApolb}) as follows
\be\label{Dpolyb}
\begin{split}
&{\cal A}_{H_2{+}T\rightarrow H_2{+}T}(s,t)={\cal A}_{Ven}(s,t)\\
%%%
&\quad\quad{1\over 2}\Bigg(\left( \zeta_{1}^{(1)}{\cdot}V_{1}^{(1)}(\partial_{\beta_{s}},\partial_{\beta_{t}})\zeta_{1}^{(4)}{\cdot}V_{1}^{(4)}(\partial_{\beta_{s}},\partial_{\beta_{t}})\right)^{2} + 2\left(\zeta_{1}^{(1)}{\cdot}\zeta_{1}^{(4)}I_{1,1}^{(1,4)}(\partial_{\beta_{s}},\partial_{\beta_{t}})\right)^{2}  \\
%%%
&\quad \quad + 4\zeta_{1}^{(1)}{\cdot}\zeta_{1}^{(4)}I_{1,1}^{(1,4)}(\partial_{\beta_{s}},\partial_{\beta_{t}})\zeta_{1}^{(1)}{\cdot}V_{1}^{(1)}(\partial_{\beta_{s}},\partial_{\beta_{t}})\zeta_{1}^{(4)}{\cdot}V_{1}^{(4)}(\partial_{\beta_{s}},\partial_{\beta_{t}}) \Bigg)\Phi_{\beta_{s},\beta_{t}}(s,t)\Big|_{\beta_{s,t}=0}
~.
\end{split}
\ee
Using the definitions of the differential operators $V_1$, $I_{1,1}$ (Eqs.\ \eqref{exprze1td}-\eqref{finI14td} in Appendix \ref{polynomials}),
\be\label{expV1N2}
V^{(1)\mu}_{1}(\partial_{\beta_{s}},\partial_{\beta_{t}})=\sqrt{2\alpha'}p^{\mu}_{2}{+}\partial_{\beta_{s}}\sqrt{2\alpha'}p^{\mu}_{3}\,,\quad V^{(4)\mu}_{1}(\partial_{\beta_{s}},\partial_{\beta_{t}})=\sqrt{2\alpha'}p^{\mu}_{1}{+}\partial_{\beta_{t}}\sqrt{2\alpha'}p^{\mu}_{2}
\, ,
\ee
\be\label{expI14N2}
I^{(1,4)}_{1,1}(\partial_{\beta_{s}},\partial_{\beta_{t}})=\partial_{\beta_{s}}
~,
\ee
one can compute the amplitude by applying (\ref{Dpolyb}). The explicit final expression of this computation is reported in Eq.\ (\ref{eqD2}).

The amplitude with states of the leading Regge trajectory at generic level $N$ can be computed similarly by the action
\be\label{DApol}
{1\over N!}\left(\zeta_1^{(1,1)}{\cdot}{\d \over\d\zeta^{(1)}_{1}}\right)^N
\left(\zeta_1^{(4,1)}{\cdot}{\d\over \d\zeta^{(4)}_{1}}\right)^N
{\cal A}_{gen}^{HHTT}(s,t)\Bigg|_{\zeta_1^{(1,4)}=0}
\ee
that yields
\be
\label{laguerre2}
\begin{split}
{\cal A}_{H_N{+}T\rightarrow H_N{+}T}(s,t) =& {\cal A}_{Ven}(s,t) %N!
\left(  \zeta_1^{(1)} {\cdot} \zeta_1^{(4)} I_{1,1}^{(1,4)}\left(\partial_{\beta_s},\partial_{\beta_t}\right) \right)^{N}
\\
%%%
&L_{N}\left(-
{\zeta_1^{(1)} \cdot V^{(1)}_{1}(\partial_{\beta_s},\partial_{\beta_t}) \zeta_1^{(4)} \cdot V^{(4)}_{1}(\partial_{\beta_s},\partial_{\beta_t})
\over \zeta_1^{(1)} {\cdot} \zeta_1^{(4)} I_{1,1}^{(1,4)}(\partial_{\beta_s},\partial_{\beta_t})}
\right)
\Phi_{\beta_{s},\beta_{t}}(s,t)\Big|_{\beta_{s,t}=0}
~,
\end{split}
\ee
where $L_{N}^{(a)}(x)$ is the Laguerre polynomial
\be
L_{N}^{(a)}(x)=\sum_{r=0}^{N} \begin{pmatrix}N+a\\N-r\end{pmatrix}{(-x)^{r}\over r!}\,,\quad L_{N}^{(a{=}0)}(x) := L_{N}(x)
~.
\ee
Expanding (\ref{laguerre2}) and computing all the derivatives one can obtain a compact expression of the form
\be
\label{afterderivativesa}
\begin{split}
{\cal A}_{H_N{+}T\rightarrow H_N{+}T}(s,t)&= {\cal A}_{Ven}(s,t)
\sum_{r=0}^{N}\begin{pmatrix}N\\N{-}r\end{pmatrix}{1\over r!} (\zeta_{1}{\cdot}\zeta_{4})^{N-r}\sum_{k_{1}+k_{2}+k_{3}+k_4=r}{r!\over k_{1}!k_{2}!k_{3}!k_4!} \\
%%%
&(\zeta_1{\cdot}p_2\zeta_4{\cdot}p_1)^{k_1}(\zeta_1{\cdot}p_3\zeta_4{\cdot}p_1)^{k_2}(\zeta_1{\cdot}p_2\zeta_4{\cdot}p_2)^{k_3}(\zeta_1{\cdot}p_3\zeta_4{\cdot}p_2)^{k_4} {\cal Q}^{[s;N+k_2+k_4]}_{[t;k_3+k_4]}(s,t)
~,
\end{split}
\ee
where we defined
\be
\label{afterderivativeb}
{\cal Q}^{[s;a]}_{[t;b]}(s,t) := {(-\alpha's-1)_{a}(-\alpha't-1)_{b} \over (-\alpha's-\alpha't-2)_{a+b}}
~.
\ee
As a check, setting $N=2$ in \eqref{afterderivativeb} recovers Eq.\ \eqref{eqD2}.

This particular amplitude (with two leading Regge trajectory states and two tachyons) was first computed in Ref.\ \cite{Mitchell:1990cu} in a particular kinematical regime. Eq.\ \eqref{afterderivativesa} provides a new compact expression for general kinematics, that exhibits a characteristic factorization of the Veneziano amplitude and clarifies the pole and polarization structure of the amplitude.

\subsubsection{Other examples of HES scattering amplitudes}
\label{otherexamplesA}

In the appendices we include further examples of HES scattering amplitudes with tachyons and photons. Amplitudes with $J=1$ string states (at $N=2$ and general $N$) are reported in App.\ \ref{AppD2}. The generating function of amplitudes with photon states is summarized in App.\ \ref{AppC}. App.\ \ref{AppD3} contains results for the scattering of two leading Regge states and two photons.

\subsection{Closed strings}
\label{closed1}

The derivation of 4-point amplitudes in the closed string sector involves similar manipulations. The worldsheet computations are performed on the sphere, where the closed string insertions are characterized by holomorphic and anti-holomorphic vertex operators. The closed string tachyon vertex operator is
\be
\label{closT1aa}
V^{(c)}_{T}(p_{j},z_{j},\overline z_{j})=V_{T}(p_{j},z_{j})\overline V_{T}(p_{j},\overline z_{j})=e^{ip_{j}{\cdot}X}(z_{j},\overline z_{j})
\ee
and the massless closed string vertex operator (for the graviton, Kalb-Ramond antisymmetric tensor and dilaton)
\be
\label{closed1abb}
V_{\xi}(p_j, z_j,\overline z_{j})
=\xi_{\mu\nu} i\partial X^\mu i\overline\partial X^\nu e^{i p_j \cdot X}(z_j,\overline z_{j})
~,
\ee
where $z$ and $\bar z$ refer to the holomorphic and anti-holomorphic worldsheet coordinates respectively. Following the double-copy notation \cite{Kawai:1985xq}, it is convenient to set $\xi_{\mu\nu}= A_{\mu}\overline A_{\nu}$. At the end of the computation, one can replace the $A_{\mu}\overline A_{\nu}$ dependence with a more general $\xi_{\mu\nu}$ dependence.

Similarly, the closed string coherent vertex operator can be written in factorized form as
\be
V^{(c)}_{{\cal C}}(p_{j},z_{j},\overline z_{j})=V_{{\cal C}}(p_{j},z_{j})\overline V_{{\cal C}}(p_{j},\overline z_{j})
~,
\ee
with $V_\CC(p,z)$ as defined in \eqref{not4ab}. We use the convention where $p_L = p_R = p$. 
We should also remark that level matching, necessary in closed strings, will be implemented at a later stage.

In this language, the generating function of the 4-point closed string scattering amplitude with two tachyons and two generic excited string states, is given by the integrated correlation function on the sphere
\be
\label{Closed1ac}
{\cal M}_{gen}^{HHTT}(s,t,u)=g_c^2
\int_{{\SS}^2}
\prod_{\ell=1}^{4}d^{2}z_{\ell}\,\Big\langle V^{(c)}_{{\cal C}}(p_{1},z_{1},\overline z_{1})\,V^{(c)}_{T}(p_{2},z_{2},\overline z_{2})\,V^{(c)}_{T}(p_{3},z_{3},\overline z_{3})\,V^{(c)}_{{\cal C}}(p_{4},z_{4},\overline z_{4}) \Big\rangle
~.
\ee
$g_c$ denotes the closed string coupling. Collecting all the Wick contractions, as described in section (\ref{open1}), using the representation (\ref{formVen}) and the sphere integral
\be
\int_{\SS^{2}}d^{2}z \,z^{a}(1{-}z)^{b}\overline z^{\overline a}(1{-}\overline z)^{\overline b}=2\pi {\Gamma(1{+}a)\Gamma(1{+}b)\over \Gamma(2{+}a{+}b)}{\Gamma(-1{-}\overline a{-}\overline b)\over \Gamma(-\overline a)\Gamma(-\overline b)}
~,
\ee
the final expression of the generating function is
\be\label{ClosedGen}
{\cal M}_{gen}^{HHTT}(s,t,u)={\cal M}_{SV}(s,t,u)e^{{\cal K}\left(\{\zeta^{(\ell)}_{n}\}; \partial_{\beta_{s}},\partial_{\beta_{t}}\right)}\Phi_{\beta_{s},\beta_{t}}(s,t)\Big|_{\beta_{s,t}{=}0}e^{\overline{\cal K}\left(\{\overline\zeta^{(\ell)}_{n}\}; \partial_{\overline\beta_{s}},\partial_{\overline\beta_{t}}\right)}\overline\Phi_{\overline\beta_{s},\overline\beta_{t}}(s,t)\Big|_{\overline\beta_{s,t}{=}0}
.
\ee
The first term is the Shapiro-Virasoro factor
\be
{\cal M}_{SV}(s,t,u)=g_{c}^{2}{\Gamma\left(-{\alpha' s\over 4}{-}1\right)\Gamma\left(-{\alpha't\over 4}{-}1\right)\Gamma\left(-{\alpha'u\over 4}{-}1{+}\sum_{\ell}N_{\ell}\right)\over \Gamma\left({\alpha's\over 4}{+}2 \right)\Gamma\left({\alpha't\over 4}{+}2 \right)\Gamma\left({\alpha'u\over 4}{+}2{-}\sum_{\ell}N_{\ell} \right)}
\ee
where $N_{\ell}$ are the excitation levels of the four states. For the HES insertions at points 1 and 4, the levels are $N_{1}=N_{4}=N$, while for the tachyon insertions at points 2 and 3 we have $N_{2}=N_{3}=0$. The remaining terms in (\ref{ClosedGen}) provide the natural generalization of (\ref{open1ai}) to the closed string case. In particular, $\KK$ and $\overline{\KK}$ are the differential operators \eqref{Kfunca} and $\Phi$, $\overline{\Phi}$ the polynomials \eqref{ffunca} evaluated, respectively, on the left- and right-sector variables $(\beta_s,\beta_t)$ and $(\overline{\beta}_s, \overline{\beta}_t)$.

Any 4-point scattering amplitude with specific external states can now be obtained by extending (\ref{open1ak}) to closed string states as follows
\be\label{ClosStrfixst}
\begin{split}
&{\cal M}_{H_N+T \rightarrow H_{N'}+T}(s,t,u) =\\
%%%
&\qquad\epsilon^{(1)}_{(\mu_{1})_{g_{1}}(\mu_{2})_{g_{2}}\ldots (\nu_{1})_{\overline{g}_{1}}(\nu_{2})_{\overline{g}_{\overline 2}}\ldots}  \prod_{n=1}^{N}{1\over \sqrt{n^{g_n}g_n!}}\prod_{r=1}^{g_n}{\partial \over \partial \lambda^{(1)}_{n\mu_{n}^{r}}}\prod_{\overline n=1}^{\overline N}{1\over \sqrt{\overline n^{\overline g_{\overline n}}\overline g_{\overline n}!}}\prod_{r=1}^{\overline{g}_n}{\partial \over \partial \overline \lambda^{(1)}_{n\nu_{n}^{r}}}\\
%%%
&\qquad\epsilon^{(4)}_{(\rho_{1})_{g_{1}}(\rho_{2})_{g_{2}} \ldots (\sigma_{1})_{\overline{g}_{1}}(\sigma_{2})_{\overline{g}_{\overline 2}}\ldots }\prod_{m=1}^{N'}{1\over \sqrt{m^{g_m}g_m!}}\prod_{v=1}^{g_m}{\partial \over \partial \lambda^{(4)}_{m\rho_{m}^{v}}}
\prod_{\overline m=1}^{\overline N'}{1\over \sqrt{\overline m^{\overline g_{\overline m}}\overline g_{\overline m}!}}\prod_{v=1}^{\overline g_{\overline m}}{\partial \over \partial \overline \lambda^{(4)}_{m\sigma_{m}^{v}}}\\
%%%
&\hspace{5cm}{\cal M}_{gen}^{HHTT}(s,t,u)\Big|_{\lambda_n^{(1,4)},\overline \lambda_n^{(1,4)}=0}
~,
\end{split}
\ee
where the level matching condition is enforced by requiring
\be
N=\sum_{n}ng_{n}=\sum_{\overline n}\overline n \,\overline g_{\overline n}=\overline N
~.
\ee
In the case of two external massless closed string states, the generating function (\ref{ClosedGen}) is modified according to the following expression
\be
{\cal M}_{gen}^{HH\xi\xi}(s,t,u)={\cal F}(\xi_{2},\xi_{3};\partial_{\beta_{s,t}},\overline\partial_{\beta_{s,t}}){\cal M}_{gen}^{HHTT}(s,t,u)
~,
\ee
with a suitable dressing factor ${\cal F}(\xi_{2},\xi_{3};\partial_{\beta_{s,t}},\overline\partial_{\beta_{s,t}})$. Assuming $\xi_{j}^{\mu\nu}=A_{j}^{\mu}\overline A_{j}^{\nu}$, the latter can be written as
\be
{\cal F}(\xi_{2},\xi_{3};\partial_{\beta_{s,t}},\overline\partial_{\beta_{s,t}})={\cal F}(A_{2},A_{3};\partial_{\beta_{s,t}})\overline{\cal F}(\overline A_{2},\overline A_{3};\overline\partial_{\overline\beta_{s,t}})
~,
\ee
where ${\cal F}(A_{2},A_{3};\partial_{\beta_{s,t}})$ is the photon dressing function in Eq.\ (\ref{Dressphot}).

A recent discussion of 4-point string scattering amplitudes with four generic external states appeared in \cite{Firrotta:2024qel}.

\section{Absorption from an arbitrarily excited string state}
\label{absorb}

\begin{figure}[t!]
\centering
\includegraphics[scale=0.4]{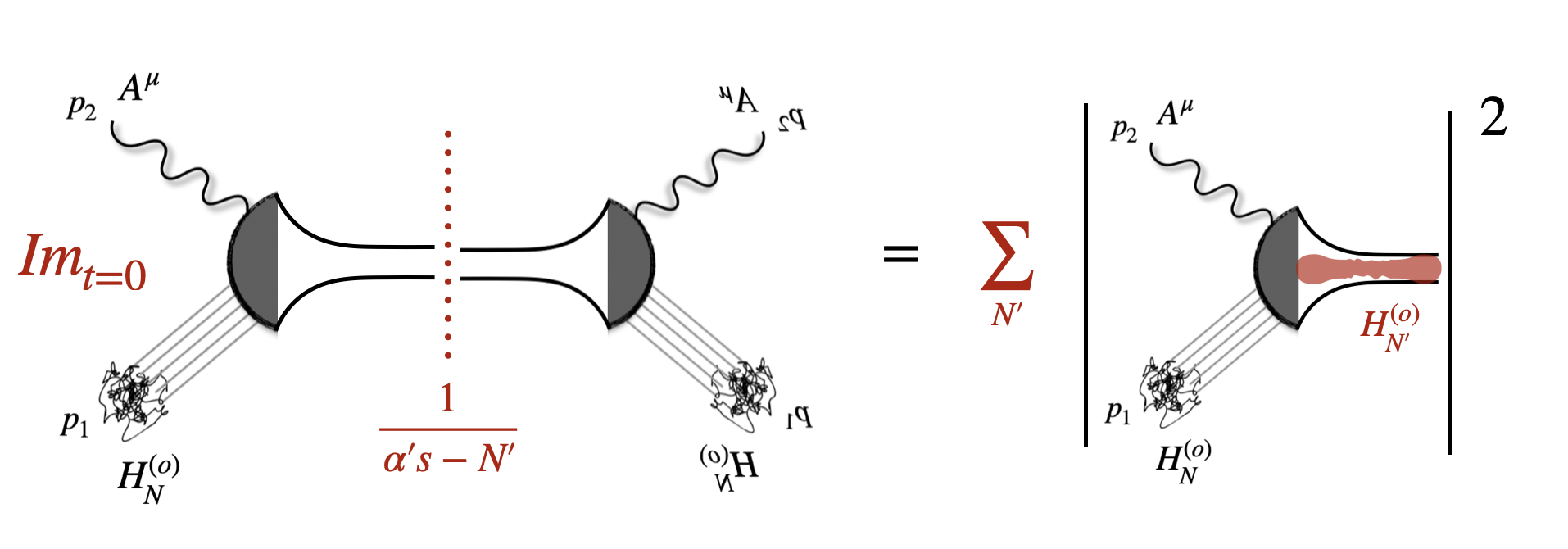}
\caption{A pictorial view of the amplitude squared, summed over final states, as a cut four-point amplitude. The four-point amplitude has forward kinematics and its final states are the complex conjugates of its initial states. We denote this pictorially by using mirrored labels of the final states.}
\label{fig:optTH}
\end{figure}

Absorption cross-sections of light states from excited string states can be obtained from the imaginary part of the forward limit of HHLL scattering amplitudes using the optical theorem (see Fig.\ \ref{fig:optTH}). This can be implemented at the level of the generating functions of the 4-point scattering amplitudes in the following manner.

We discuss separately the open and closed string cases.

\subsection{Open strings}
\label{open2}

We consider first the absorption of an open string light state from an {open} HES state. According to the optical theorem
\beq
\label{open2aa}
\sigma^{H_N+ L\rightarrow \text{any}\,H_N'}_{abs}(s) = \frac{\alpha'^{\frac{d-2}{2}}}{\alpha'F_{\phi}^{(HL)}}\im \AA_{H_N+L \to H_N+L}(s,t) \Big |_{t=0}
~,
\eeq
where $F_{\phi}^{(HL)}$ is the relative M\o{}ller flux factor of the initial heavy and light states. In the lab frame outlined in Appendix \ref{kinetach}
\be
\label{open2ab}
F_{\phi}^{(HL)}
=2 M_H \sqrt{E_{L}^{2}-M_{L}^{2}}
~,
\ee
with $M_{H,L}, E_{H,L}$ the masses and energies of the initial heavy and light states, respectively. When specialized to the light tachyon and photon states
\be
\label{open2ac}
\alpha'F_{\phi}^{(HT)}=2\sqrt{N{-}1} \sqrt{\alpha'E^{2}+1}\,,\quad \alpha'F_{\phi}^{(H\gamma)}=2\sqrt{N{-}1}\sqrt{\alpha'}\,\omega
~.
\ee

The forward limit of the imaginary part of the 4-point amplitude can be studied using the generating amplitude $\AA_{gen}^{HHLL}(s,t)$. Accordingly, we define
\be
\label{open2ad}
\widetilde\sigma^{H_N+L\rightarrow {\rm any}\,H_{N'}}_{abs,gen}(s)= \im \,{\cal A}_{gen}^{HHLL}(s,t)\Big|_{t=0}
~.
\ee

In the tachyon case,
\be
\label{absGenOp}
\widetilde\sigma^{H_N+ T\rightarrow {\rm any}\,H_{N'}}_{abs,gen}(s)= \im \,{\cal A}_{Ven}(s,t)\,e^{{\cal K}\left(\{\zeta^{(\ell)}_{n}\};\partial_{\beta_{s}},\partial_{\beta_{t}}\right)}\Phi_{\beta_{s},\beta_{t}}\left(s,t\right)\Big|_{\beta_{s,t},t{=}0}
~.
\ee
The imaginary part is contained in the Veneziano factor and can be extracted by selecting an $s$-channel pole
\be
{\cal A}_{Ven}(s,t)\Big|_{\alpha's=N'-1}=g_{o}^{2}{(-1)^{\alpha's}\over \Gamma(\alpha's{+}2)} {1\over \alpha' s - N'+1} {\Gamma(-\alpha't-1)\over \Gamma(-\alpha's-\alpha't-2)}
~.
\ee
Shifting the pole by an infinitesimal $i\epsilon$ parameter
\be
{1\over \alpha' s - N'+1+i\epsilon} = {\alpha' s - N'+1\over (\alpha' s - N'+1)^{2}+\epsilon^{2}}-{i\epsilon\over (\alpha' s - N'+1)^{2}+\epsilon^{2}}
~,
\ee
and taking the $\epsilon\rightarrow 0$ limit, leads to the desired imaginary part
\be
\im\, {\cal A}_{Ven}(s,t)=g_{o}^{2}\pi{\Gamma(3+\alpha's+\alpha't)\over \Gamma(2+\alpha's)\Gamma(2+\alpha't)}\delta(\alpha's-N'+1)
~.
\ee
The forward limit of (\ref{absGenOp}) contains three components:
\be
\label{immFven}
\im\, {\cal A}_{Ven}(s,t=0)=g_{o}^{2}\pi(2+\alpha's)\delta(\alpha's-N'+1)
~,
\ee
\be
{\cal K}\left(\{\zeta^{(\ell)}_{n}\};\partial_{\beta_{s}},\partial_{\beta_{t}}\right)\Big|_{t=0}:={\cal K}_{forw}\left(\{\zeta^{(\ell)}_{n}\};\partial_{\beta_{s}},\partial_{\beta_{t}}\right)
~,
\ee
(the latter following from the kinematics \ref{forwtachkine}), and finally
\be
\begin{split}
\Phi_{\beta_{s},\beta_{t}}(s,t{=}0)=\sum_{r,v=0}{\beta^{r}_{s}\over r!}{\beta_{t}^{v}\over v!}{(-\alpha's{-}1)_{r}({-}1)_{v}\over (-\alpha' s{-}2)_{r+v}}=\sum_{r=0}{\beta^{r}_{s}\over r!}\left(1-{r\over \alpha's{+}2}+ {\beta_{t}\over \alpha's{+}2} \right)
~.
\end{split}
\ee
The various terms, here summarised, are discussed and classified in appendix \ref{SForLimAmp}.

The combination of these contributions yields
\be
\widetilde\sigma^{H_N+ T\rightarrow {\rm any}\,H_{N'}}_{abs, gen}(s)=\pi g_{o}^{2}{(2{+}\alpha's)}e^{{\cal K}_{forw}\left(\{\zeta^{(\ell)}_{n}\};\partial_{\beta_{s}},\partial_{\beta_{t}}\right)}e^{\beta_{s}}\left(1-{\beta_{s}-\beta_{t}\over \alpha's{+}2}\right)\Big|_{\beta_{s,t}{=}0}
~.
\ee
As in the case of the generating function of scattering amplitudes (\ref{open1ak}), one can compute the absorption cross-section of a tachyon from a specific highly excited state using the formula
\be
\label{open1akCS}
\begin{split}
&\sigma_{abs}^{H_N+ T\rightarrow {\rm any}\,H_{N'}}(s)= \\
%%%
&\qquad\frac{\alpha'^{\frac{d-2}{2}}}{\alpha'F_{\phi}^{(HT)}}\epsilon_{(\mu_{1})_{g_{1}}(\mu_{2})_{g_{2}}\ldots}
\epsilon^{*}_{(\nu_{1})_{g_{1}}(\nu_{2})_{g_{2}}\ldots}
\prod_{n=1}^{N}{1\over n^{g_n}g_n!}\prod_{r=1}^{g_n}{\partial \over \partial \lambda^{(1)}_{n\mu_{n}^{r}}}
{\partial \over \partial \lambda^{(4)}_{n\nu_{n}^{r}}}\widetilde{\sigma}_{abs,gen}^{H_N+T\rightarrow H_{N'}}(s)\Big|_{\lambda^{(1,4)}_n=0}
.
\end{split}
\ee
where as a consequence of the forward limit
\be
\epsilon^{(4)}_{(\nu_{1})_{g_{1}}(\nu_{2})_{g_{2}}\ldots}=\epsilon^{*}_{(\nu_{1})_{g_{1}}(\nu_{2})_{g_{2}}\ldots}\,,\quad \epsilon^{(1)}_{(\mu_{1})_{g_{1}}(\mu_{2})_{g_{2}}\ldots}=\epsilon_{(\mu_{1})_{g_{1}}(\mu_{2})_{g_{2}}\ldots}\,.
\ee

Similarly, in the case of photon absorption, we can use the generating amplitude \eqref{AgenPhot} to obtain
\be\label{finCStach}
\widetilde\sigma_{abs,gen}^{H_N+\gamma\rightarrow {\rm any}\,H_N'}(s)=\pi g_{o}^{2}{(2{+}\alpha's)}\,e^{{\cal K}_{forw}\left(\{\zeta^{(\ell)}_{n}\};\partial_{\beta_{s}},\partial_{\beta_{t}}\right)}e^{\beta_{s}}\left(1-{\beta_{s}{-}\beta_{t}{+}1\over \alpha's{+}2}\right)\Big|_{\beta_{s,t}{=}0}
~.
\ee

\subsection{Closed strings}
\label{closed2}

For closed strings, the analogous formula for absorption is
\beq
\label{closed2aa}
\Sigma^{H_N+L \rightarrow {\rm any}\,H_{N'}}_{abs}(s) = \frac{\alpha'^{\frac{d-2}{2}}}{\alpha'F_{\phi}^{(HL)}}\im \MM_{H_N+L \to H_N+L}(s,t) \Big |_{t=0}
~.
\eeq
Using the generating function of 4-point amplitudes \eqref{ClosedGen}, we can define the forward limit of the imaginary part as
\be\label{ClosedGenCS}
\widetilde\Sigma_{abs,gen}^{H_N +L\rightarrow {\rm any}\,H_{N'}}(s)={\im\,{\cal M}_{gen}^{HHLL}(s,t,u)}\Big|_{t=0}
~.
\ee
For tachyon states, one finds
\be
\begin{split}
\widetilde\Sigma_{abs,gen}^{H+T\rightarrow {\rm any}\,H_{N'}}(s)=\pi g_{c}^{2}{(2{+}\alpha's)^{2}}&e^{{\cal K}_{forw}\left(\{\zeta^{(\ell)}_{n}\};\partial_{\beta_{s}},\partial_{\beta_{t}}\right)}e^{\beta_{s}}\left(1-{\beta_{s}-\beta_{t}\over \alpha's{+}2}\right)\Big|_{\beta_{s,t}{=}0}\\
%%%
\times~ &e^{\overline{\cal K}_{forw}\left(\{\overline\zeta^{(\ell)}_{\overline n}\};\overline\partial_{\overline\beta_{s}},\overline\partial_{\overline\beta_{t}}\right)}e^{\overline\beta_{s}}\left(1-{\overline\beta_{s}-\overline\beta_{t}\over \alpha's{+}2}\right)\Big|_{\overline\beta_{s,t}{=}0}
~.
\end{split}
\ee
Similarly, for massless closed string states (that we henceforth denote collectively as $\xi$, see (\ref{closed1abb})),
\be
\begin{split}
\widetilde\Sigma_{abs,gen}^{H_N+\xi\rightarrow {\rm any}\,H_{N'}}(s)=\pi g_{c}^{2}{(2{+}\alpha's)^{2}}&e^{{\cal K}_{forw}\left(\{\zeta^{(\ell)}_{n}\};\partial_{\beta_{s}},\partial_{\beta_{t}}\right)}e^{\beta_{s}}\left(1-{\beta_{s}{-}\beta_{t}{+}1\over \alpha's{+}2}\right)\Big|_{\beta_{s,t}{=}0}\\
%%%
\times~ &e^{\overline{\cal K}_{forw}\left(\{\overline\zeta^{(\ell)}_{\overline n}\};\overline\partial_{\overline\beta_{s}},\overline\partial_{\overline\beta_{t}}\right)}e^{\overline\beta_{s}}\left(1-{\beta_{s}{-}\beta_{t}{+}1\over \alpha's{+}2}\right)\Big|_{\beta_{s,t}{=}0}
~.
\end{split}
\ee
Note that for massless open and closed string states, the flux factor is the same, $F_{\phi}^{(H\xi)}=F_{\phi}^{(H\gamma)}$.

\subsection{Absorption universality}
\label{absorbexamples}

\begin{figure}[t!]
\centering
\includegraphics[scale=0.35]{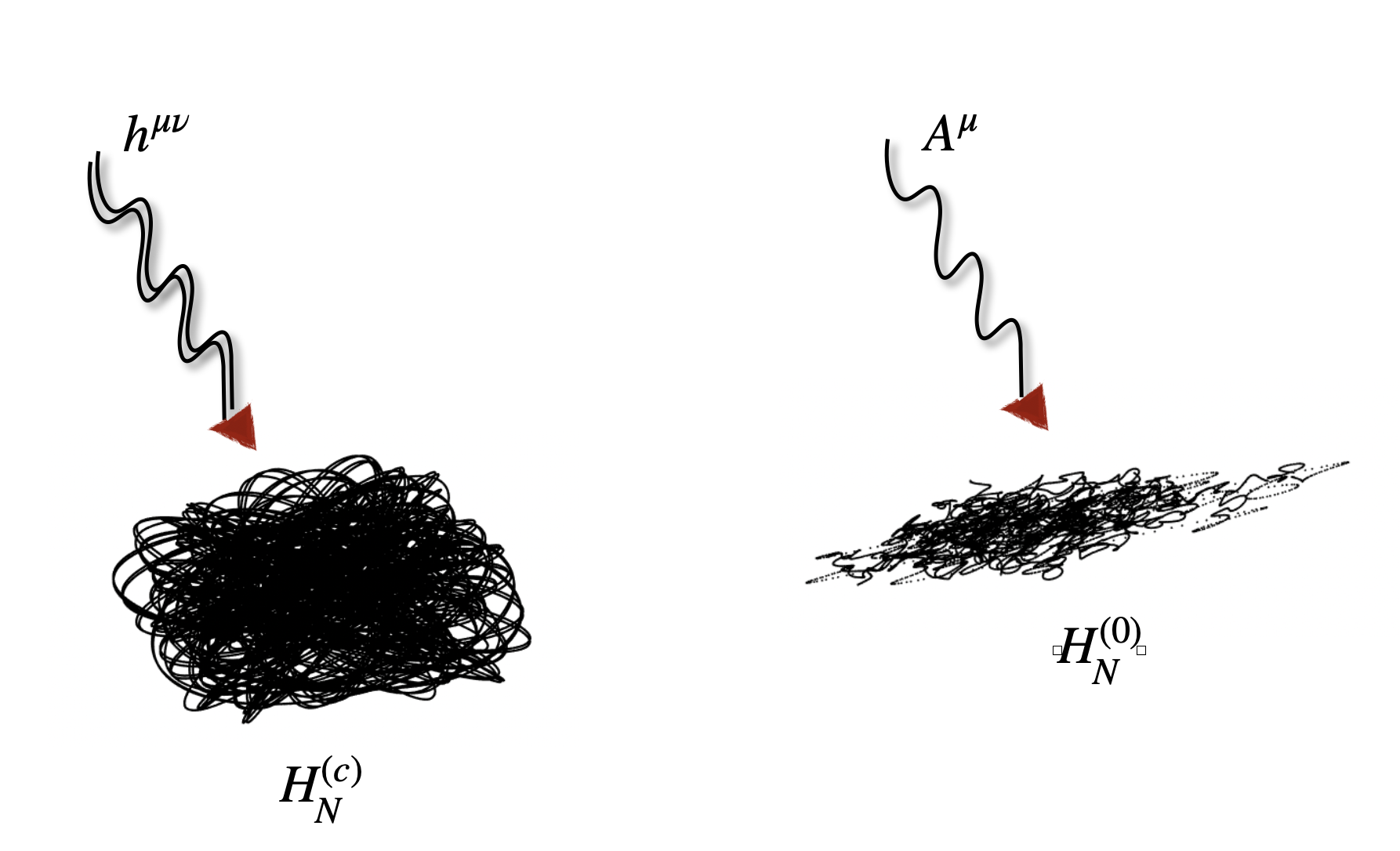}
\caption{From left to right, graviton/antisymmetric tensor/dilaton and photon absorption from open and closed HES respectively.}
\label{fig:abslight}
\end{figure}

We now show that the absorption cross section, Fig.\ \ref{fig:abslight}, is independent of the details (eg.  the spin) of the heavy states. We first discuss a few simple explicit cases and then generalize to the generic heavy state.

\subsubsection{Example: Absorption from leading Regge states }
\label{absorbregge}

We first  consider the absorption of a tachyon from a leading Regge state of level $N=2$. Then, \eqref{open1akCS} takes the following form
\be
\sigma_{abs}^{H_2+T\rightarrow {\rm any}\,H' }(s) =
{\alpha'^{\frac{d-2}{2}}\over 2\alpha'F_{\phi}^{(HT)}}\epsilon_{(\mu_{1}^{1}\mu_{1}^{2})}{\partial \over \partial \lambda^{(1)}_{n\mu_{1}^{1}}}{\partial \over \partial \lambda^{(1)}_{n\mu_{1}^{2}}}
\epsilon^{*}_{(\nu_{1}^{1}\nu_{1}^{2})}{\partial \over \partial \lambda^{(4)}_{n\nu_{1}^{1}}}{\partial \over \partial \lambda^{(4)}_{n\nu_{1}^{2}}}\widetilde{\sigma}_{abs,gen}^{(H_N+T\rightarrow H_{N'})}(s)\Bigg|_{\zeta^{(1,4)}_n=0}
~,
\ee
which leads to
\be
\begin{split}
&\sigma_{abs}^{H_2+T\rightarrow {\rm any}\,H'}(s)=\\
%%%
&\qquad\pi \alpha'^{\frac{d-2}{2}}g_{o}^{2}{(2{+}\alpha's)\over \alpha'F_{\phi}^{(HT)}}{1\over 2}\left(\Tr({\cal E}^{T}{\cal E}^{*}){+}\Tr({\cal E}\,{\cal E}^{*})\right)\left(I_{1,1}^{(1,4)}(\partial_{\beta_{s}},\partial_{\beta_{t}})\right)^{2}e^{\beta_{s}}\left(1-{\beta_{s}{-}\beta_{t}\over \alpha's{+}2}\right)\Big|_{\beta_{s,t}{=}0}\,.
\end{split}
\ee
where the tensors ${\cal E}$ were defined in (\ref{RotPol}).

Using \eqref{expI14N2} one can compute the derivative action with respect to the $\beta_{s}$ parameter finding
\be\label{protobohh}
\begin{split}
\sigma_{abs}^{H_2+T\rightarrow {\rm any}\,H'}(s)&={1\over 2}{\left(\Tr({\cal E}^{T}{\cal E}^{*}){+}\Tr({\cal E}\,{\cal E}^{*})\right)}\pi \alpha'^{\frac{d-2}{2}}g_{o}^{2}{(2{+}\alpha's)\over \alpha'F_{\phi}^{(HT)}}\left(1-{2\over \alpha's+2}\right)\\
&=\pi \alpha'^{\frac{d-2}{2}}g_{o}^{2} \frac{\sqrt{\alpha'}E}{\sqrt{1+\alpha' E^2}}
~,
\end{split}
\ee
where in the last equality we implemented the normalization conditions $\Tr({\cal E} {\cal E}^{*})=\Tr({\cal E}^{T} {\cal E}^{*})=1$ and substituted the value of the flux factor from (\ref{open2ac}).

The generalization to a leading Regge state at arbitrary level $N$ follows from the general expression \eqref{open1akCS}
\be
\begin{split}
\sigma_{abs}^{H_N+T\rightarrow {\rm any}\,H_{N'}}(s)=\pi \alpha'^{\frac{d-2}{2}}g_{o}^{2}{(2{+}\alpha's)\over \alpha'F_{\phi}^{H,T}}\left(I_{1,1}^{(1,4)}(\partial_{\beta_{s}},\partial_{\beta_{t}})\right)^{N}e^{\beta_{s}}\left(1-{\beta_{s}{-}\beta_{t}\over \alpha's{+}2}\right)\Big|_{\beta_{s,t}{=}0}\, ,
\end{split}
\ee
by noting that the contribution of the polarization tensors is precisely canceled by the normalization factor, according to the identity
\be
{\cal E}_{(\mu_{1}^{1}.....\mu_{1}^{N})}{\cal E}^{*}_{(\nu_{1}^{1}.....\nu_{1}^{N})}\left(\delta^{\mu_{1}^{1}\nu_{1}^{1}}...\delta^{\mu_{1}^{N}\nu_{1}^{N}}+ {\rm perm}_{N}(\mu^{k}_{1},\nu^{k}_{1})\right)=N!\, \Tr({\cal E}{\cal E}^{*})=N! \, .
\ee
Taking the derivatives with respect to $\beta_{s}$ produces the factor
\be
1{-}{N\over \alpha's{+}2}
~,
\ee
which combines with the prefactor to give the final result
\be\label{LreggCSop}
\sigma_{abs}^{H_N+T\rightarrow {\rm any} \,H_{N'}}(s)=\pi \alpha'^{\frac{d-2}{2}}g_{o}^{2}{(2{+}\alpha's{-}N)\over \alpha'F_{\phi}^{(HT)}}
~.
\ee

As a check, we note that the same result can be obtained from \eqref{afterderivativesa},\footnote{One can check that in the forward-limit kinematics, \eqref{forwtachkine}, the only non-zero term arises from $r=0$.} which involves
\be
{\cal Q}^{[s;N]}_{t;0}(s,t)\Big|_{t=0}={(-\alpha's-1)_{N}(-1)_{0}\over (-\alpha's-2)_{N}}=1-{N\over \alpha's+2}.
\ee
In this approach, we computed the absorption cross-section starting from the computation of the full 4-point amplitude and then implemented the forward limit of the imaginary part. This was possible because of the relative simplicity of the 4-point amplitude for leading Regge states. In more general situations, however, this approach would have been very difficult to implement. In contrast, the expression \eqref{finCStach} allows us to carry out the computation of the absorption cross-section without these technical difficulties, as we shall see in the next subsection.

\subsubsection{Absorption of light states from generic states}

Following the previous comment, we shall now  compute the absorption cross-section for generic heavy states. As a warmup, we consider first a state with level $N=3$ and two modes, $n=1$ and $n=2$. This state includes features (like the $W_{n,m}^{(\ell)}$ polynomials) that are absent in the leading Regge state computation.

Employing \eqref{open1akCS} we obtain
\be
\sigma_{abs}^{H_3+T\rightarrow {\rm any}\,H'}(s) =
{\alpha'^{\frac{d-2}{2}}\over 2 \alpha' F_{\phi}^{(HT)}}\epsilon_{\mu_{1}^{1}\mu_{2}^{1}}{\partial \over \partial \lambda^{(1)}_{1 \mu_{1}^{1}}}{\partial \over \partial \lambda^{(1)}_{2\mu_{2}^{1}}}\epsilon^{*}_{\nu_{1}^{1}\nu_{2}^{1}}
{\partial \over \partial \zeta^{(4)}_{1\nu_{1}^{1}}}{\partial \over \partial \zeta^{(4)}_{2\nu_{2}^{1}}}\widetilde{\sigma}_{abs,gen}^{(H_N+T\rightarrow H_{N'})}(s)\Bigg|_{\zeta^{(1)}_n,\zeta^{(4)}_n=0}
\ee
which leads to
\be\label{CSN3}
\begin{split}
\sigma_{abs}^{H_3+T\rightarrow {\rm any}\, H'}(s) =&\pi \alpha'^{\frac{d-2}{2}}g_{o}^{2}{(2{+}\alpha's)\over 2\alpha'F_{\phi}^{(HT)}}{\cal E}_{\mu_{1}^{1}\mu_{2}^{1}}{\cal E}^{*}_{\nu_{1}^{1}\nu_{2}^{1}}
\Big(\delta^{\mu_{1}^{1}\nu_{1}^{1}}\delta^{\mu_{2}^{1}\nu_{2}^{1}}I_{1,1}^{(1,4)}(\partial_{\beta_{s}},\partial_{\beta_{t}})I_{2,2}^{(1,4)}(\partial_{\beta_{s}},\partial_{\beta_{t}})\\
%%%
&+\delta^{\mu_{1}^{1}\nu_{2}^{1}}\delta^{\mu_{2}^{1}\nu_{1}^{1}}I_{1,2}^{(1,4)}(\partial_{\beta_{s}},\partial_{\beta_{t}})I_{2,1}^{(1,4)}(\partial_{\beta_{s}},\partial_{\beta_{t}})\\
%%%
& + \delta^{\mu_{1}^{1}\mu_{2}^{1}}\delta^{\nu_{1}^{1}\nu_{2}^{1}}W_{1,2}^{(1)}(\partial_{\beta_{s}},\partial_{\beta_{t}})W_{1,2}^{(4)}(\partial_{\beta_{s}},\partial_{\beta_{t}})\Big)e^{\beta_{s}}\left(1-{\beta_{s}-\beta_{t}\over \alpha's{+}2}\right)\Big|_{\beta_{s,t}{=}0}
\end{split}
\ee
with polynomial differential operators
\be
I_{1,1}^{(1,4)}(\partial_{\beta_{s}},\partial_{\beta_{t}})=\partial_{\beta_{s}}
\ee
\be
I_{1,2}^{(1,4)}(\partial_{\beta_{s}},\partial_{\beta_{t}})=-\partial_{\beta_{s}}\left(\partial_{\beta_{s}}+2\alpha'q_{4}{\cdot}p_{1}\partial_{\beta_{s}}+\partial_{\beta_{t}}-2\alpha'q_{4}{\cdot}p_{3}\partial_{\beta_{t}} \right)
\ee
\be
I_{2,1}^{(1,4)}(\partial_{\beta_{s}},\partial_{\beta_{t}})=2\partial_{\beta_{s}}\left( \partial_{\beta_{s}}{+}2\alpha'q_{1}{\cdot}p_{4}\partial_{\beta_{s}}-2\alpha'q_{1}{\cdot}p_{2}\partial_{\beta_{t}} \right)
\ee
\be
I_{2,2}^{(1,4)}(\partial_{\beta_{s}},\partial_{\beta_{t}})=-\partial_{\beta_{s}}\Big(2\partial_{\beta_{s}}{+}4\left(\partial_{\beta_{s}}{+}2\alpha'q_{1}{\cdot}p_{4}\partial_{\beta_{s}}{-}2\alpha'q_{1}{\cdot}p_{2}\partial_{\beta_{t}}\right)\left(\partial_{\beta_{s}}{+}2\alpha'q_{4}{\cdot}p_{1}\partial_{\beta_{s}}{-}2\alpha'q_{4}{\cdot}p_{3}\partial_{\beta_{t}}\right)\Big)
\ee
\be
W_{1,1}^{(1)}(\partial_{\beta_{s}},\partial_{\beta_{t}})={1\over 2}(1{+}2\alpha'q_{1}{\cdot}p_{4})2\alpha'q_{1}{\cdot}p_{4}\partial_{\beta_{s}}^{2}+{1\over 2}(1{+}2\alpha'q_{1}{\cdot}p_{2})2\alpha'q_{1}{\cdot}p_{2}\partial_{\beta_{t}}^{2}-2\alpha'q_{1}{\cdot}p_{2}2\alpha'q_{1}{\cdot}p_{4}\partial_{\beta_{s}}\partial_{\beta_{t}}
\ee
\be
W_{1,1}^{(4)}(\partial_{\beta_{s}},\partial_{\beta_{t}})={1\over 2}(1{+}2\alpha'q_{4}{\cdot}p_{1})2\alpha'q_{4}{\cdot}p_{1}\partial_{\beta_{s}}^{2}+{1\over 2}(1{+}2\alpha'q_{4}{\cdot}p_{3})2\alpha'q_{4}{\cdot}p_{3}\partial_{\beta_{t}}^{2}-2\alpha'q_{4}{\cdot}p_{1}2\alpha'q_{4}{\cdot}p_{3}\partial_{\beta_{s}}\partial_{\beta_{t}}
\ee
\be
\begin{split}
W_{1,2}^{(1)}(\partial_{\beta_{s}},\partial_{\beta_{t}})&={2\over 3}2\alpha'q_{1}{\cdot}p_{4}(1{+}2\alpha'q_{1}{\cdot}p_{4})(1{+}4\alpha'q_{1}{\cdot}p_{4})\partial_{\beta_{s}}^{3}-3\alpha'q_{1}{\cdot}p_{2}2\alpha'q_{1}{\cdot}p_{4}(1{+}4\alpha'q_{1}{\cdot}p_{4})\partial_{\beta_{s}}^{2}\partial_{\beta_{t}}\\
%%%
&+3\alpha'q_{1}{\cdot}p_{2}2\alpha'q_{1}{\cdot}p_{4}(1{+}4\alpha'q_{1}{\cdot}p_{2})\partial_{\beta_{s}}\partial_{\beta_{t}}^{2}-{2\over 3}2\alpha'q_{1}{\cdot}p_{2}(1{+}2\alpha'q_{1}{\cdot}p_{2})(1{+}4\alpha'q_{1}{\cdot}p_{2})\partial_{\beta_{t}}^{3}
\end{split}
\ee
\be
\begin{split}
W_{1,2}^{(4)}(\partial_{\beta_{s}},\partial_{\beta_{t}})&={2\over 3}2\alpha'q_{4}{\cdot}p_{3}(1{+}2\alpha'q_{4}{\cdot}p_{3})(1{+}4\alpha'q_{4}{\cdot}p_{3})\partial_{\beta_{t}}^{3}-3\alpha'q_{4}{\cdot}p_{1}2\alpha'q_{4}{\cdot}p_{3}(1{+}4\alpha'q_{4}{\cdot}p_{3})\partial_{\beta_{t}}^{2}\partial_{\beta_{s}}\\
%%%
&+3\alpha'q_{4}{\cdot}p_{1}2\alpha'q_{4}{\cdot}p_{3}(1{+}4\alpha'q_{4}{\cdot}p_{1})\partial_{\beta_{t}}\partial_{\beta_{s}}^{2}-{2\over 3}2\alpha'q_{4}{\cdot}p_{1}(1{+}2\alpha'q_{4}{\cdot}p_{1})(1{+}4\alpha'q_{4}{\cdot}p_{1})\partial_{\beta_{s}}^{3}
\, .
\end{split}
\ee
Since the derivative polynomials in \eqref{CSN3} act on a function which is linear in $\beta_{t}$, all the derivative terms with degree greater than one do not contribute. This simplifying observation, together with the kinematics \eqref{forwtachkine}, imply that the only non zero-term in \eqref{CSN3} is given by
\be\label{remT}
I_{1,1}^{(1,4)}(\partial_{\beta_{s}},\partial_{\beta_{t}})I_{2,2}^{(1,4)}(\partial_{\beta_{s}},\partial_{\beta_{t}})e^{\beta_{s}}\left(1-{\beta_{s}-\beta_{t}\over \alpha's{+}2}\right)\Big|_{\beta_{s,t}{=}0}=-2\partial_{\beta_{s}}^{3}e^{\beta_{s}}\left(1-{\beta_{s}-\beta_{t}\over \alpha's{+}2}\right)\Big|_{\beta_{s,t}{=}0}
~.
\ee
All the remaining terms start from $\partial_{\beta_{t}}^{2}$. Consequently, the final result for the absorption cross-section can be obtained by inserting \eqref{remT} into \eqref{CSN3} to obtain
\be
\sigma_{abs}^{H_3+T\rightarrow {\rm any}\,H'}(s) =\pi \alpha'^{\frac{d-2}{2}}g_{o}^{2}\, \Tr({\cal E}^{T}{\cal E}^{*}){(\alpha's{-}1)\over \alpha'F_{\phi}^{(HT)}}=\pi \alpha'^{\frac{d-2}{2}}g_{o}^{2}{(\alpha's{-}1)\over \alpha'F_{\phi}^{(HT)}}\,.
\ee
This form is identical to the absorption cross-section of a leading Regge state, \eqref{LreggCSop}, at level $N=3$. In fact, as we now show, the result is more generally independent of the details of the heavy state.

For a generic state of the form \eqref{notad}, the tachyon absorption cross-section takes the form
\be
\begin{split}
{\sigma_{abs}^{H_N+T\rightarrow {\rm any}\,H_{N'}}(s)} =\, & \pi \alpha'^{\frac{d-2}{2}}g_{o}^{2} {(2{+}\alpha's)\over \alpha'F_{\phi}^{(HT)}}
{\cal E}_{(\mu_{1})_{g_{1}}(\mu_{2})_{g_{2}}\ldots}{\cal E}^{*}_{(\nu_{1})_{g_{1}}(\nu_{2})_{g_{2}}\ldots}
%\left(\prod_{n}{1\over n^{g_{n}}g_{n}!}\right)
\\
%%%
&
{\rm Poly}^{(\mu_{1})_{g_{1}}(\mu_{2})_{g_{2}}\ldots (\nu_{1})_{g_{1}}(\nu_{2})_{g_{2}}\ldots} ~
e^{\beta_{s}}\left(1{-}{\beta_{s}{-}\beta_{t}\over \alpha's{+}2}\right)\Big|_{\beta_{s,t}{=}0}
~,
\end{split}
\ee
where
\be
\label{polydefa}
{\rm Poly}^{(\mu_{1})_{g_{1}}(\mu_{2})_{g_{2}}\ldots (\nu_{1})_{g_{1}}(\nu_{2})_{g_{2}}\ldots} =
\prod_{n}{1\over n^{g_{n}}g_{n}!}
\prod_{r=1}^{g_n} \frac{\d}{\d\lambda_{n\mu_n^r}} \frac{\d}{\d\lambda_{n\nu_n^r} }
e^{\KK_{forw}\left(\{ \zeta_n^{(\ell)};\d_{\beta_s}, \d_{\beta_t} \right)}
\ee
is a polynomial of the $\beta$-derivatives that involves products of the polynomials $I$, $W$, in analogy to the expressions in \eqref{CSN3}. From the explicit expressions \eqref{finWz1td}-\eqref{finI14td}, one can isolate the non-zero contributions noting that pairs of polynomials of the form
\beq
I_{i,j}^{(1,4)}I_{j,i}^{(1,4)}\propto {\rm Poly}[\partial_{\beta_{s}},\partial_{\beta_{t}}]\,\partial_{\beta_{t}}^{2}\,, \quad {\rm for}~ i\neq j
\eeq
and
\be
W^{(1)}_{i,j}W^{(4)}_{k,\ell}\propto {\rm Poly}[\partial_{\beta_{s}},\partial_{\beta_{t}}]\,\partial_{\beta_{t}}^{2}
\ee
start with a second derivative with respect to $\beta_{t}$, leading to zero contribution to the absorption cross-section. The only term of the polynomial in \eqref{polydefa} that contributes to the absorption cross section is given by
\be
\prod_{n}\left(\prod_{r=1}^{g_{n}}\delta^{\mu^{r}_{n}\nu^{r}_{n}}\right)\left(I^{(1,4)}_{n,n}(\partial_{\beta_{s}},\partial_{\beta_{t}})\right)^{g_{n}}
\subset
{\rm Poly}^{(\mu_{1})_{g_{1}}(\mu_{2})_{g_{2}}\ldots (\nu_{1})_{g_{1}}(\nu_{2})_{g_{2}}\ldots}
~,
\ee
leading to the expression
\be\label{IntAbsSt}
\begin{split}
\sigma_{abs}^{H_N+T\rightarrow {\rm any}\,H_{N'}}(s)=& \pi \alpha'^{\frac{d-2}{2}}g_{o}^{2} {(2{+}\alpha's)\over \alpha'F_{\phi}^{(HT)}}
{\cal E}_{(\mu_{1})_{g_{1}}(\mu_{2})_{g_{2}}\ldots}{\cal E}^{*}_{(\nu_{1})_{g_{1}}(\nu_{2})_{g_{2}}\ldots}
\left(\prod_{n}{1\over n^{g_{n}}g_{n}!}\right)\\
&\prod_{n}\left(\prod_{r=1}^{g_{n}}\delta^{\mu^{r}_{n}\nu^{r}_{n}}\right)\left(I^{(1,4)}_{n,n}(\partial_{\beta_{s}},\partial_{\beta_{t}})\right)^{g_{n}}e^{\beta_{s}}\left(1{-}{\beta_{s}{-}\beta_{t}\over \alpha's{+}2}\right)\Big|_{\beta_{s,t}}~.
\end{split}
\ee

From the tensorial structure one obtains
\be\label{TensContr}
{\cal E}_{(\mu_{1})_{g_{1}}(\mu_{2})_{g_{2}}\ldots}{\cal E}^{*}_{(\nu_{1})_{g_{1}}(\nu_{2})_{g_{2}}\ldots}\prod_{n}\left(\prod_{r=1}^{g_{n}}\delta^{\mu^{r}_{n}\nu^{r}_{n}}\right)=\Tr({\cal E}^{T}{\cal E}^{*})\prod_{n}g_{n}!
\ee
which includes a multiplicity factor of $g_{n}!$ for any set of $g_{n}$ symmetric indices.
On the other hand, the derivative polynomial $I^{(1,4)}_{n,n}(\partial_{\beta_{s}},\partial_{\beta_{t}})$, defined in \ref{finI14td}, takes the following form in the forward limit \eqref{forwtachkine}
\be
I^{(1,4)}_{n,n}(\partial_{\beta_{s}},\partial_{\beta_{t}})=n\, \partial_{\beta_{s}}^{n}+\partial_{\beta_{t}}^{2} {\rm Poly}[\partial_{\beta_{s}},\partial_{\beta_{t}}]
~.
\ee
Only the first term gives a non-zero contribution,
\be\label{frosm}
\prod_{n}\left(n\, \partial_{\beta_{s}}^{n}\right)^{g_{n}}e^{\beta_{s}}\left(1{-}{\beta_{s}{-}\beta_{t}\over \alpha's{+}2}\right)\Big|_{\beta_{s,t}{=}0}=\left(\prod_{n}n^{g_{n}}\right)\left(1{-}{\sum_{n}ng_{n}\over\alpha's{+}2}\right)=\left(\prod_{n}n^{g_{n}}\right)\left(1{-}{N\over\alpha's{+}2}\right)\,.
\ee

Putting all the contributions together, we find that the absorption cross-section of a tachyon from a generic state at level $N$ is
\be
\sigma_{abs}^{H_N+T\rightarrow {\rm any}\,H_{N'}}(s) =\pi \alpha'^{\frac{d-2}{2}}g_{o}^{2}\,\Tr({\cal E}^{T}{\cal E}^{*})\,{(2{+}\alpha's{-}N)\over \alpha'F_{\phi}^{(HT)}}=\pi \alpha'^{\frac{d-2}{2}}g_{o}^{2}{(2{+}\alpha's{-}N)\over \alpha'F_{\phi}^{(HT)}}
~.
\ee
where in the last equality we used \eqref{NormCond} and \eqref{NormRot}.

With a similar computation, the absorption cross-section of a photon with polarization $A^{\mu}$ from a generic state at level $N$ is given by the formula
\be
\sigma_{abs}^{H_N+\gamma\rightarrow {\rm any}\,H_{N'}}(s) =\pi \alpha'^{\frac{d-2}{2}}g_{o}^{2}\, \Tr({\cal E}^{T}{\cal E}^{*})\, |A|^{2}{(1{+}\alpha's{-}N)\over \alpha'F_{\phi}^{(H\gamma)}}=\pi \alpha'^{\frac{d-2}{2}}g_{o}^{2}{(1{+}\alpha's{-}N)\over \alpha'F_{\phi}^{(H\gamma)}}
%%%
~.
\ee
where the last equality incorporated the normalization of the HES and photon states. In general, properly normalized states contribute in the same way to the absorption cross section, since all the non zero contractions of the polarization tensors restore exactly their normalization conditions. Moreover, the forward limit \eqref{forwtachkine} imposes strong constraints on the differential polynomials\ref{finI14td}, as discussed in appendix \ref{SForLimAmp}. The interplay of these two features gives rise to the universality of the absorption cross section.

The absorption cross-sections in the closed string case can be obtained with similar techniques. For the tachyon and massless states, respectively, we find
\be
\Sigma_{abs}^{H_N+T\rightarrow {\rm any}\, H_{N'}}(s) =\pi \alpha'^{\frac{d-2}{2}}g_{c}^{2}{(2{+}\alpha's{-}N)^{2}\over \alpha'F_{\phi}^{(HT)}}\,,\quad \Sigma_{abs}^{H_N+\xi\rightarrow H_{N'} }(s) =\pi \alpha'^{\frac{d-2}{2}}g_{c}^{2}{(1{+}\alpha's{-}N)^{2}\over \alpha'F_{\phi}^{(H\xi)}}
~.
\ee

Finally, using the explicit energy-mass dependence of the Mandelstam variable $s$
\be
\alpha's\big|_{T}=N-2+2\sqrt{N{-}1}\sqrt{\alpha'}\,E\,,\quad \alpha's\big|_{\gamma\, {\rm or}\, \xi}=N-1+2\sqrt{N{-}1}\sqrt{\alpha'}\,\omega
\ee
one can also express the absorption cross-sections (in the open and closed string cases, respectively) as follows
\be\label{opabsfinL}
\sigma_{abs}^{H_N+T\rightarrow {\rm any}\,H_{N'}}(E) =\pi \alpha'^{\frac{d-2}{2}} g_{o}^{2}{\sqrt{\alpha'}E\over \sqrt{1{+}\alpha'E^2}}\,,\quad \sigma_{abs}^{H_N+\gamma\rightarrow H_{N'}}(\omega) =\pi \alpha'^{\frac{d-2}{2}}g_{o}^{2}
\ee
\be\label{clabsfinL}
\Sigma_{abs}^{H_N+T\rightarrow {\rm any}\,H_{N'}}(E) =2\pi \alpha'^{\frac{d-2}{2}} g_{c}^{2}\sqrt{N{-}1}{\alpha'E^2\over\sqrt{1{+}\alpha'E^2}}\,,\quad \Sigma_{abs}^{H_N+\xi\rightarrow H_{N'}}(\omega) =2\pi \alpha'^{\frac{d-2}{2}}g_{c}^{2}\sqrt{N{-}1}\sqrt{\alpha'}\,\omega
~.
\ee
$E$ is the energy of the tachyon and $\omega$ the energy of the massless open or closed string state.

\subsubsection{Absorption of generic states from generic states}

The absorption cross-section of a generic string state at level $N_2$ from a generic string state at level $N_1$ can be computed by applying techniques similar to the ones discussed in the previous section.

More specifically, in the open string case one can write the imaginary part of the forward scattering generating function as
\be
\widetilde\sigma_{abs,gen}^{H_{N_{1}}+H_{N_{2}}\rightarrow {\rm any}\,H_{N_{3}}}(s)=\pi g_{o}^{2}{(2{+}\alpha's)}e^{{\cal K}_{forw}\left(\{\zeta^{(\ell)}_{n}\};\partial_{\beta_{s}},\partial_{\beta_{t}}\right)}e^{\beta_{s}}\left(1-{\beta_{s}-\beta_{t}\over \alpha's{+}2}\right)\Big|_{\beta_{s,t}{=}0}
~,
\ee
All the relevant contributions in these expressions are described in appendix \ref{4HESsc} and appendix \ref{SForLimAmp} . By defining the derivative operators
\be
{\cal D}_{N_{1}}^{\{g_n\}}\left(\epsilon^{(1)},\epsilon^{*(1)} \right):=\epsilon^{(1)}_{(\mu_{1})_{g_{1}}(\mu_{2})_{g_{2}}\ldots}\epsilon^{(1)*}_{(\nu_{1})_{g_{1}}(\nu_{2})_{g_{2}}\ldots}\prod_{n=1}^{N_{1}}{1\over n^{g_n}g_n!}\prod_{r=1}^{g_n}{\partial \over \partial \lambda^{(1)}_{n \mu_{n}^{r}}}
{\partial \over \partial \lambda^{(4)}_{n\nu_{n}^{r}}}
~,
\ee
\be
{\cal D}_{N_{2}}^{\{g'_n\}}\left(\epsilon^{(2)},\epsilon^{*(2)} \right):=\epsilon^{(2)}_{(\alpha_{1})_{g'_{1}}(\alpha_{2})_{g'_{2}}\ldots}\epsilon^{(2)*}_{(\beta_{1})_{g'_{1}}(\beta_{2})_{g'_{2}}\ldots}\prod_{n=1}^{N_{2}}{1\over n^{g'_n}g'_n!}\prod_{r=1}^{g'_n}{\partial \over \partial \lambda^{(2)}_{n \alpha_{n}^{r}}}
{\partial \over \partial \lambda^{(3)}_{n\beta_{n}^{r}}}
~,
\ee
the absorption cross-section at a level-$N_2$ state from a level-$N_1$ state can be written as
\be
\label{open1akCS4p}
\sigma_{abs}^{H_{N_{1}}+H_{N_{2}}\rightarrow {\rm any}\,H_{N_{3}}}(s) =
{\cal D}_{N_{1}}^{\{g_n\}}\left(\epsilon^{(1)},\epsilon^{*(1)} \right){\cal D}_{N_{2}}^{\{g'_n\}}\left(\epsilon^{(2)},\epsilon^{*(2)} \right){1\over F_{\phi}^{(H_1H_2)}}{\widetilde{\sigma}_{abs,gen}^{H_{N_{1}}+H_{N_{2}}\rightarrow H_{N_{3}}}(s)}\Big|_{\zeta^{(\ell)}_n=0}
~,
\ee
where the relative flux factor, in the rest frame of the state $H_1$, is given by
\be
\label{open2abHH}
\alpha'F_{\phi}^{(H_1 H_2)} =2 \sqrt{N_1{-}1} \sqrt{\alpha'E_{2}^{2}{-}N_2{+}1}\,,\quad \alpha'M_{H_2}^2=N_2{-}1=\alpha'E_{2}^2-\alpha'\vec{p}_{2}^{\,2}
~.
\ee
Following the steps of the previous section, one can isolate the non-zero contributions, which come from the action of the differential operators
\be
I^{(1,4)}_{n,n}(\partial_{\beta_{s}},\partial_{\beta_{t}})=n\, \partial_{\beta_{s}}^{n}+\partial_{\beta_{t}}^{2} {\rm Poly}^{(1,4)}[\partial_{\beta_{s}},\partial_{\beta_{t}}]
\ee
and
\be
I^{(2,3)}_{n,n}(\partial_{\beta_{s}},\partial_{\beta_{t}})=n\, \partial_{\beta_{s}}^{n}+\partial_{\beta_{t}}^{2} {\rm Poly}^{(2,3)}[\partial_{\beta_{s}},\partial_{\beta_{t}}]
~.
\ee
This yields the generalization of \eqref{IntAbsSt}
\be\label{IntAbsStA}
\begin{split}
&\sigma_{abs}^{H_{N_1}+H_{N_2}\rightarrow {\rm any}\,H_{N_3}}(s)=\alpha'^{\frac{d-2}{2}}g_{o}^{2} { \pi(2{+}\alpha's)\over \alpha'F_{\phi}^{(H_{1}H_{2})}}\\
%%%
&\qquad \left(\prod_{n=1}^{N_{1}}{1\over n^{g_{n}}g_{n}!}\right)\left(\prod_{m=1}^{N_{2}}{1\over m^{g'_{m}}g'_{m}!}\right){\cal E}^{(1)}_{(\mu_{1})_{g_{1}}(\mu_{2})_{g_{2}}\ldots}{\cal E}^{(1)*}_{(\nu_{1})_{g_{1}}(\nu_{2})_{g_{2}}\ldots}{\cal E}^{(2)}_{(\alpha_{1})_{g'_{1}}(\alpha_{2})_{g'_{2}}\ldots}{\cal E}^{(2)*}_{(\beta_{1})_{g'_{1}}(\beta_{2})_{g'_{2}}\ldots}\\
%%%
&\prod_{n=1}^{N_{1}}\left(\prod_{r=1}^{g_{n}}\delta^{\mu^{r}_{n}\nu^{r}_{n}}\right)\left(I^{(1,4)}_{n,n}(\partial_{\beta_{s}},\partial_{\beta_{t}})\right)^{g_{n}}\prod_{m=1}^{N_{2}}\left(\prod_{r=1}^{g'_{m}}\delta^{\alpha^{r}_{m}\beta^{r}_{m}}\right)\left(I^{(2,3)}_{m,m}(\partial_{\beta_{s}},\partial_{\beta_{t}})\right)^{g'_{m}}e^{\beta_{s}}\left(1{-}{\beta_{s}{-}\beta_{t}\over \alpha's{+}2}\right)\Big|_{\beta_{s,t}}~.
\end{split}
\ee
In this case, the relevant tensorial contributions are given by the direct generalization of \eqref{TensContr}
\be
{\cal E}^{(1)}_{(\mu_{1})_{g_{1}}(\mu_{2})_{g_{2}}\ldots}{\cal E}^{(1)*}_{(\nu_{1})_{g_{1}}(\nu_{2})_{g_{2}}\ldots}\prod_{n}\left(\prod_{r=1}^{g_{n}}\delta^{\mu^{r}_{n}\nu^{r}_{n}}\right)=\Tr({\cal E}^{(1)T}{\cal E}^{(1)*})\prod_{n}g_{n}!
\ee
\be
{\cal E}^{(2)}_{(\alpha_{1})_{g'_{1}}(\alpha_{2})_{g'_{2}}\ldots}{\cal E}^{(2)*}_{(\beta_{1})_{g'_{1}}(\beta_{2})_{g'_{2}}\ldots}\prod_{m}\left(\prod_{r=1}^{g'_{m}}\delta^{\alpha^{r}_{m}\beta^{r}_{m}}\right)=\Tr({\cal E}^{(2)T}{\cal E}^{(2)*})\prod_{m}g'_{m}!
\ee
while the analog of \ref{frosm}, now is given by
\be
\prod_{n=1}^{N_1}\left(n\, \partial_{\beta_{s}}^{n}\right)^{g_{n}}\prod_{m=1}^{N_2}\left(m\, \partial_{\beta_{s}}^{m}\right)^{g'_{m}}e^{\beta_{s}}\left(1{-}{\beta_{s}{-}\beta_{t}\over \alpha's{+}2}\right)\Big|_{\beta_{s,t}{=}0}=\left(\prod_{n=1}^{N_{1}}n^{g_{n}} \right)\left(\prod_{m=1}^{N_{2}}m^{g'_{m}} \right)\left(1{-}{N_1{+}N_2\over\alpha's{+}2}\right)
~.
\ee
These allow us to obtain the final result
\be\label{finABSopHH}
\sigma_{abs}^{H_{N_1}+H_{N_2}\rightarrow {\rm any}\,H_{N_3}}(s) =\pi \alpha'^{\frac{d-2}{2}}g_{o}^{2}\,\Tr({\cal E}^{(1)T}{\cal E}^{(1)*}) \Tr({\cal E}^{(2)T}{\cal E}^{(2)*}){(2{+}\alpha's{-}N_1{-}N_2)\over \alpha'F_{\phi}^{(H_1H_2)}}={\pi \alpha'^{\frac{d-2}{2}}g_{o}^{2}\over \sqrt{1-{N_2-1\over \alpha'E_2^{2}}}}
~.
\ee
The last equality follows from the proper normalization of both the $N_{1}$- and $N_{2}$-states, the laboratory frame kinematics and the relation
\be\label{schan}
\alpha's=\alpha'M_{H_1}^2+\alpha'M_{H_2}^2+2\sqrt{\alpha'}M_{H_1}\sqrt{\alpha'}E_2=N_1{+}N_2{-}2+2\sqrt{N_1{-}1}\sqrt{\alpha'}E_2
~.
\ee

The absorption cross-section in the closed string case can be obtained similarly, and the resulting expression is
\be\label{finABSclHH}
\Sigma_{abs}^{H_{N_1}+H_{N_2}\rightarrow {\rm any} \,H_{N_3}}(s) =\pi \alpha'^{\frac{d-2}{2}}g_{c}^{2}{(2{+}\alpha's{-}N_1{-}N_2)^2\over \alpha'F_{\phi}^{(H_1H_2)}}={2\pi \alpha'^{\frac{d-2}{2}}g_{c}^{2}\sqrt{N_1{-}1}\sqrt{\alpha'}E_2\over \sqrt{1-{N_2-1\over \alpha'E_2^{2}}}}
~,
\ee
where the last equality is obtained in the laboratory frame using \eqref{schan}.

The results \eqref{finABSopHH},\eqref{finABSclHH} are the direct generalization of  \eqref{opabsfinL} and  \eqref{clabsfinL} respectively. For completeness, we also present here the absorption cross-sections \eqref{finABSopHH} and \eqref{finABSclHH} in the center-of-mass frame
\be
\sigma_{abs}^{H_{N_1}+H_{N_2}\rightarrow {\rm any}\,H_{N_3}}(E_{cm})\Big|_{CoM}=\pi \alpha'^{\frac{d-2}{2}}g_{o}^{2}{(2{+}\alpha'E_{cm}^2{-}N_1{-}N_2)\over \sqrt{\alpha'}|\vec{p}_{cm}| \sqrt{\alpha'}E_{cm}}
~,
\ee
\be
\Sigma_{abs}^{H_{N_1}+H_{N_2}\rightarrow {\rm any}\,H_{N_3}}(E_{cm})\Big|_{CoM}=\pi \alpha'^{\frac{d-2}{2}}g_{c}^{2}{(2{+}\alpha'E_{cm}^2{-}N_1{-}N_2)^2\over \sqrt{\alpha'}|\vec{p}_{cm}| \sqrt{\alpha'}E_{cm}}
~.
\ee
The center-of-mass spacelike momentum is defined as
\be
|\vec{p}_{cm}|={1\over 2 E_{cm}}\sqrt{\lambda(E_{cm}^2,M_{H_1}^2,M_{H_2}^2)} \,,\quad E_{cm}=E_1+E_2=M_3
\ee
where $\lambda(E_{cm}^2,M_{H_1}^2,M_{H_2}^2)$ is the K\"allen triangle function
\be
\lambda(E_{cm}^2,M_{H_1}^2,M_{H_2}^2)=\left(E_{cm}^2-(M_{H_1}{+}M_{H_2})^2\right)\left(E_{cm}^2-(M_{H_1}{-}M_{H_2})^2\right)
~.
\ee

\section{Emission rates of excited string states}
\label{emission}

Given the absorption cross-section, the differential emission rate can be computed using the time reversal symmetry of the scattering matrix, as discussed in appendix \ref{AbsEmOpt}. In this section, we apply this methodology to tree-level string theory. First, we compute the emission rate of arbitrary string states from general excited string states. Then, we focus on the emission of massless modes from highly excited strings and conclude with some remarks on the features of the resulting rates and their comparison with existing results in the literature.

\subsection{General emission rates}
\label{emissionGen}

First we start from the general absorption process (in open or closed string theory separately)
\be
\label{absprocess}
H_{N_1} + H_{N_2} \rightarrow H_{N'}
\ee
where two string excitations ($H_{N_1}$, $H_{N_2}$), at levels $N_1$ and $N_2$, respectively, produce a final string excitation $H_{N'}$ at level $N'$. Each of these states belongs in a vector space of degenerate states,  with dimension (degeneracy) $\rho(N)$ at level $N$. In $d$ space-time dimensions, the multiplicity $\rho(N)$, asymptotes to
\be\label{Ldeg}
\rho(N)={1\over 2\pi i }\oint {dw\over w^{N+1}} w^{{d-2\over 24}}\left(-{\log w\over 2\pi} \right)^{d-2\over 2}\eta \left(-{2\pi i\over \log w}\right)^{2-d}\Big|_{N \gg 1}\simeq N^{-{d+1\over 4}}e^{2\pi \sqrt{{(d{-}2)\over 6}N}}
~,
\ee
where $\eta(z)$ is the standard Dedekind $\eta$-function.

Applying Eqs.\ \eqref{absEGaoFlu} and \eqref{bahe} to the present context, we obtain the absorption cross-section in terms of squared 3-point scattering amplitudes in the following form
\be\label{AbsStrin3pt}
\sigma^{H_{N_1}+H_{N_2} \rightarrow H_{N'}}_{abs}
={1\over \phi^{(N_1,N_2)}_{flux}}{2\pi\, \delta\left(E_{N_{1}}{+}E_{N_{2}}{-}M_{N'}\right)\over M_{N'}^{3} \Big(1- \left({M_{N_{1}}^{2}-M_{N_{2}}^{2}\over M_{N'}^{2}}\right)^{2}\Big)}\Big|{\cal{S}}_{H_{N_1}+H_{N_2}\rightarrow H_{N'}}\Big|^{2}
~.
\ee
where $H_{N_1}$, $H_{N_2}$ and $H_{N'}$ are three individual excited string states, and no summation over states is implicit above.

The definition of the relative flux factor $\phi^{(N_1,N_2)}_{flux}$ appears in Eq.\ \eqref{Rflux}. For example, at leading order in open string perturbation theory, the 3-point scattering matrix element is $O(g_{o})$
\be
{\cal{S}}_{H_{N_1}+H_{N_2} \rightarrow H_{N'}}= g_{o} \,{\cal A}
_{H_{N_1}+H_{N_2} \rightarrow H_{N'}} + O(g_{o}^{3})
\ee
and can be expressed in terms of the corresponding tree-level 3-point scattering amplitude $\AA$. Accordingly, the leading order form of the absorption cross-section in \eqref{AbsStrin3pt} reads
\be
\sigma^{H_{N_1}+H_{N_2} \rightarrow H_{N'}}_{abs}={g_{o}^{2}\over \phi^{(N_1,N_2)}_{flux}}{2\pi\, \delta\left(E_{N_{1}}{+}E_{N_{2}}{-}M_{N'}\right)\over M_{N'}^{3} \Big(1- \left({M_{N_{1}}^{2}-M_{N_{2}}^{2}\over M_{N'}^{2}}\right)^{2}\Big)}
\Big|{\cal{A}}_{H_{N_1}+H_{N_2} \rightarrow H_{N'}}\Big|^{2} +O(g_{o}^{4})
~.
\ee

Consequently, the inclusive absorption cross-section, can be computed by taking the sum over all the final states at the same mass
\be\label{abs3ptdiff}
\sigma^{H_{N_1}+H_{N_2} \rightarrow {\rm any}\, H_{N'}}_{abs}={g_{o}^{2}\over \phi^{1,2}_{flux}}{2\pi\, \delta\left(E_{N_{1}}{+}E_{N_{2}}{-}M_{N'}\right)\over M_{N'}^{3} \Big(1- \left({M_{N_{1}}^{2}-M_{N_{2}}^{2}\over M_{N'}^{2}}\right)^{2}\Big)}
\sum_{H_{N'}} \Big|{\cal{A}}_{H_{N_1}+H_{N_2} \rightarrow H_{N'}}\Big|^{2} +O(g_{o}^{4})
~.
\ee
In this equation (as well as in the rest of this section and App.\ \ref{AbsEmOpt}) the sum over states involves implicitly all the continuum of states in the Hilbert space of degenerate states at a fixed mass, including sums over polarizations. It is a symbolic way of writing the first integral  in equation (\ref{g3}) in appendix \ref{ortho}.
 When expressed in terms of a basis of states, the sum contains the inverse matrix of state inner-products and will not be diagonal if the basis is not orthonormal\footnote{In appendix \ref{ortho} we present a simple example as a reminder of the proper formula for summing over non-orthogonal states. In string theory, all convenient bases of states are non-orthogonal.}. We also note in passing, that the use of the DDF formalism is a particularly convenient tool in expressing the relevant sums of squared amplitudes, over independent polarizations, reproducing the results of the covariant formalism \cite{Firrotta:2022cku}.

Performing the exact analytical summation on the RHS, by summing all the relevant 3-point amplitudes, is challenging even at tree-level. Alternatively, one can compute the absorption cross-section using the optical theorem. This bypasses the subtleties related to the non-orthogonality of bases of states.  This is precisely what we did in the previous Section \ref{absorb}. In accordance with Eqs.\ \eqref{Opexpgen} and \eqref{absLead}, at leading order in the string coupling, one has
\be
\sigma^{H_{N_1}+H_{N_2} \rightarrow {\rm any}\,H_{N'}}
_{abs}={1\over  F^{(N_1,N_2)}_{\phi}} {\rm Im} {\cal A}^{\rm forward}_{H_{N_1}+H_{N_2} \rightarrow H_{N_1} + H_{N_2}}
~.
\ee
As we have shown in Section \ref{absorb}, the absorption cross-section is independent of the details of the initial states, although we have not averaged over initial states,
\be
\sigma^{H_{N_1}+H_{N_2} \rightarrow {\rm any}\,H_{N'}}_{abs}
\equiv\sigma^{N_{1}+N_{2}\rightarrow N'}_{abs}
\ee
Hence, an average over the (degenerate in energy) initial states leaves the result unchanged
\be\label{externInd}
{1\over \rho(N_{1})}\sum_{H'_{N_1}} {1\over \rho(N_{2})}\sum_{H'_{N_2}}
\sigma^{H'_{N_1}+H'_{N_2} \rightarrow {\rm any}\,H_{N'}}_{abs}
=\sigma^{N_{1}+N_{2}\rightarrow N'}_{abs}\;.
\ee

Next, we turn our attention to the emission of two string states from a single string state. We analyse the kinematics of this process in appendix \ref{AbsEmOpt}, where the reader is referred to before continuing below. Emission is the inverse process of \eqref{absprocess}
\be
\label{emitaa}
H_{N'} \to H_{N_1} + H_{N_2}
~.
\ee
According to \eqref{ems}, the emission rate corresponding to \eqref{emitaa} is given by
\be\label{emsstri}
\begin{split}
{\Delta\Gamma_{em}\over \Delta E_{N_{1}}}^{\hspace{-0.5cm}\,^{H_{N'}\rightarrow  H_{N_1}+H_{N_2}}}=&{g_{s}^{2}\Omega^{(d-2)}_{solid}\over 8 (2\pi)^{d-2} } {E_{N_{1}}^{d-3}\over E_{N_{2}} }M_{N'}^{2}\left(1{-}{M_{N_{1}}^{2}\over E_{N_{1}}^{2}}\right)^{d-3\over2}\times\\
%%%
&\times\quad \quad\delta(E_{N_1}{+}E_{N_2}{-}E_{N'}) \Big|{\cal{A}}_{H_{N'}\rightarrow H_{N_1}+H_{N_2}}\Big|^{2}\,,
\end{split}
\ee
where we used $\Delta$ (instead of a derivative) to draw attention to the fact that the variation of the emission with respect to $E_{1}$ involves finite differences, since $E_1$ is discrete
\be
\label{energyA}
E_{N_{1}}={M_{N_{1}}^{2} + M_{N'}^{2}-M_{N_{2}}^{2}\over 2M_{N'}}\,\quad \Rightarrow\,\quad \sqrt{\alpha'}E_{N_{1}}\in\Big[{N_{1}\over 2\sqrt{N'{-}1}},{N_1 {+}N'{-}2\over 2\sqrt{N'{-}1}}\Big]
~.
\ee
This point is further discussed in Appendix \ref{AbsEmOpt} around Eq.\ \eqref{discreteEnergy}.

These observations allow us to connect emission and absorption if we take the average over initial states and the sum over final states and use the time-reversal symmetry.\footnote{
In principle, our techniques allow us to compute the emission cross section without averaging over the initial state, but we will not do this in this paper.}
\be
{\cal{A}}_{H_{N'}\rightarrow H_{N_1}+ H_{N_2}}={\cal{A}^*}_{H_{N_1}+H_{N_2}\rightarrow H_{N'}}
~.
\ee
Therefore,
\be
\sum_{H_{N_1}, H_{N_2}, H_{N'}}
\Big|{\cal{A}}_{H_{N'}\rightarrow H_{N_1}+ H_{N_2}}\Big|^{2}
=\sum_{H_{N_1}, H_{N_2}, H_{N'}}
\Big|{\cal{A}}_{H_{N_1}+H_{N_2}\rightarrow H_{N'}}\Big|^{2}
~.
\ee
Defining
\be
{\Delta\Gamma_{em}\over \Delta E_{N_{1}}}^{\hspace{-0.5 cm}^{N'\rightarrow N_{1}+N_{2}}}\hspace{-0.4cm} := \frac{1}{\rho(N')} \sum_{H_{N'}, H_{N_1} ,  H_{N_2}}
{\Delta\Gamma_{em}\over \Delta E_{N_{1}}}^{\hspace{-0.5cm}\,^{H_{N'}\rightarrow  H_{N_1}+H_{N_2}}}
\ee
and using the independence on external states of the absorption cross-section \eqref{externInd}, one arrives in this manner at the following relation between emission and absorption for string states
\be\label{emfixstr}
{\Delta\Gamma_{em}\over \Delta E_{N_{1}}}^{\hspace{-0.5 cm}^{N'\rightarrow N_{1}+N_{2}}}\hspace{-0.4cm}={\Omega^{(d-2)}_{solid}\over 2 (2\pi)^{d-1} } {E_{N_{1}}^{d-2}}\left(1{-}{M_{N_{1}}^{2}\over E_{N_{1}}^{2}}\right)^{d-3\over2}{\rho(N_{1})\rho(N_{2})\over \rho(N')} \phi^{1,2}_{flux}\sigma^{\,^{N_{1}+N_{2}\rightarrow N'}}_{abs}
~.\ee
We used
\be
E_{N_{2}}=M_{N'}-E_{N_{1}}\,,\quad E_{N_{1}}E_{N_{2}}={M_{N'}^{2}\over 4}\left(1-\left({M_{N_{1}}^{2}-M_{N_{2}}^{2}\over M_{N'}^{2}}\right)^{2}\right)
~.
\ee
In Section \ref{absorb} we computed the absorption cross sections
\be
\sigma_{abs}^{N_{1}+N_{2}\rightarrow {\rm any}\,N'}(E_{1})={g_{o}^{2}\pi \alpha'^{\frac{d-2}{2}} \over \sqrt{1-{M_{N_{1}}^{2}\over E_{1}^{2}}}}\,,\quad \Sigma_{abs}^{N_{1}+N_{2}\rightarrow {\rm any}\, N'}(E_{1})={2 g_{c}^{2} \pi \alpha'^{\frac{d}{2}} \over \sqrt{1-{M_{N_{1}}^{2}\over E_{1}^{2}}}}
M_{N'} E_{N_{1}}
~.
\ee
Inserting the explicit form of the flux factor in terms of $M_{N'}$ and $E_{N_{1}}$
\be
\phi_{flux}^{(N_1,N_2)}={M_{N'}\over M_{N'}-E_{N_{1}}} \sqrt{1-{M_{N_{1}}^{2}\over E_{1}^{2}}}
\ee
one can therefore find the corresponding emission rates in open and closed string theory respectively
\be\label{opEmissgen}
{\Delta\Gamma_{em}\over \Delta E_{N_{1}}}^{\hspace{-0.5 cm}^{N'\rightarrow N_{1}+N_{2}}}\Bigg|_{\rm open}=g_{o}^{2}{\pi\alpha'^{\frac{d-2}{2}}}{\Omega^{(d-2)}_{solid}\over 2 (2\pi)^{d-1} } {E_{N_{1}}^{d-2}\over 1{-}{E_{N_{1}} \over M_{N'}}}\left(1{-}{M_{N_{1}}^{2}\over E_{N_{1}}^{2}}\right)^{d-3\over2}{\rho(N_{1})\rho(N_{2})\over \rho(N')}
~,
\ee
\be\label{clEmissgen}
{\Delta\Gamma_{em}\over \Delta E_{N_{1}}}^{\hspace{-0.5 cm}^{N'\rightarrow N_{1}+N_{2}}}\Bigg|_{\rm closed}=g_{c}^{2}{\pi\alpha'^{\frac{d}{2}}}
{\Omega^{(d-2)}_{solid}\over  (2\pi)^{d-1} } {
M_{N'}E_{N_{1}}^{d-1}\over 1{-}{E_{N_{1}} \over M_{N'}}}\left(1{-}{M_{N_{1}}^{2}\over E_{N_{1}}^{2}}\right)^{d-3\over2}\left({\rho\left({N_{1}\over 4}\right)\rho\left({N_{2}\over 4}\right)\over \rho\left({N'\over 4}\right)}\right)_{L,R}
~,
\ee
where we introduced the following compact notation for the closed string degeneracy in the respective holomorphic and anti-holomorphic sectors
\be
\left({\rho\left({N_{1}\over 4}\right)\rho\left({N_{2}\over 4}\right)\over \rho\left({N'\over 4}\right)}\right)_{L,R}\equiv {\rho\left({N_{1}\over 4}\right)\rho\left({N_{2}\over 4}\right)\over \rho\left({N'\over 4}\right)} {\rho\left({\overline{N_{1}}\over 4}\right)\rho\left({\overline{N_{2}}\over 4}\right)\over \rho\left({\overline{N'}\over 4}\right)}
~.
\ee
By level matching $N_1=\overline{N}_1$, $N_2=\overline{N}_2$, $N'=\overline{N}'$.
The rates in (\ref{opEmissgen}) and (\ref{clEmissgen})  are correct to leading order in string perturbation theory, but are otherwise exact.

In what follows, we shall be especially interested in the absorption and emission of highly excited string states, where $N' \gg 1$. In this limit, the finite-difference emission rate $\frac{\Delta \Gamma_{em}}{\Delta E}$ is converted to a differential emission rate with an extra factor of $2\sqrt{N'}$ (as explained in Eq.\ \eqref{diffvsderiv} in Appendix \ref{AbsEmOpt}), yielding the highly-excited string limit of Eq.\ \eqref{emfixstr}
\be\label{emfixstra}
{d\Gamma_{em}\over  dE_{N_{1}}}^{\hspace{-0.5 cm}^{N'\rightarrow N_{1}+N_{2}}}\hspace{-0.4cm}=\sqrt{N'}{\Omega^{(d-2)}_{solid}\over  (2\pi)^{d-1} } {E_{N_{1}}^{d-2}}\left(1{-}{M_{N_{1}}^{2}\over E_{N_{1}}^{2}}\right)^{d-3\over2}{\rho(N_{1})\rho(N_{2})\over \rho(N')} \phi^{1,2}_{flux}\sigma^{\,^{N_{1}+N_{2}\rightarrow N'}}_{abs}
~.
\ee

\subsection{Massless emission from highly excited string states}
\label{emissionMassless}

In this subsection, we specialize the above discussion to the case of massless emission from highly excited string states. We consider first the case of open string theory. The emission process of interest
\be
\label{emitphotona}
H_{N'} \rightarrow \gamma + H_N
\eeq
involves an open string state at level $N'$ decaying into a massless open string mode, e.g.\ a photon, and another state at level $N$.

At any $N'$ and photon energy $\omega$, the emission rate is
\be\label{opEmissgen}
{\Delta\Gamma_{em}\over \Delta \omega}^{\hspace{-0.5 cm}^{N'\rightarrow \gamma+N}}
=g_{o}^{2}{\pi\alpha'^{\frac{d-2}{2}}}{\Omega^{(d-2)}_{solid}\over 2 (2\pi)^{d-1} } {\omega^{d-2}\over 1{-}{\sqrt{\alpha'}\omega \over\sqrt{N'{-}1}}}{\rho\Big(N'{-}2\sqrt{N'{-}1}\sqrt{\alpha'}\omega\Big)\over \rho(N')}+{\cal O}(g_o^4)
\ee
where
\be
\sqrt{\alpha'}\omega\in\Bigg[{1\over 2\sqrt{N'{-}1}},{\sqrt{N'{-}1}\over 2}\Bigg]
~.
\ee
as shown in appendix \ref{AbsEmOpt}.

We are mainly interested, however, in the regime, where the decaying string is highly excited $(N' \gg 1)$ and the energy of the emitted photon is much smaller than the mass of the decaying string
\be
\label{emitphotonb}
\omega \ll M_{N'} \sim \sqrt{\frac{N'}{\alpha'}}
~.
\ee
In this limit, we can appropriately expand Eq.\ \eqref{opEmissgen} (using \eqref{Ldeg}) to obtain the differential emission rate
\be
\label{emitphotonc}
{d\Gamma_{em}\over  d\omega}^{\hspace{-0.5 cm}^{N'\rightarrow \gamma+N}}
\simeq g_{o}^{2}{\pi\alpha'^{\frac{d-1}{2}}}{\Omega^{(d-2)}_{solid}\over (2\pi)^{d-1} } M_{N'} \, \omega^{d-2}\, e^{-\frac{\omega}{T_H}}
~,
\ee
where $T_H$ is the Hagedorn temperature
\beq
\label{emiab}
T_H = \frac{1}{2\pi\sqrt{\alpha'}}\sqrt{\frac{6}{d-2}}
~.
\eeq
In particular, the Boltzmann factor emerges from the approximation to $\rho(n)$ for large $n$ and  will be discussed further in the next subsection.

Similarly, in closed string theory, the emission rate of a massless mode $\xi$, e.g.\ a graviton, from a highly excited state, can be obtained from \eqref{clEmissgen},
\be
\label{a1}
{\Delta\Gamma_{em}\over \Delta \omega}^{\hspace{-0.5 cm}^{N'\rightarrow \xi+N}}=g_{c}^{2}{\pi\alpha'^{\frac{d}{2}}}
{\Omega^{(d-2)}_{solid}\over  (2\pi)^{d-1} } {
M_{N'}\omega^{d-1}\over 1{-}{\omega \over M_{N'}}}\left({\rho\Big(N'/4{-}2\sqrt{N'{-}1}\sqrt{\alpha'}\omega/4\Big)\over \rho\left({N'\over 4}\right)}\right)_{L,R}+{\cal O}(g_c^4)
\ee
A suitable expansion in the regime \eqref{emitphotonb} and $N'\gg 1$ yields
\be
\label{emitgravitona}
{d\Gamma_{em}\over  d\omega}^{\hspace{-0.5 cm}^{N'\rightarrow \xi+N}}
\simeq 2g_{c}^{2}{\pi\alpha'^{\frac{d+1}{2}}}{\Omega^{(d-2)}_{solid}\over (2\pi)^{d-1} }
M_{N'}^2 \, \omega^{d-1}\, e^{-\frac{\omega}{T_H}}
~.
\ee

\subsection{Comments}
\label{comment}

The emission rates (\ref{opEmissgen}), (\ref{a1}) and their approximations \eqref{emitphotonc}, \eqref{emitgravitona}, as well as the cross-sections \eqref{opabsfinL}, \eqref{clabsfinL}, are some of the main results of the paper.
(\ref{opEmissgen}) and (\ref{a1}) are exact at tree level string theory, while \eqref{emitphotonc}, \eqref{emitgravitona} are exact in the limit $N'\gg 1$, $\omega \ll M_{N'}$.
Finally, \eqref{opabsfinL}, \eqref{clabsfinL}
 are exact in all regimes in tree-level string theory.

For the emission rates in the limit $N'\gg 1$, $\omega \ll M_{N'}$, we observe that the results are proportional to a Boltzmann factor both in the open and closed string theory. This is reminiscent of the thermal radiation pattern in black body physics. In fact, {had we applied the detailed balance condition (which is familiar from black hole physics and  assumes thermal equilibrium)}
\be
\label{commaa}
\frac{d\Gamma}{d\omega} = \sigma\, \Omega_{solid}^{(d-2)}\, \omega^{d-2} \frac{1}{e^{\frac{\omega}{T}}-1}
\ee
to the cross-sections \eqref{opabsfinL}, \eqref{clabsfinL} at the Hagedorn temperature $T=T_H$, we would obtain for open strings
\beq
\label{commab}
{d\Gamma_{em}\over  d\omega}^{\hspace{-0.5 cm}^{N'\rightarrow \gamma+N}}
=g_{o}^{2}{\pi\alpha'^{\frac{d-1}{2}}}{\Omega^{(d-2)}_{solid}\over (2\pi)^{d-1} } M_{N'} \, \omega^{d-2}\, \frac{1}{e^{\frac{\omega}{T_H}}-1}
\eeq
and for closed strings
\beq
\label{commac}
{d\Gamma_{em}\over  d\omega}^{\hspace{-0.5 cm}^{N'\rightarrow \xi+N}}
=2g_{c}^{2}{\pi\alpha'^{\frac{d+1}{2}}}{\Omega^{(d-2)}_{solid}\over (2\pi)^{d-1} }
M_{N'}^2 \, \omega^{d-1}\, \frac{1}{e^{\frac{\omega}{T_H}}-1}
~.
\eeq

The emission rate in \eqref{commab} is exactly the same as that of a black body at the Hagedorn temperature. It is also identical to the formula for emission rates in Ref.\ \cite{Amati:1999fv}. The result \eqref{emitphotonc} obtained in this paper using an exact computation of the tree-level cross-section and time-reversal symmetry deviates from \eqref{commab} when $\omega\lesssim T_H$.

In closed string theory, the emission rate \eqref{commac} exhibits a non-trivial grey-body factor, which vanishes at small energies $\omega$ (unlike the corresponding result in black hole physics, which asymptotes to a constant when the black hole has a regular horizon). In addition, \eqref{commac} is also significantly different when compared with the corresponding expression of massless closed string emission rates in Ref.\ \cite{Amati:1999fv}. More specifically, in the closed string case, the expressions \eqref{emitgravitona}, \eqref{commac} and the corresponding emission rates in \cite{Amati:1999fv} are all different when $\omega\lesssim T_H$. However, they are all in agreement, when $T_H \ll \omega \ll M_{N'}$ for $N' \gg 1$. In this regime, they all exhibit the thermal emission rates which are familiar from black holes.

The emission rates and absorption cross-sections, derived in this paper through a straightforward exact computation, suggest that the correct expressions at all energies $\omega$ are given by Eqs.\ \eqref{emitphotonc}, \eqref{emitgravitona}, \eqref{opabsfinL}, \eqref{clabsfinL} in tree-level string theory. {In particular, no equilibrium assumptions were made, and only $T$-invariance was used that in our setup is exact}. It would be interesting, however, to understand better the deviations that arise at $\omega \lesssim T_H$ in the light-cone gauge computation of Ref.\ \cite{Amati:1999fv} and the expressions \eqref{commab}, \eqref{commac} following from the detailed balance condition.

We note that the calculation of emission rates in Ref.\  \cite{Amati:1999fv} was based on the average of the inclusive rate of emission of massless states using a direct diagonal sum of squared 3-point amplitudes in the light-cone gauge. The average over the decaying states at a fixed level is defined in Eq.\ (2.14) in \cite{Amati:1999fv}\footnote{This equation applies to the open string case. Similar statements extend to the closed string case as well.} with respect to the Fock space basis appearing implicitly in Eq.\ (2.13) in the same paper. In (2.15) of \cite{Amati:1999fv}, the resulting expression is claimed to be equivalent to a trace (and, therefore, basis-independent), but this statement does not seem to be correct, since the Fock space basis in the light-cone gauge is not orthonormal\footnote{See the discussion in appendix \ref{ortho}.}. As a result, the computation of the average emission rates in Ref.\ \cite{Amati:1999fv} is missing contributions with respect to the basis-independent formula (2.15).

The correct computation of (2.15) would require a mass-level-dependent transformation to an orthonormal basis, which is not straightforward to incorporate. It would be interesting to see if a careful analysis of the missing contributions in Ref.\ \cite{Amati:1999fv} explains the deviation from our results in the substringy regime $\omega \lesssim T_H$, leaving the Boltzmann, superstringy, behaviour of the emission rates intact.

\section*{Acknowledgments}
We would like to thank M.\ Bianchi, P.\ Di Vecchia, S.\ Mathur, K.\ Papadodimas, F.\ Quevedo, J.\ Russo, B.\ Sundborg, A.\ Tseytlin, G.\ Veneziano and G.\ Villa for useful discussions and comments. This work was partially supported by the H.F.R.I call ``Basic research Financing (Horizontal support of all Sciences)'' under the National Recovery and Resilience Plan ``Greece 2.0'' funded by the European Union – NextGenerationEU (H.F.R.I. Project Number: 15384). MF is supported by the European MSCA grant HORIZON-MSCA- 2022-PF-01-01 ``BlackHoleChaos'' No.101105116. E. K. was partially supported by a Siemens-Humboldt research award. He thanks the Arnold Sommerfeld Institute for hospitality.

\begin{appendix}

\section{Polynomial contributions to generating functions}
\label{polynomials}

In the main text we have seen that the generating function of 4-point amplitudes receives several contributions. These can be expressed in term of the Jacobi polynomials
\be
P_{N}^{(\alpha,\beta)}(x)=\sum_{r=0}^{N}\begin{pmatrix}N+\alpha\\N-r\end{pmatrix}\begin{pmatrix}N+\beta\\r\end{pmatrix}\left({x-1\over 2}\right)^{r}\left({x+1\over 2}\right)^{N-r}
\ee
via the functions
\be\label{exprze1t}
V^{(1)\mu}_{n}(z)=\sqrt{2\alpha'}p^{\mu}_{2}P_{n-1}^{\hspace{0 mm}^{\hspace{0 mm}^{(\alpha_{1}^{(n)},\beta_{1}^{(n)})}}}\hspace{-9mm}(1{-}2 z)+z\sqrt{2\alpha'}p^{\mu}_{3} P_{n-1}^{\hspace{0 mm}^{\hspace{0 mm}^{(\alpha_{1}^{(n)}{+}1,\beta_{1}^{(n)})}}}\hspace{-12mm}(1{-}2z)
\ee
\be\label{exprz4t}
V^{(4)\mu}_{n}(z)=\sqrt{2\alpha'}p^{\mu}_{1}P_{n-1}^{\hspace{0 mm}^{\hspace{0 mm}^{(\alpha_{4}^{(n)},\beta_{4}^{(n)})}}}\hspace{-9mm}(2z{-}1)+(1{-}z)\sqrt{2\alpha'}p^{\mu}_{2} P_{n-1}^{\hspace{0 mm}^{\hspace{0 mm}^{(\alpha_{4}^{(n)}{+}1,\beta_{4}^{(n)})}}}\hspace{-12mm}(2z{-}1)
\ee
\be\label{finWz1t}
W^{(1)}_{n,m}(z)=\sum_{r=1}^{m}r\,P_{n+r}^{\hspace{0 mm}^{\hspace{0 mm}^{(\alpha_{1}^{(n)}-r,\beta_{1}^{(n)}-1)}}}\hspace{-14.5mm}(1{-}2z)\,\,P_{m-r}^{\hspace{0 mm}^{\hspace{0 mm}^{(\alpha_{1}^{(m)}+r,\beta_{1}^{(m)}-1)}}}\hspace{-15mm}(1{-}2 z)
\ee
\be\label{finWz4t}
W^{(4)}_{n,m}(z)=\sum_{r=1}^{m}r\,P_{n+r}^{\hspace{0 mm}^{\hspace{0 mm}^{(\alpha_{4}^{(n)}-r,\beta_{4}^{(n)}-1)}}}\hspace{-14.5mm}(2z{-}1)\,\,P_{m-r}^{\hspace{0 mm}^{\hspace{0 mm}^{(\alpha_{4}^{(m)}+r,\beta_{4}^{(m)}-1)}}}\hspace{-15mm}(2z{-}1)
\ee
\be\label{finI14t}
{ I}^{(1,4)}_{n_{1},m_{4}}(z)=(-)^{m_{4}{+}1}\sum_{r,s=0}^{n_{1},m_{4}}z^{r+s+1}P_{n_{1}{-}r{-}s{-}1}^{\hspace{0 mm}^{\hspace{0 mm}^{(\alpha_{1}^{(n_{1})}{+}r{+}s{+}1;\,\beta_{1}^{(n_{1})}{-}1)}}}\hspace{-15mm}(1{-}2 z)\,\,P_{m_{4}{-}r{-}s{-}1}^{\hspace{0 mm}^{\hspace{0 mm}^{(\beta_{4}^{(m_{4})}+r+s+1;\,\alpha_{4}^{(m_{4})}-1)}}}\hspace{-16mm}(1{-}2z)
~.
\ee
The parameters $\alpha_1, \alpha_4, \beta_1, \beta_4$ are related to the momenta as follows
\be\label{coeffJacze1}
\alpha_{1}^{(n)}=-n-2\alpha'nq_{1}{\cdot}p_{2}\,,\quad \beta_{1}^{(n)}=-n-2\alpha'nq_{1}{\cdot}p_{4}
~,
\ee
\be\label{coeffJacz4}
\alpha_{4}^{(n)}=-n-2\alpha'nq_{4}{\cdot}p_{1}\,,\quad \beta_{4}^{(n)}=-n-2\alpha'nq_{4}{\cdot}p_{3}
~.
\ee

Notice that by using the formula
\be\label{formVen}
e^{\sum_{n}a_{n}z^{n}}e^{\sum_{n}b_{n}(1{-}z)^{n}}=\exp\left(\sum_{n}a_{n}\partial^{n}_{\beta_{s}}\right)\exp\left(\sum_{n}b_{n}\partial^{n}_{\beta_{t}}\right)e^{\beta_{s}z+\beta_{t}\,(1{-}z)}\Big|_{\beta_{s,t}=0}
\ee
all the dependence on $z$ and $1{-}z$ of the functions $V_{n}^{(\ell)}$, $W_{n,m}^{(\ell)}$ and $I_{n,m}^{(v,f)}$  can be re-parametrized by making the identifications
\be\label{idVen}
z^{n}\rightarrow \partial^{n}_{\beta_{s}}\,,\quad (1{-}z)^{n}\rightarrow \partial^{n}_{\beta_{t}}
\ee
that converts the polynomials \eqref{exprze1t}-\eqref{finI14t} to differential operators. For $n=1$, this implies $1 \rightarrow \partial_{\beta_{s}}+\partial_{\beta_{t}}$. In this manner, the Jacobi polynomials are mapped to the differential operators
\bea\label{Jpppoly}
&&P_{N}^{(\alpha,\beta)}(1{-}2z)=\sum_{r=0}^{N}(-)^{r}\begin{pmatrix}N{+}\alpha\\N{-}r\end{pmatrix}\begin{pmatrix}N{+}\beta\\r\end{pmatrix}z^{r}\left(1{-}z\right)^{N-r}
\nonumber\\
\longrightarrow
&&P_{N}^{(\alpha,\beta)}(\d_{\beta_t} - \d_{\beta_s}) =
\sum_{r=0}^{N}(-)^{r}\begin{pmatrix}N{+}\alpha\\N{-}r\end{pmatrix}\begin{pmatrix}N{+}\beta\\r\end{pmatrix}\partial_{\beta_{s}}^{r}\partial_{\beta_{t}}^{N-r}
\eea
and the polynomials \eqref{exprze1t}-\eqref{finI14t} to
\be\label{exprze1td}
V^{(1)\mu}_{n}(\partial_{\beta_{s}},\partial_{\beta_{t}})=\sqrt{2\alpha'}p^{\mu}_{2}P_{n-1}^{\hspace{0 mm}^{\hspace{0 mm}^{(\alpha_{1}^{(n)},\beta_{1}^{(n)})}}}\hspace{-9mm}(\partial_{\beta_{t}}{-}\partial_{\beta_{s}})+\partial_{\beta_{s}}\sqrt{2\alpha'}p^{\mu}_{3} P_{n-1}^{\hspace{0 mm}^{\hspace{0 mm}^{(\alpha_{1}^{(n)}{+}1,\beta_{1}^{(n)})}}}\hspace{-12mm}(\partial_{\beta_{t}}{-}\partial_{\beta_{s}})
\ee
\be\label{exprz4td}
V^{(4)\mu}_{n}(\partial_{\beta_{s}},\partial_{\beta_{t}})=\sqrt{2\alpha'}p^{\mu}_{1}P_{n-1}^{\hspace{0 mm}^{\hspace{0 mm}^{(\alpha_{4}^{(n)},\beta_{4}^{(n)})}}}\hspace{-9mm}(\partial_{\beta_{s}}{-}\partial_{\beta_{t}})+\partial_{\beta_{t}}\sqrt{2\alpha'}p^{\mu}_{2} P_{n-1}^{\hspace{0 mm}^{\hspace{0 mm}^{(\alpha_{4}^{(n)}{+}1,\beta_{4}^{(n)})}}}\hspace{-12mm}(\partial_{\beta_{s}}{-}\partial_{\beta_{t}})
\ee
\be\label{finWz1td}
W^{(1)}_{n,m}(\partial_{\beta_{s}},\partial_{\beta_{t}})=\sum_{r=1}^{m}r\,P_{n+r}^{\hspace{0 mm}^{\hspace{0 mm}^{(\alpha_{1}^{(n)}-r,\beta_{1}^{(n)}-1)}}}\hspace{-14.5mm}(\partial_{\beta_{t}}{-}\partial_{\beta_{s}})\,\,P_{m-r}^{\hspace{0 mm}^{\hspace{0 mm}^{(\alpha_{1}^{(m)}+r,\beta_{1}^{(m)}-1)}}}\hspace{-15mm}(\partial_{\beta_{t}}{-}\partial_{\beta_{s}})
\ee
\be\label{finWz4td}
W^{(4)}_{n,m}(\partial_{\beta_{s}},\partial_{\beta_{t}})=\sum_{r=1}^{m}r\,P_{n+r}^{\hspace{0 mm}^{\hspace{0 mm}^{(\alpha_{4}^{(n)}-r,\beta_{4}^{(n)}-1)}}}\hspace{-14.5mm}(\partial_{\beta_{s}}{-}\partial_{\beta_{t}})\,\,P_{m-r}^{\hspace{0 mm}^{\hspace{0 mm}^{(\alpha_{4}^{(m)}+r,\beta_{4}^{(m)}-1)}}}\hspace{-15mm}(\partial_{\beta_{s}}{-}\partial_{\beta_{t}})
\ee
\be\label{finI14td}
{ I}^{(1,4)}_{n_{1},m_{4}}(\partial_{\beta_{s}},\partial_{\beta_{t}})=(-)^{m_{4}{+}1}\sum_{r,s=0}^{n_{1},m_{4}}\partial_{\beta_{s}}^{r+s+1}P_{n_{1}{-}r{-}s{-}1}^{\hspace{0 mm}^{\hspace{0 mm}^{(\alpha_{1}^{(n_{1})}{+}r{+}s{+}1;\,\beta_{1}^{(n_{1})}{-}1)}}}\hspace{-15mm}(\partial_{\beta_{t}}{-}\partial_{\beta_{s}})\,\,P_{m_{4}{-}r{-}s{-}1}^{\hspace{0 mm}^{\hspace{0 mm}^{(\beta_{4}^{(m_{4})}+r+s+1;\,\alpha_{4}^{(m_{4})}-1)}}}\hspace{-16mm}(\partial_{\beta_{t}}{-}\partial_{\beta_{s}})
.
\ee

With these specifications, the generating functions of the 4-point amplitudes
\be\label{intamp}
\begin{split}
{\cal A}_{gen}(s,t)&=g_{o}^{2}\int_{0}^{1}dz\, z^{-{\alpha' s}{-}2}(1{-}z)^{-{\alpha' t}{-}2}\\
%%%
&\exp{\left(\sum_{\ell}\sum_{n_{\ell}}\zeta_{n_{\ell}}^{(\ell)}{\cdot}V^{(\ell)}_{n_{\ell}}(z)+\sum_{\ell}\sum_{n_{\ell},m_{\ell}}\zeta_{n_{\ell}}^{(\ell)}{\cdot}\zeta_{m_{\ell}}^{(\ell)}W^{(\ell)}_{n_{\ell},m_{\ell}}(z)+\sum_{v,f}\sum_{n_{v},m_{f}}\zeta_{n_{v}}^{(v)}{\cdot}\zeta_{m_{f}}^{(f)} I_{n_{v},m_{f}}^{(v,f)}(z)\right)}
\end{split}
\ee
can be reformulated as
\be\label{intamp}
\begin{split}
&{\cal A}_{gen}(s,t)=g_{o}^{2}\\
%%%
& \exp\Bigg(\sum_{\ell=1,4}\left(\sum_{n}\zeta_{n}^{(\ell)}{\cdot}V^{(\ell)}_{n}(\partial_{\beta_{s}},\partial_{\beta_{s}}){+}\sum_{n,m}\zeta_{n}^{(\ell)}{\cdot}\zeta_{m}^{(\ell)} W_{n,m}^{(\ell)}(\partial_{\beta_{s,}},\partial_{\beta_{t}})\right)+\sum_{n,m}\zeta_{n}^{(1)}{\cdot}\zeta_{m}^{(4)} I_{n,m}^{(1,4)}(\partial_{\beta_{s}},\partial_{\beta_{t}}) \Bigg) \\
%%%
&\quad \quad \quad \quad \quad  \int_{0}^{1}dz\, z^{-{\alpha' s}{-}2}(1{-}z)^{-{\alpha' t}{-}2}e^{\beta_{s}z+\beta_{t}\,(1{-}z)}\Big|_{\beta_{s,t}=0}
~.
\end{split}
\ee
Performing the $z$-integration on the last line, and defining
\be\label{Kfunc}
\begin{split}
{\cal K}\left(\{\zeta^{(\ell)}_{n}\}; \partial_{\beta_{s}},\partial_{\beta_{t}}\right):=&\sum_{\ell=1,4}\left(\sum_{n}\zeta_{n}^{(\ell)}{\cdot}V^{(\ell)}_{n}(\partial_{\beta_s},\d_{\beta_t}){+}\sum_{n,m} \zeta_{n}^{(\ell)}{\cdot}\zeta_{m}^{(m)}W_{n,m}^{(\ell)}(\partial_{\beta_s}, \d_{\beta_t})\right)\\
%%%
&\quad\quad+\sum_{n,m}\zeta^{(1)}_{n}{\cdot}\zeta^{(4)}_{m} I_{n,m}^{(1,4)}(\partial_{\beta_s}, \d_{\beta t})\,,
\end{split}
\ee
one gets
\be
{\cal A}_{gen}(s,t)={\cal A}_{Ven}(s,t)\,e^{{\cal K}\left(\{\zeta^{(\ell)}_{n}\};\partial_{\beta_{s}},\partial_{\beta_{t}}\right)}\Phi_{\beta_{s},\beta_{t}}\left(s,t\right)\Big|_{\beta_{s,t}{=}0}\ee
with
\be\label{ffunc}
\Phi_{\beta_{s},\beta_{t}}(s,t)\equiv {1\over A_{Ven}(s,t)} \int_{0}^{1}dz\, z^{-{\alpha' s}{-}2}(1{-}z)^{-{\alpha' t}{-}2}e^{\beta_{s}z+\beta_{t}\,(1{-}z)}=
\ee
$$= \sum_{r=0}^{\infty}\sum_{v=0}^{\infty}{\beta_{s}^{r}\over r!}{\beta_{t}^{v}\over v!}{(-\alpha' s{-}1)_{r}(-\alpha' t{-}1)_{v}\over (-\alpha' s{-}\alpha' t{-}2)_{r+v}}.
$$
We clearly have the identity
\be\label{phiID}
(\partial_{\beta_{s}}+\partial_{\beta_{t}})\Phi_{\beta_{s}\beta_{t}}(s,t)=\Phi_{\beta_{s}\beta_{t}}(s,t)
\ee

\section{Kinematical relations}
\label{kinetach}

In the main text, we adopt the following kinematical frame for the scattering amplitudes of excited string states with tachyons
\be
p_{1}=(M_{N},\vec{0})\,,\quad p_{2}=(E,0,p,\vec{0})~,
\ee
\be
p_{3}=-(E',p'\sin\theta,p'\cos\theta,\vec{0})\,,\quad p_{4}=-(M_{N}{+}E{-}E',-p'\sin\theta,p{-}p'\cos\theta,\vec{0})
~.
\ee
$N$ refers to the level of the excited string states with momenta $p_1$ and $p_4$. The corresponding Mandelstam variables $s$ and $t$ are
\be
s=-(p_{1}+p_{2})^{2}=M_{N}^{2}+M_{T}^{2}+2M_{N}E
~,
\ee
\bea
t &=& -(p_{2}{+}p_{3})^{2} = -2M_T^2-2EE'\left(1-{pp'\over EE'}\cos\theta\right)
\nonumber\\
&=&-(p_{1}{+}p_{4})^{2}=2M_{N}(E'{-}E)
~.
\eea
We also choose the null DDF reference momenta as
\be
q_{1}=-{1\over 2\alpha' M_{N}}(1,0,1,\vec{0})\,, \quad q_{4}={(1,0,1,\vec{0})\over 2\alpha' (M_{N}{+}E{-}E'{-}p {+}p'\cos\theta)}
~,
\ee
which satisfy the relations $2\alpha' q_1\cdot p_1 = 1$, $2\alpha' q_4\cdot p_4 = 1$.

The energies and momenta of the tachyons in terms of the Mandelstam invariants are therefore
\be
p=\sqrt{E^{2}-M_{T}^{2}}\,,\quad p'=\sqrt{E'^{2}-M_{T}^{2}}
~,
\ee
\be
E={s-M_{N}^{2}-M_{T}^{2}\over 2M_{N}}\,,\quad E'={s+t-M_{N}^{2}-M_{T}^{2}\over 2M_{N}}
~,
\ee
and the cosine of the scattering angle $\theta$
\be
\cos\theta={E E'\over p p'}-{M_{N}(E-E')\over p p'}-{M_{T}^{2}\over p p'}
~.
\ee
The scalar products between the momenta of the scattering states and the DDF reference momenta are
\be\label{q12}
2\alpha'q_{1}{\cdot}p_{2}={E{-}p\over M_{N}}\,,\quad 2\alpha'q_{1}{\cdot}p_{3}={p'\cos\theta{-}E'\over M_{N}}\,,\quad 2\alpha'q_{1}{\cdot}p_{4}={-}1-{E{-}E'\over M_{N}} + {p{-}p'\cos\theta\over M_{N}}
\ee
and
\be\label{q4all}
2\alpha'q_{4}{\cdot}p_{1}=-1-2\alpha'q_{4}{\cdot}p_{2}-2\alpha'q_{4}{\cdot}p_{3}={1\over 2\alpha'q_{1}{\cdot}p_{4}}\,, \quad 2\alpha'q_{4}{\cdot}p_{2}={q_{1}{\cdot}p_{2}\over q_{1}{\cdot}p_{4}}\,, \quad 2\alpha'q_{4}{\cdot}p_{3}={q_{1}{\cdot}p_{3}\over q_{1}{\cdot}p_{4}}
\ee

The transverse polarizations
\be
\zeta_{n}^{(1,4)\mu}=\lambda_{n}^{(1,4)\mu}-2\alpha'\lambda^{(1,4)}_{n}{\cdot}p_{1,4}\,q_{1,4}^{\mu}
\ee
are constructed from DDF polarizations $\lambda_{n}^{(1,4)}$, which are only constrained by the conditions $\lambda_{n}^{(1,4)}{\cdot}q_{1,4}=0$. An independent choice is given by
\be
\lambda_{n}^{(1,4)\mu}=(0,1,0,\vec\Lambda^{(1,4)}_{n})\,;\quad \vec\Lambda^{(1,4)}_{n}{\cdot}\vec\Lambda^{(1,4)}_{n}=0
~.
\ee

In the forward limit (where $\theta{=}0$ and $t=0$), which implies in our convention
\be
p_{1}^{\mu}=-p_{4}^{\mu}\,,\quad p_{2}^{\mu}=-p_{3}^{\mu}\, ,
\ee
 the relevant scalar products are given by
\be
\label{forwtachkine}
 2\alpha'q_{1}{\cdot}p_{2}=-2\alpha'q_{1}{\cdot}p_{3}={E-\sqrt{E^{2}{-}M_{T}^{2}}\over M_{N}}\,,\quad 2\alpha'q_{1}{\cdot}p_{4}={-}1
~,
\ee
for $j=1,2,3,4$, in addition with everything else obtained by using eqs.\ (\ref{q4all}).

For the scattering with photon states we consider similar kinematics by setting
\be
p_{1}=(M_{N},\vec{0})\,,\quad p_{2}=(\omega,0,\omega,\vec{0})
~,
\ee
\be
p_{3}=-(\omega',\omega'\sin\theta,\omega'\cos\theta,\vec{0})\,,\quad p_{4}=-(M_{N}{+}\omega{-}\omega',-\omega'\sin\theta,\omega{-}\omega'\cos\theta,\vec{0})
\ee
and
\be
q_{1}=-{(1,1,0,\vec{0})\over 2\alpha' M_{N}}\,,\, q_{2}=-{(1,1,0,\vec{0})\over 2\alpha' \omega} \,,\,  q_{3}={(1,1,0,\vec{0})\over 2\alpha'(\omega'{-}\omega'\sin\theta)}    \,,\, q_{4}={(1,1,0,\vec{0})\over 2\alpha' (M_{N}{+}\omega{-}\omega'{+}\omega'\sin\theta)}
.
\ee

\section{Generating function of scattering amplitudes with photons}
\label{AppC}

In this appendix we consider in more detail the 4-point scattering amplitudes of string states with photons. The discussion follows closely the computations of Section \ref{4point} with the necessary modifications induced by the special features of the photon states. Accordingly, the generating function of interest takes the general form
\be
\label{open1acph}
{\cal A}_{gen}^{HHAA}(s,t)=g_o^2
\int_{{\DD}_2}
\prod_{\ell=1}^{4}dz_{\ell}\,\Big\langle V_{{\cal C}}(p_{1},z_{1})\,V_{\gamma}(p_{2},z_{2})\,V_{\gamma}(p_{3},z_{3})\,V_{{\cal C}}(p_{4},z_{4}) \Big\rangle
~,
\ee
where $V_\CC$ and $V_\gamma$ are, respectively, the coherent vertex operators of excited open string states and the vertex operators of the photons (as defined in eqs.\ \eqref{not4ab} and \eqref{open1ab} in the main text).

Taking into account the extra Wick contractions of the photon operators with the coherent vertex operators, one can show that
\be
\label{AgenPhot}
{\cal A}_{gen}^{HHAA}(s,t)={\cal F}(A_{2,3};\partial_{\beta_{s,t}}){\cal A}_{gen}^{HHTT}(s,t)
~,
\ee
with the dressing factor
\be\label{Dressphot}
\begin{split}
{\cal F}(A_{2,3};\partial_{\beta_{s,t}})&\equiv\Bigg(A_{2}{\cdot}A_{3}\partial_{\beta_{s}}+\left(\sqrt{2\alpha'}A_{2}{\cdot}p_{3}+\sqrt{2\alpha'}A_{2}{\cdot}p_{4}\partial_{\beta_{t}}\right)\left(\sqrt{2\alpha'}A_{3}{\cdot}p_{4}+\sqrt{2\alpha'}A_{3}{\cdot}p_{1}\partial_{\beta_{s}}\right)\\
%%%
&+A_{2}{\cdot}\zeta^{(1)}_{n}(-)^{n+1}\partial_{\beta_{t}}P_{n-1}^{\beta_{1}+1,\alpha_{1}-1}(\partial_{\beta_{s}}{-}\partial_{\beta_{t}})\left(\sqrt{2\alpha'}A_{3}{\cdot}p_{4}+\sqrt{2\alpha'}A_{3}{\cdot}p_{1}\partial_{\beta_{s}}\right)\\
%%%
&+A_{2}{\cdot}\zeta^{(4)}_{n}\partial^{n}_{\beta_{s}}\partial_{\beta_{t}}P_{n-1}^{\alpha_{4}+1,\sigma_{4}-1}\left(1{+}2{\partial_{\beta_{t}}\over \partial_{\beta_{s}}}\right)\left(\sqrt{2\alpha'}A_{3}{\cdot}p_{4}+\sqrt{2\alpha'}A_{3}{\cdot}p_{1}\partial_{\beta_{s}}\right)\\
%%%
&+A_{3}{\cdot}\zeta^{(1)}_{n}\partial^{n}_{\beta_{t}}\partial_{\beta_{s}}P_{n-1}^{\alpha_{1}+1,\sigma_{1}-1}\left(1{+}2{\partial_{\beta_{s}}\over \partial_{\beta_{t}}}\right)\left(\sqrt{2\alpha'}A_{2}{\cdot}p_{3}+\sqrt{2\alpha'}A_{2}{\cdot}p_{4}\partial_{\beta_{t}}\right)\\
%%%
&+A_{3}{\cdot}\zeta_{n}^{(4)}\partial_{\beta_{t}}P_{n-1}^{\alpha_{4}+1,\beta_{4}-1}(\partial_{\beta_{s}}{-}\partial_{\beta_{t}})\left(\sqrt{2\alpha'}A_{2}{\cdot}p_{3}+\sqrt{2\alpha'}A_{2}{\cdot}p_{4}\partial_{\beta_{t}}\right)\\
%%%
&+A_{2}{\cdot}\zeta_{n}^{(1)}A_{3}{\cdot}\zeta_{n}^{(1)}(-)^{n+1}\partial_{\beta_{t}}^{n+1}\partial_{\beta_{s}}P_{n-1}^{\alpha_{1}+1,\sigma_{1}-1}\left(1{+}2{\partial_{\beta_{s}}\over \partial_{\beta_{t}}}\right)P_{n-1}^{\beta_{1}+1,\alpha_{1}-1}(\partial_{\beta_{s}}{-}\partial_{\beta_{t}})\\
%%%
&+A_{2}{\cdot}\zeta_{n}^{(4)}A_{3}{\cdot}\zeta_{n}^{(4)}\partial^{n}_{\beta_{s}}\partial^{2}_{\beta_{t}} P_{n-1}^{\alpha_{4}+1,\sigma_{4}-1}\left(1{+}2{\partial_{\beta_{t}}\over \partial_{\beta_{s}}}\right)P_{n-1}^{\alpha_{4}+1,\beta_{4}-1}(\partial_{\beta_{s}}{-}\partial_{\beta_{t}})\\
%%%
&+A_{2}{\cdot}\zeta_{n}^{(1)}A_{3}{\cdot}\zeta_{n}^{(4)}(-)^{n+1}\partial_{\beta_{t}}^{2}P_{n-1}^{\alpha_{4}+1,\beta_{4}-1}(\partial_{\beta_{s}}{-}\partial_{\beta_{t}})P_{n-1}^{\beta_{1}+1,\alpha_{1}-1}(\partial_{\beta_{s}}{-}\partial_{\beta_{t}})\\
%%%
&+A_{2}{\cdot}\zeta_{n}^{(4)}A_{3}{\cdot}\zeta_{n}^{(1)}\partial_{\beta_{s}}^{n+1}\partial_{\beta_{t}}^{n+1} P_{n-1}^{\alpha_{4}+1,\sigma_{4}-1}\left(1{+}2{\partial_{\beta_{t}}\over \partial_{\beta_{s}}}\right)P_{n-1}^{\alpha_{1}+1,\sigma_{1}-1}\left(1{+}2{\partial_{\beta_{s}}\over \partial_{\beta_{t}}}\right)\Bigg)
~.
\end{split}
\ee
Eq.\ \eqref{AgenPhot} provides a direct relation between the tachyon-amplitude and the photon-amplitude generating functions.

\section{Generating function of general 4-point scattering amplitudes}\label{4HESsc}

As described in \cite{Firrotta:2024qel}, the generating function of four arbitrarily excited string states is given by
\be
{\cal A}_{gen}^{4HES}(s,t)={\cal A}_{Ven}(s,t)\,e^{{\cal K}_4\left(\{\zeta^{(\ell)}_{n}\};\partial_{\beta_{s}},\partial_{\beta_{t}}\right)}\Phi_{\beta_{s},\beta_{t}}\left(s,t\right)\Big|_{\beta_{s,t}{=}0}\ee
with
\be
{\cal A}_{Ven}(s,t)=g_{o}^{2}{\Gamma({-}\ell_{s}^{2}s{-}1)\Gamma({-}\ell_{s}^{2}t{-}1)\over \Gamma({-}\ell_{s}^{2}s{-}\ell_{s}^{2}t{-}2)}\,,
\ee
the exponential dressing factor with exponent
\be\label{Kfunc}
\begin{split}
{\cal K}_4\left(\{\zeta^{(\ell)}_{n}\}; \partial_{\beta_{s}},\partial_{\beta_{t}}\right)=&\sum_{\ell=1}^{4}\left(\sum_{n}\zeta_{n}^{(\ell)}{\cdot}V^{(\ell)}_{n}(\partial_{\beta_s},\d_{\beta_t}){+}\sum_{n,m} \zeta_{n}^{(\ell)}{\cdot}\zeta_{m}^{(m)}W_{n,m}^{(\ell)}(\partial_{\beta_s}, \d_{\beta_t})\right)\\
%%%
&\quad\quad+\sum_{v<f=1}^{4}\sum_{n,m}\zeta^{(v)}_{n}{\cdot}\zeta^{(f)}_{m} I_{n,m}^{(v,f)}(\partial_{\beta_s}, \d_{\beta t})\,,
\end{split}
\ee
and the pole-free function
\be\label{ffunc}
\Phi_{\beta_{s},\beta_{t}}(s,t)=\sum_{r=0}^{\infty}\sum_{v=0}^{\infty}{\beta_{s}^{r}\over r!}{\beta_{t}^{v}\over v!}{(-\ell_{s}^{2}s{-}1)_{r}(-\ell_{s}^{2}t{-}1)_{v}\over (-\ell_{s}^{2}s{-}\ell_{s}^{2}t{-}2)_{r+v}}
\ee

The contributions present in the amplitude can be summarized as follows:
\begin{itemize}
\item Terms linear in the HES polarizations
\be\label{exprV2}
V^{(\ell)\mu}_{n}(z)=\sqrt{2}\ell_{s}p^{\mu}_{\ell+1}P_{n-1}^{\hspace{0 mm}^{\hspace{0 mm}^{(\alpha_{\ell}^{(n)},\beta_{\ell}^{(n)})}}}\hspace{-9mm}(1{-}2 R_{\ell})+R_{\ell}\sqrt{2}\ell_{s}p^{\mu}_{\ell+2} P_{n-1}^{\hspace{0 mm}^{\hspace{0 mm}^{(\alpha_{\ell}^{(n)}{+}1,\beta_{\ell}^{(n)})}}}\hspace{-12mm}(1{-}2 R_{\ell})
\ee
\item Terms bilinear in the HES polarizations of the same state
\be\label{finW}
W^{(\ell)}_{n_{\ell},m_{\ell}}(z)=\sum_{r=1}^{m_{\ell}}r\,P_{n_{\ell}+r}^{\hspace{0 mm}^{\hspace{0 mm}^{(\alpha_{\ell}^{(n_{\ell})}-r,\beta_{\ell}^{(n_{\ell})}-1)}}}\hspace{-16.5mm}(1{-}2 R_{\ell})\,\,P_{m_{\ell}-r}^{\hspace{0 mm}^{\hspace{0 mm}^{(\alpha_{\ell}^{(m_{\ell})}+r,\beta_{\ell}^{(m_{\ell})}-1)}}}\hspace{-17mm}(1{-}2 R_{\ell})
\ee
\item Terms bilinear in HES polarizations of adjacent states
\be\label{Ibv}
\begin{split}
I^{(\ell,\ell-1)}_{n_{\ell},m_{\ell-1}}(z)=&(-)^{m_{\ell-1}{+}1}\sum_{r,s=0}^{n_{\ell},m_{\ell-1}}R_{\ell}^{r+s+1}
%%%
P_{n_{\ell}{-}r{-}s{-}1}^{\hspace{0 mm}^{\hspace{0 mm}^{(\alpha_{\ell}^{(n_{\ell})}{+}r{+}s{+}1;\,\beta_{\ell}^{(n_{\ell})}{-}1)}}}\hspace{-16mm}(1{-}2 R_{\ell})\,\,P_{m_{\ell-1}{-}r{-}s{-}1}^{\hspace{0 mm}^{\hspace{0 mm}^{(\beta_{\ell-1}^{(m_{\ell-1})}+r+s+1;\,\alpha_{\ell-1}^{(m_{\ell-1})}-1)}}}\hspace{-20mm}(1{-}2 R_{\ell})
\end{split}
\ee
\item Terms bilinear in HES polarizations of non-adjacent states
\be\label{Ibl}
\begin{split}
I^{(\ell,\ell+2)}_{n_{\ell},m_{\ell+2}}(z)=&(R_{\ell+1})^{m_{\ell+2}{+}n_{\ell}}\sum_{r,s=0}^{n_{\ell},m_{\ell-1}}\left({R_{\ell}\over R_{\ell+1}}\right)^{r+s+1}\\
%%%
&P_{n_{\ell}{-}r{-}s{-}1}^{\hspace{0 mm}^{\hspace{0 mm}^{(\alpha_{\ell}^{(n_{\ell})}{+}r{+}s{+}1;\,\sigma_{\ell}^{(n_{\ell})}{-}1)}}}\hspace{-16mm}(1{+}2 {R_{\ell}/ R_{\ell+1}})\,\,P_{m_{\ell+2}{-}r{-}s{-}1}^{\hspace{0 mm}^{\hspace{0 mm}^{(\alpha_{\ell+2}^{(m_{\ell+2})}+r+s+1;\,\sigma_{\ell+2}^{(m_{\ell+2})}-1)}}}\hspace{-20mm}(1{+}2 R_{\ell}/R_{\ell+1})
\end{split}
\ee
\end{itemize}
with coefficients
\be
R_{1}={z_{21}z_{34}\over z_{24}z_{31}}=z\,,\quad R_{2}={z_{32}z_{41}\over z_{31}z_{42}}=1{-}z\,,\quad R_{3}={z_{43}z_{12}\over z_{42}z_{13}}=z\,,\quad R_{4}={z_{14}z_{23}\over z_{13}z_{24}}=1{-}z
\ee
and
\be\label{coeffJac}
\alpha_{\ell}^{(n)}=-n-2\ell_{s}^{2}nq_{\ell}{\cdot}p_{\ell+1}\,,\quad \beta_{\ell}^{(n)}=-n-2\ell_{s}^{2}nq_{\ell}{\cdot}p_{\ell-1}\,,\quad  \sigma_{\ell}^{(n)}=-n-2\ell_{s}^{2}n q_{\ell}{\cdot}p_{\ell+2}
~.
\ee
Every contribution is expressed in terms of Jacobi polynomials
\be
P_{N}^{(\alpha,\beta)}(x)=\sum_{r=0}^{N}\begin{pmatrix}N+\alpha\\N-r\end{pmatrix}\begin{pmatrix}N+\beta\\r\end{pmatrix}\left({x-1\over 2}\right)^{r}\left({x+1\over 2}\right)^{N-r}
~.
\ee
and with the identification \eqref{idVen}, in analogy with \eqref{finI14td}-\eqref{exprze1td}, one can express \eqref{exprV2}-\eqref{Ibl} in terms of polynomial derivative actions.

\section{Examples of 4-point scattering amplitudes}\label{exemAmplLR}

\subsection{Leading Regge states at level $N{=}2$ and two tachyons}

Computing the derivative action
\be\label{DApolD}
{1\over 2!}\left(\zeta_1^{(1,1)}{\cdot}{\d \over\d\zeta^{(1)}_{1}}\right)^2
\left(\zeta_1^{(4,1)}{\cdot}{\d\over \d\zeta^{(4)}_{1}}\right)^2
{\cal A}_{gen}^{HHTT}(s,t)\Bigg|_{\zeta_1^{(1,4)}=0}
\ee
we obtain the following result for the scattering amplitude of two tachyons and two level-2 string states
\bea\label{eqD2}
&&{\cal A}_{H_2{+}T\rightarrow H_2{+}T}(s,t)={{\cal A}_{Ven}(s,t)\over ( \alpha's {+} \alpha't{-}1) (\alpha's {+} \alpha't) (1 {+} \alpha's {+}
      \alpha't) (2 {+} \alpha's {+} \alpha't)}\nonumber\\
      %%%
    &&\Bigg(2 \left(\zeta_{1}^{(1)}{\cdot}\zeta_{1}^{(4)}\right)^2 \alpha's (1{+}\alpha's) (\alpha's {+} \alpha't{-}1) (\alpha's {+} \alpha't)  +
   2\alpha'\left(\zeta^{(1)}_{1}{\cdot}p_{3}\right)^2 \alpha's (1 {+} \alpha's)\nonumber\\
   %%%
  && \Big(2\alpha'\left(\zeta^{(4)}_{1}{\cdot}p_{2}\right)^2 \alpha't (1 {+} \alpha't)
 +4 \alpha'  \zeta^{(4)}_{1}{\cdot}p_{1}\zeta^{(4)}_{1}{\cdot}p_{2}(1 {+} \alpha't) (\alpha's {+} \alpha't{-}1)\nonumber\\
      %%%
      && + 2\alpha'\left(\zeta^{(4)}_{1}{\cdot}p_{1}\right)^2 ( \alpha's {+} \alpha't{-}1) (\alpha's {+} \alpha't)\Big)+ 4\alpha' \zeta^{(1)}_{1}{\cdot}p_{2} \zeta^{(1)}_{1}{\cdot}p_{3} (1 {+} \alpha's) ( \alpha's {+} \alpha't{-}1)\nonumber \\
   %%%
   &&\Big(2\alpha'\left(\zeta^{(4)}_{1}{\cdot}p_{2}\right)^2 \alpha't (1 {+} \alpha't) +
      4\alpha' \zeta^{(4)}_{1}{\cdot}p_{1} \zeta^{(4)}_{1}{\cdot}p_{2} (1 {+} \alpha't) (\alpha's {+} \alpha't){+} 2\alpha'\left(\zeta^{(4)}_{1}{\cdot}p_{1}\right)^2\nonumber\\
      %%%
   &&(\alpha's {+} \alpha't) (1 {+} \alpha's {+} \alpha't)\Big)    +2\alpha' \left( \zeta^{(1)}_{1}{\cdot}p_{2}\right)^2 (\alpha's {+}\alpha' t{-}1) (\alpha's {+} \alpha't) \Big(2\alpha'\left(\zeta^{(4)}_{1}{\cdot}p_{2}\right)^2 \alpha't (1 {+} \alpha't)\nonumber \\
   %%%
   &&+4\alpha' \zeta^{(4)}_{1}{\cdot}p_{1} \zeta^{(4)}_{1}{\cdot}p_{2} (1 {+} \alpha't) (1 {+} \alpha's {+} \alpha't) + 2\alpha'\left(\zeta^{(4)}_{1}{\cdot}p_{1}\right)^2 (1 {+} \alpha's {+} \alpha't) (2 {+} \alpha's {+} \alpha't)\Big)\nonumber\\
      %%%
      && +
   4 \zeta_{1}^{(1)}{\cdot}\zeta_{1}^{(4)} (1 {+} \alpha's) ( \alpha's {+}
     \alpha' t{-}1) \Big(\sqrt{2\alpha'}\zeta^{(1)}_{1}{\cdot}p_{3} \alpha's \left(\sqrt{2\alpha'}\zeta^{(4)}_{1}{\cdot}p_{2} {+} \sqrt{2\alpha'}\zeta^{(4)}_{1}{\cdot}p_{1} \alpha's {-}\sqrt{2\alpha'}\zeta_{1}^{(4)}{\cdot}p_{3}\alpha't\right)\nonumber \\
      %%%
      &&+ \sqrt{2\alpha'} \zeta^{(1)}_{1}{\cdot}p_{2} (\alpha's {+} \alpha't) \left(-\sqrt{2\alpha'}\zeta^{(4)}_{1}{\cdot}p_{3}  {+}
         \sqrt{2\alpha'}\zeta^{(4)}_{1}{\cdot}p_{1} \alpha's {-}\sqrt{2\alpha'}\zeta_{1}^{(4)}{\cdot}p_{3} \alpha't\right)\Big)\Bigg)
         ~.
\eea

\subsection{Two $J{=}1$ states and two tachyons}\label{AppD2}

In this subsection we are considering the scattering amplitude of two tachyons with two string states of occupation number $J=1$. As a warmup, we first analyse the case of $J=1$ states at level $N=2$. The relevant amplitude is generated by the following derivative action
\be\label{DApolJ1}
{\cal A}_{H^{1}_2{+}T\rightarrow H^{1}_2{+}T}(s,t)={1\over 2}\left(\zeta_2^{(1,1)}{\cdot}{\d \over\d\zeta^{(1)}_{2}}\right)
\left(\zeta_2^{(4,1)}{\cdot}{\d\over \d\zeta^{(4)}_{2}}\right)
{\cal A}_{gen}^{HHTT}(s,t)\Bigg|_{\zeta_2^{(1,4)}=0}
~,
\ee
which produces
\be
{{\cal A}_{Ven}(s,t)\over 2}\Big( \zeta_{2}^{(1)}{\cdot}V_{2}^{(1)}(\partial_{\beta_{s}},\partial_{\beta_{t}})\zeta_{2}^{(4)}{\cdot}V_{2}^{(4)}(\partial_{\beta_{s}},\partial_{\beta_{t}})+\zeta_{2}^{(1)}{\cdot}\zeta_{2}^{(4)}I_{2,2}^{(1,4)}(\partial_{\beta_{s}},\partial_{\beta_{t}})\Big)\Phi_{\beta_{s},\beta_{t}}(s,t)\Big|_{\beta_{s,t}=0}
~.
\ee
Using the explicit form of the polynomial differential operators
\bea
V_{2}^{(1)\mu}(\partial_{\beta_{s}},\partial_{\beta_{t}})&=&-\sqrt{2\alpha'}p^{\mu}_{2}\left(\partial_{\beta_{s}}+4\alpha'q_{1}{\cdot}p_{4}\partial_{\beta_{s}}+\partial_{\beta_{t}}+4\alpha'q_{1}{\cdot}p_{2}\partial_{\beta_{t}}\right)
~,\nonumber\\
%%%
&&-\sqrt{2\alpha'}p^{\mu}_{3}\partial_{\beta_{s}}\left(\partial_{\beta_{s}}+4\alpha'q_{1}{\cdot}p_{4}\partial_{\beta_{s}}+4\alpha' q_{1}{\cdot}p_{2}\partial_{\beta_{t}}\right)
~,\\
V_{2}^{(4)\mu}(\partial_{\beta_{s}},\partial_{\beta_{t}})&=&-\sqrt{2\alpha'}p^{\mu}_{1}\left(\partial_{\beta_{s}}+4\alpha'q_{4}{\cdot}p_{1}\partial_{\beta_{s}}+\partial_{\beta_{t}}+4\alpha'q_{4}{\cdot}p_{3}\partial_{\beta_{t}}\right)
\nonumber\\
%%%
&&-\sqrt{2\alpha'}p^{\mu}_{2}\partial_{\beta_{s}}\left(\partial_{\beta_{t}}+4\alpha'q_{4}{\cdot}p_{1}\partial_{\beta_{s}}+4\alpha' q_{4}{\cdot}p_{3}\partial_{\beta_{t}}\right)
~, \\
I_{2,2}^{(1,4)}(\partial_{\beta_{s}},\partial_{\beta_{t}})&=&-\Big(2 \partial_{\beta_{s}}^2 + 4 \partial_{\beta_{s}} \Big((1 {+}  2\alpha'q_{1}{\cdot}p_{4}) \partial_{\beta_{s}} +2\alpha'q_{1}{\cdot}p_{2} \partial_{\beta_{t}}\Big)
\nonumber\\
%%%
&&\hspace{3.5cm} \Big((1 +  2\alpha'q_{4}{\cdot}p_{1}) \partial_{\beta_{s}}+  2 \alpha' q_{4}{\cdot}p_{3} \partial_{\beta_{t}}\Big)\Big)
\eea
one obtains the explicit result
\be
\begin{split}
&{\cal A}_{H^{1}_2{+}T\rightarrow H^{1}_2{+}T}(s,t)={{\cal A}_{Ven}(s,t)\over ( \alpha's {+} \alpha't{-}1) (\alpha's {+} \alpha't) (1 {+} \alpha's {+}
      \alpha't) (2 {+} \alpha's {+} \alpha't)}\\
%%%
&\Big(-2 \zeta_{2}^{(1)}{\cdot}\zeta_{2}^{(4)} (1{+}\alpha's) (\alpha' s {+}
     \alpha' t{-}1) \Big((3 {+} 4\alpha' q_{4}{\cdot}p_{1} + 4\alpha' q_{1}{\cdot}p_{4} (1 + 2\alpha'q_{4}{\cdot}p_{1})) (\alpha's)^2 \\
      %%%
      &+
      4\alpha' q_{1}{\cdot}p_{2} q_{4}{\cdot}p_{3} \alpha't (1 {+} \alpha't) {+}
      \alpha's \Big(-2 {-} 4\alpha' q_{4}{\cdot}p_{1} {+} 4\alpha' q_{4}{\cdot}p_{3} {+} \alpha't {+} 4\alpha' q_{4}{\cdot}p_{3} \alpha't \\
      %%%
      &+
         4\alpha' q_{1}{\cdot}p_{2} (1 {+} 2\alpha' q_{4}{\cdot}p_{1}) (1 {+} \alpha't) -
         4\alpha' q_{1}{\cdot}p_{4} (1 + 2\alpha'q_{4}{\cdot}p_{1} -2\alpha' q_{4}{\cdot}p_{3} (1 {+} \alpha't))\Big)\Big) \\
         %%%
         &+ \sqrt{2\alpha'}\zeta_{2}^{(1)}{\cdot}p_{3}(1 {+}\alpha' s) \Big(\sqrt{2\alpha'}\zeta_{2}^{(4)}{\cdot}p_{2} (1 {+}\alpha'
         t)\Big((1 {+} 4\alpha' q_{4}{\cdot}p_{3}) (\alpha's {+} 4 \alpha'q_{1}{\cdot}p_{4} \alpha's \\
         %%%
         &+ 4\alpha' q_{1}{\cdot}p_{2} (-1 {+}\alpha' t)) \alpha't +
         4\alpha' q_{4}{\cdot}p_{1} \alpha's (-1 + 4\alpha' q_{1}{\cdot}p_{4} (-1 {+} \alpha's) + \alpha's + 4 q_{1}{\cdot}p_{2} \alpha't)\Big)\\
         %%%
         & +\sqrt{2\alpha'} \zeta_{2}^{(4)}{\cdot}p_{1}( \alpha's {+}
         \alpha't{-}1) \Big(4\alpha' q_{1}{\cdot}p_{2} (1 {+} \alpha't) (\alpha's {+} 4\alpha' q_{4}{\cdot}p_{1} \alpha's {+} \alpha't {+} 4\alpha' q_{4}{\cdot}p_{3}\alpha't)\\
         %%%
         & + (1 {+}
            4\alpha' q_{1}{\cdot}p_{4}) \alpha's \Big(4\alpha' q_{4}{\cdot}p_{1} ( \alpha's{-}1) {+} \alpha's {+} \alpha't {+} 4\alpha' q_{4}{\cdot}p_{3} (1 {+} \alpha't)\Big)\Big)\Big)\\
            %%%
            & + \sqrt{2\alpha'}\zeta_{2}^{(1)}{\cdot}p_{2}( \alpha's {+}\alpha' t{-}1) \Big(\sqrt{2\alpha'}\zeta_{2}^{(4)}{\cdot}p_{2}(1 {+}\alpha'
         t) \Big((1 {+} 4\alpha' q_{4}{\cdot}p_{3}) \alpha't\Big(\alpha's + 4 q_{1}{\cdot}p_{4} (1 {+} \alpha's) \\
         %%%
         & + 4\alpha' q_{1}{\cdot}p_{2} (\alpha't{-}1) + \alpha't\Big) + 4\alpha' q_{4}{\cdot}p_{1} (1 {+} \alpha's) (\alpha's {+} 4\alpha' q_{1}{\cdot}p_{4} \alpha's {+} \alpha't {+} 4\alpha' q_{1}{\cdot}p_{2} \alpha't)\Big) \\
         %%%
         &+ \sqrt{2\alpha'}\zeta_{2}^{(4)}{\cdot}p_{1} (\alpha's {+}
         \alpha't) \Big(4\alpha' q_{1}{\cdot}p_{2} (1 {+} \alpha't) (1 {+} \alpha's {+} 4\alpha' q_{4}{\cdot}p_{1} (1 {+} \alpha's) {+} \alpha't {+} 4 \alpha' q_{4}{\cdot}p_{3} \alpha't)\\
         %%%
         & + 4\alpha' q_{1}{\cdot}p_{4} (1 {+} \alpha's) (1 {+} \alpha's {+} 4\alpha' q_{4}{\cdot}p_{1} \alpha's {+} \alpha't {+} 4\alpha' q_{4}{\cdot}p_{3} (1 {+} \alpha't)) {+} (1 {+}
             \alpha's {+} \alpha't) \\
             %%%
             &(2 {+} \alpha's {+} 4\alpha' q_{4}{\cdot}p_{1} (1 {+} \alpha's) {+} \alpha't {+}
            4\alpha' q_{4}{\cdot}p_{3} (1 {+} \alpha't))\Big)\Big)\Big)
            ~.
            \end{split}
\ee

More generally, the scattering amplitude with a state of occupation number 1 at level $N$ and a tachyon can be computed from
\be\label{DApolJ1}
{\cal A}_{H^{1}_N{+}T\rightarrow H^{1}_N{+}T}(s,t)={1\over N}\left(\zeta_N^{(1,1)}{\cdot}{\d \over\d\zeta^{(1)}_{N}}\right)
\left(\zeta_N^{(4,1)}{\cdot}{\d\over \d\zeta^{(4)}_{N}}\right)
{\cal A}_{gen}^{HHTT}(s,t)\Bigg|_{\zeta_N^{(1,4)}=0}
\ee
yielding the result
\be\label{ampJ1}
\begin{split}
&{\cal A }_{H_{N}^{1}{+}T\rightarrow H_{N}^{1}{+}T}=A_{Ven}(s,t)\\
%%%
&\Bigg( 2\alpha' \zeta_{N}^{(1)}{\cdot}p_{2}\zeta_{N}^{(4)}{\cdot}p_{1} \sum_{r_{1},r_{4}=0}^{N-1}\begin{pmatrix}N{-}1{+}\alpha_{1}\\N{-}1{-}r_{1} \end{pmatrix}\begin{pmatrix}N{-}1{+}\alpha_{4}\\N{-}1{-}r_{4} \end{pmatrix}\begin{pmatrix}N{-}1{+}\beta_{1}\\r_{1} \end{pmatrix}\begin{pmatrix}N{-}1{+}\beta_{4}\\r_{4} \end{pmatrix} {\cal Q}^{[s;N-1+r_{1}-r_{4}]}_{[t;N-1+r_{4}-r_{1}]}\\
%%%
&+2\alpha' \zeta_{N}^{(1)}{\cdot}p_{2}\zeta_{N}^{(4)}{\cdot}p_{2} \sum_{r_{1},r_{4}=0}^{N-1}\begin{pmatrix}N{-}1{+}\alpha_{1}\\N{-}1{-}r_{1} \end{pmatrix}\begin{pmatrix}N{+}\alpha_{4}\\N{-}1{-}r_{4} \end{pmatrix}\begin{pmatrix}N{-}1{+}\beta_{1}\\r_{1} \end{pmatrix}\begin{pmatrix}N{-}1{+}\beta_{4}\\r_{4} \end{pmatrix} {\cal Q}^{[s;N-1+r_{1}-r_{4}]}_{[t;N+r_{4}-r_{1}]}\\
%%%
&+2\alpha' \zeta_{N}^{(1)}{\cdot}p_{3}\zeta_{N}^{(4)}{\cdot}p_{1} \sum_{r_{1},r_{4}=0}^{N-1}\begin{pmatrix}N{+}\alpha_{1}\\N{-}1{-}r_{1} \end{pmatrix}\begin{pmatrix}N{-}1{+}\alpha_{4}\\N{-}1{-}r_{4} \end{pmatrix}\begin{pmatrix}N{-}1{+}\beta_{1}\\r_{1} \end{pmatrix}\begin{pmatrix}N{-}1{+}\beta_{4}\\r_{4} \end{pmatrix} {\cal Q}^{[s;N+r_{1}-r_{4}]}_{[t;N-1+r_{4}-r_{1}]}\\
%%%
&+2\alpha' \zeta_{N}^{(1)}{\cdot}p_{3}\zeta_{N}^{(4)}{\cdot}p_{2}\sum_{r_{1},r_{4}=0}^{N-1}\begin{pmatrix}N{+}\alpha_{1}\\N{-}1{-}r_{1} \end{pmatrix}\begin{pmatrix}N{+}\alpha_{4}\\N{-}1{-}r_{4} \end{pmatrix}\begin{pmatrix}N{-}1{+}\beta_{1}\\r_{1} \end{pmatrix}\begin{pmatrix}N{-}1{+}\beta_{4}\\r_{4} \end{pmatrix} {\cal Q}^{[s;N+r_{1}-r_{4}]}_{[t;N+r_{4}-r_{1}]}\\
%%%
&+\zeta_{N}^{(1)}{\cdot}\zeta_{N}^{(4)}(-)^{N-1}\sum_{v_{1},v_{4}=0}^{N-1}\sum_{r_{1},r_{4}=0}^{N-1}\begin{pmatrix}N{+}\alpha_{1}\\N{-}v_{1}{-}v_{4}{-}1{-}r_{1} \end{pmatrix}\begin{pmatrix}N{-}v_{1}{-}v_{4}{-}2{+}\beta_{1}\\r_{1} \end{pmatrix}\\
%%%
&\hspace{4.3cm}\begin{pmatrix}N{+}\beta_{4}\\N{-}v_{1}{-}v_{4}{-}1{-}r_{4} \end{pmatrix}\begin{pmatrix}N{-}v_{1}{-}v_{4}{-}2{+}\alpha_{4}\\r_{4} \end{pmatrix}{\cal Q}^{[s;N+r_{1}+r_{4}]}_{[t;2(N-1-v_{1}-v_{4})+r_{1}+r_{4}]}\Bigg)
\end{split}
\ee
where
\be
{\cal Q}^{[s;c_{s}]}_{[t;c_{t}]}:= {(-\alpha's{-}1)_{c_{s}}(-\alpha't{-}1)_{c_{t}}\over (-\alpha's{-}\alpha't{-}2)_{c_{s}+c_{t}}}
\ee
and
\be
\alpha_{1}=-N-2\alpha'Nq_{1}{\cdot}p_{2}\,,\quad \beta_{1}=-N-2\alpha'Nq_{1}{\cdot}p_{4}
~,
\ee
\be
\alpha_{4}=-N-2\alpha'Nq_{4}{\cdot}p_{1}\,,\quad \beta_{4}=-N-2\alpha'Nq_{4}{\cdot}p_{3}
~.
\ee

\subsection{Two Leading Regge states and two photons}\label{AppD3}
The amplitude with two leading Regge states and two photons is given by
\bea
&&{\cal A}_{H_N{+}A\rightarrow H_N{+}A}(s,t)=\Bigg(A_{2}{\cdot}A_{3}{\cal R}_{N[t;0]}^{(0)[s;1]}(\zeta_{1},\zeta_{4};{\cal Q})+2\alpha'A_{2}{\cdot}p_{3}A_{3}{\cdot}p_{4}{\cal R}_{N[t;0]}^{(0)[s;0]}(\zeta_{1},\zeta_{4};{\cal Q})
\nonumber\\
%%%
&&+2\alpha'A_{2}{\cdot}p_{3}A_{3}{\cdot}p_{1}{\cal R}_{N[t;0]}^{(0)[s;1]}(\zeta_{1},\zeta_{4};{\cal Q})+2\alpha'A_{2}{\cdot}p_{4}A_{3}{\cdot}p_{4}{\cal R}_{N[t;1]}^{(0)[s;0]}(\zeta_{1},\zeta_{4};{\cal Q})
\nonumber\\
%%%
&&+2\alpha'A_{2}{\cdot}p_{4}A_{3}{\cdot}p_{1}{\cal R}_{N[t;1]}^{(0)[s;1]}(\zeta_{1},\zeta_{4};{\cal Q})+ {2\alpha'\over N}A_{2}{\cdot}\zeta_{1}\zeta_{4}{\cdot}p_{1}A_{3}{\cdot}p_{4}{\cal R}_{N-1[t;0]}^{(1)[s;-1]}(\zeta_{1},\zeta_{4};{\cal Q})
\nonumber\\
%%%
&&+ {2\alpha'\over N}A_{2}{\cdot}\zeta_{1}\zeta_{4}{\cdot}p_{2}A_{3}{\cdot}p_{4}{\cal R}_{N-1[t;1]}^{(1)[s;-1]}(\zeta_{1},\zeta_{4};{\cal Q})+{2\alpha'\over N}A_{2}{\cdot}\zeta_{1}\zeta_{4}{\cdot}p_{1}A_{3}{\cdot}p_{1}{\cal R}_{N-1[t;0]}^{(1)[s;0]}(\zeta_{1},\zeta_{4};{\cal Q})
\nonumber\\
%%%
&&+ {2\alpha'\over N}A_{2}{\cdot}\zeta_{1}\zeta_{4}{\cdot}p_{2}A_{3}{\cdot}p_{4}{\cal R}_{N-1[t;1]}^{(1)[s;0]}(\zeta_{1},\zeta_{4};{\cal Q})+ {2\alpha'\over N}A_{2}{\cdot}\zeta_{4}\zeta_{1}{\cdot}p_{2}A_{3}{\cdot}p_{4}{\cal R}_{N-1[t;1]}^{(1)[s;0]}(\zeta_{1},\zeta_{4};{\cal Q})
\nonumber\\
%%%
&&+ {2\alpha'\over N}A_{2}{\cdot}\zeta_{4}\zeta_{1}{\cdot}p_{3}A_{3}{\cdot}p_{4}{\cal R}_{N-1[t;1]}^{(1)[s;1]}(\zeta_{1},\zeta_{4};{\cal Q})+ {2\alpha'\over N}A_{2}{\cdot}\zeta_{4}\zeta_{1}{\cdot}p_{2}A_{3}{\cdot}p_{1}{\cal R}_{N-1[t;1]}^{(1)[s;1]}(\zeta_{1},\zeta_{4};{\cal Q})
\nonumber\\
%%%%
&&+ {2\alpha'\over N}A_{2}{\cdot}\zeta_{4}\zeta_{1}{\cdot}p_{3}A_{3}{\cdot}p_{1}{\cal R}_{N-1[t;1]}^{(1)[s;2]}(\zeta_{1},\zeta_{4};{\cal Q})+{2\alpha'\over N}A_{3}{\cdot}\zeta_{1}\zeta_{4}{\cdot}p_{1}A_{2}{\cdot}p_{3}{\cal R}_{N-1[t;1]}^{(1)[s;0]}(\zeta_{1},\zeta_{4};{\cal Q})
\nonumber\\
%%%
&&+{2\alpha'\over N}A_{3}{\cdot}\zeta_{1}\zeta_{4}{\cdot}p_{2}A_{2}{\cdot}p_{3}{\cal R}_{N-1[t;2]}^{(1)[s;0]}(\zeta_{1},\zeta_{4};{\cal Q})+{2\alpha'\over N}A_{3}{\cdot}\zeta_{1}\zeta_{4}{\cdot}p_{1}A_{2}{\cdot}p_{4}{\cal R}_{N-1[t;2]}^{(1)[s;0]}(\zeta_{1},\zeta_{4};{\cal Q})
\nonumber\\
%%%
&&+{2\alpha'\over N}A_{3}{\cdot}\zeta_{1}\zeta_{4}{\cdot}p_{2}A_{2}{\cdot}p_{4}{\cal R}_{N-1[t;3]}^{(1)[s;0]}(\zeta_{1},\zeta_{4};{\cal Q})+ {2\alpha'\over N}A_{3}{\cdot}\zeta_{4}\zeta_{1}{\cdot}p_{2}A_{2}{\cdot}p_{3}{\cal R}_{N-1[t;1]}^{(1)[s;-1]}(\zeta_{1},\zeta_{4};{\cal Q})
\nonumber\\
%%%
&&+ {2\alpha'\over N}A_{3}{\cdot}\zeta_{4}\zeta_{1}{\cdot}p_{3}A_{2}{\cdot}p_{3}{\cal R}_{N-1[t;1]}^{(1)[s;0]}(\zeta_{1},\zeta_{4};{\cal Q})+ {2\alpha'\over N}A_{3}{\cdot}\zeta_{4}\zeta_{1}{\cdot}p_{2}A_{2}{\cdot}p_{4}{\cal R}_{N-1[t;2]}^{(1)[s;-1]}(\zeta_{1},\zeta_{4};{\cal Q})
\nonumber\\
%%%
&&+ {2\alpha'\over N}A_{3}{\cdot}\zeta_{4}\zeta_{1}{\cdot}p_{2}A_{2}{\cdot}p_{4}{\cal R}_{N-1[t;2]}^{(1)[s;0]}(\zeta_{1},\zeta_{4};{\cal Q})+ {2\alpha'A_{2}{\cdot}\zeta_{1}A_{3}{\cdot}\zeta_{1}\over N(N{-}1)}\zeta_{4}{\cdot}p_{1}\zeta_{4}{\cdot}p_{1}{\cal R}_{N-2[t;2]}^{(2)[s;-1]}(\zeta_{1},\zeta_{4};{\cal Q})
\nonumber\\
%%%
&&+ {4\alpha'A_{2}{\cdot}\zeta_{1}A_{3}{\cdot}\zeta_{1}\over N(N{-}1)}\zeta_{4}{\cdot}p_{1}\zeta_{4}{\cdot}p_{2}{\cal R}_{N-2[t;3]}^{(2)[s;-1]}(\zeta_{1},\zeta_{4};{\cal Q})+ {2\alpha'A_{2}{\cdot}\zeta_{1}A_{3}{\cdot}\zeta_{1}\over N(N{-}1)}\zeta_{4}{\cdot}p_{2}\zeta_{4}{\cdot}p_{2}{\cal R}_{N-2[t;4]}^{(2)[s;-1]}(\zeta_{1},\zeta_{4};{\cal Q})
\nonumber\\
%%%
&&+{2\alpha'A_{2}{\cdot}\zeta_{4}A_{3}{\cdot}\zeta_{4}\over N(N{-}1)}\zeta_{1}{\cdot}p_{2}\zeta_{1}{\cdot}p_{2}{\cal R}_{N-2[t;2]}^{(2)[s;-1]}(\zeta_{1},\zeta_{4};{\cal Q})+{4\alpha'A_{2}{\cdot}\zeta_{4}A_{3}{\cdot}\zeta_{4}\over N(N{-}1)}\zeta_{1}{\cdot}p_{2}\zeta_{1}{\cdot}p_{3}{\cal R}_{N-2[t;2]}^{(2)[s;0]}(\zeta_{1},\zeta_{4};{\cal Q})
\nonumber\\
%%%
&&+{2\alpha'A_{2}{\cdot}\zeta_{4}A_{3}{\cdot}\zeta_{4}\over N(N{-}1)}\zeta_{1}{\cdot}p_{3}\zeta_{1}{\cdot}p_{3}{\cal R}_{N-2[t;2]}^{(2)[s;1]}(\zeta_{1},\zeta_{4};{\cal Q})+ {A_{2}{\cdot}\zeta_{1}A_{3}{\cdot}\zeta_{4}\over N}{\cal R}_{N-1[t;2]}^{(0)[s;-1]}(\zeta_{1},\zeta_{4};{\cal Q})
\nonumber\\
%%%
&&+{A_{2}{\cdot}\zeta_{4}A_{3}{\cdot}\zeta_{1}\over N}{\cal R}_{N-1[t;2]}^{(0)[s;1]}(\zeta_{1},\zeta_{4};{\cal Q})\Bigg){\cal A}_{Ven}(s,t)
\eea
with
\be
\begin{split}
{\cal R}_{{{N}}[t;{{v_{t}}}]}^{({{a}})[s; {{v_{s}}}]}(\zeta_{1},\zeta_{4};{\cal Q})&=
\sum_{r=0}^{{{N}}}\begin{pmatrix}{{N}}+{{a}}\\{{N}}{-}r\end{pmatrix}{1\over r!} (\zeta_{1}{\cdot}\zeta_{4})^{{{N}}-r}\sum_{k_{1}+k_{2}+k_{3}+k_4=r}{r!\over k_{1}!k_{2}!k_{3}!k_4!} \\
%%%
&(\zeta_1{\cdot}p_2\zeta_4{\cdot}p_1)^{k_1}(\zeta_1{\cdot}p_3\zeta_4{\cdot}p_1)^{k_2}(\zeta_1{\cdot}p_2\zeta_4{\cdot}p_2)^{k_3}(\zeta_1{\cdot}p_3\zeta_4{\cdot}p_2)^{k_4} {\cal Q}^{[s;{{N}}+k_2+k_4+{{v_{s}}}]}_{[t;k_3+k_4+{{v_{t}}}]}
 \end{split}
\ee

\section{Absorption, emission and time reversal symmetry}\label{AbsEmOpt}

In this appendix we review the relation between absorption and emission implied by time reversal symmetry in a quantum theory.

In four space-time dimensions (for concreteness), the decay $a\to b+c$ of a single-particle state $|a\rangle$ into two single-particle states $|b\rangle$ and $|c\rangle$ can be written as
\be
\Gamma_{em}^{\,a {\rightarrow} b + c}={1\over 2 E_{a}}\int {d^{3}\vec{p}_{b}\over (2\pi)^{3} 2E_{b}}\int {d^{3}\vec{p}_{c}\over (2\pi)^{3} 2E_{c}} (2\pi)^{4}\delta^{(4)}(p_{b}{+}p_{c}{-}p_{a})  \Big|{\cal{S}}_{a\rightarrow b+c}\Big|^{2}
\ee
with
\be
{\cal{S}}_{a\rightarrow b+c}=\Big\langle b(p_{b}),c(p_{c})|{\widehat{\cal S}}|a(p_{a})\Big\rangle
\ee
the relevant $\widehat{S}$ scattering matrix element. Performing the integration over the spacelike momentum $\vec p_c$ yields
\be
\label{Gamma_S}
\Gamma_{em}^{a{\rightarrow} b+c}={1\over 8 (2\pi)^{2} E_{a}}\int {d^{3}\vec{p}_{b}\over E_{b}E_{c}} \delta(E_{b}{+}E_{c}{-}E_{a})  \Big|{\cal{S}}_{a\rightarrow b+c}\Big|^{2}
~.
\ee
In the rest frame of the decaying state
\be
p_{a}=(M_{a},\vec{0})\,,\quad p_{b}=(E_{b},\vec{p_{b}}=\vec{p})\,,\quad p_{c}=(E_{c},\vec{p_{c}}=-\vec{p})
\ee
\be
E_{b}={M_{a}^{2}+M_{b}^{2}-M_{c}^{2}\over 2M_{a}}\,,\quad E_{c}={M_{a}^{2}-M_{b}^{2}+M_{c}^{2}\over 2M_{a}}\,,\quad |\vec{p}|\equiv p={\sqrt{\lambda(M_{a}^{2},M_{b}^{2},M_{c}^{2})}\over 2M_{a}}
\ee
\be
\lambda(M_{a}^{2},M_{b}^{2},M_{c}^{2}) :=\left(M_{a}^{2}-(M_{b}{+}M_{c})^2\right)\left(M_{a}^{2}-(M_{b}{-}M_{c})^2\right)
~.
\ee
Therefore, in spherical coordinates
\be
d^{3}\vec{p}_{b}=dp_{b}\, p_{b}^{2}\, d\Omega_{solid} = dE_{b}\, E_{b}\, p\, d\Omega_{solid}
\ee
and the differential emission rate can be expressed in the following form
\be
{d\Gamma_{em}^{a{\rightarrow} b+c}\over dE_{b}}={\Omega_{solid}\over 8 (2\pi)^{2} } {p\over M_{a}E_{c} }\delta(E_{b}{+}E_{c}{-}E_{a})  \Big|{\cal{S}}_{a\rightarrow b+c}\Big|^{2}
~.
\ee

In a general number of space-time dimensions $d$, the corresponding expression for the differential emission rate becomes
\be\label{ems}
{d\Gamma_{em}^{a \rightarrow b+c }\over dE_{b}}={\Omega^{(d-2)}_{solid}\over 8 (2\pi)^{d-2} } {p^{d-3}\over M_{a}E_{c} }\delta(E_{b}{+}E_{c}{-}E_{a})  \Big|{\cal{S}}_{a \rightarrow b+c}\Big|^{2}
\ee
with
\be
\Omega^{(d-2)}_{solid}={2\pi^{d-1\over 2}\over \Gamma({d{-}1\over 2})}
~.
\ee

The inclusive differential emission rate can be obtained by summing over all the possible final states allowed by global symmetries
\be
\label{difEmissionA}
\sum_{{\rm allowed}\,\, b,c}
{d\Gamma_{em}^{a \rightarrow b+c} \over dE_{b}}={\Omega^{(d-2)}_{solid}\over 8 (2\pi)^{d-2} } {p^{d-3}\over M_{a}E_{c} }\delta(E_{b}{+}E_{c}{-}E_{a})
\sum_{{\rm allowed}\,\, b,c}
\Big|{\cal{S}}_{a\rightarrow b+c}\Big|^{2}
~.
\ee

So far, we treated $E_b$ as a continuous variable. However, in some cases, including the case of strings, the spectrum is discrete. For example, using an expression like the one appearing in Eq.\ \eqref{energyA} for string states, we can write $E_b$ in terms of an integer $n$ in the form
\beq
\label{discreteEnergy}
E_b(n) = \varepsilon + \delta\, n
~,
\eeq
where $\varepsilon$ and $\delta$ are constants. In the example of Eq.\ \eqref{energyA}, $n$ would be the shifted level $N_2-1$ and the constant $\delta = - \frac{\sqrt{\alpha'}}{2\sqrt{N'-1}}$ at fixed $N'$. In such cases, the integrals in equations like \eqref{Gamma_S} involve a sum over the discrete energies $E_b(n)$ instead of an integral. We express this fact by denoting the differential emission rate as a ratio of finite differences
\be
\label{discreteDifEmis}
\frac{\Delta \Gamma_{em}^{a \rightarrow b+c}}{\Delta E_b}
\nonumber
~.
\ee
In continuous limits, where $\delta \to 0$, the conversion of the sum to an integral involves an extra factor $\delta^{-1}$, which converts the above ratio of finite differences to a derivative
\be
\label{diffvsderiv}
{d\Gamma_{em}^{a \rightarrow b+c} \over dE_{b}} = \delta^{-1}
\frac{\Delta \Gamma_{em}^{a \rightarrow b+c}}{\Delta E_b}
~.
\ee
In the string theory example of Eq.\ \eqref{energyA}, this means that we have to include an extra factor of $2\sqrt{N'-1}\sim 2\sqrt{N'}$ (for $N'\gg 1$) in expressions like Eq.\ \eqref{ems}.

We can think about the absorption cross section of a generic process $b+c \rightarrow a$ in a similar manner. We can write the absorption cross section as
\be\label{absEGaoFlu}
\sigma^{b+c\rightarrow a}_{abs}={\Gamma^{b+c\rightarrow a}_{abs}\over \phi^{(b,c)}_{flux}}
~,
\ee
where $ \phi^{(b,c)}_{flux}$ is the relative flux of the incoming particles $b,c$
\be\label{Rflux}
\phi^{(b,c)}_{flux}={|\vec{v}_{b}{-}\vec{v}_{c}|\over {\rm Volume}}
\ee
and $\Gamma^{b+c \rightarrow a}_{abs}$ is the product of a phase space factor times the squared modulus of the corresponding scattering matrix element
\be
\Gamma^{b+c\rightarrow a}_{abs}={1\over 2 E_{b} 2 E_{c}}\int{d^{d-1}\vec{p}_{a}\over (2\pi)^{d-1}2E_{a}}(2\pi)^{d}\delta^{d}(p_{a}{-}p_{b}{-}p_{c})\Big|{\cal{S}}_{b+c \to a}\Big|^{2}
\ee
where now
\be
{\cal{S}}_{b+c \rightarrow a}=\Big\langle a(p_{a})|{\widehat{\cal S}}| b(p_{b}),c(p_{c})\Big\rangle
~.
\ee
Implementing the relation
\be
\int {d^{d-1}\vec{p}\over 2E}=\int {d^{d}p\over 2 E}\delta(p^{0}-E)=\int d^{d}p\,\delta(p^{2}-M^{2})\theta(p^{0})
\ee
in the center-of-mass frame of the colliding particles $(b,c)$, one obtains
\be\label{bahe}
\Gamma^{b+c \rightarrow a}_{abs}={2\pi \delta\left(E_{b}{+}E_{c}{-}M_{a}\right)\over M_{a}^{3} \Big(1- \left({M_{b}^{2}-M_{c}^{2}\over M_{a}^{2}}\right)^{2}\Big)}\Big|{\cal{S}}_{b+c \rightarrow a}\Big|^{2}
~.\ee

Accordingly, the inclusive absorption cross section is obtained by summing over the allowed final states
\be\label{baho}
\sum_{{\rm allowed}\,\, a} \sigma^{b+c \rightarrow a}_{abs}={1\over \phi^{(b,c)}_{flux}}{2\pi \delta\left(E_{b}{+}E_{c}{-}M_{a}\right)\over M_{a}^{3} \Big(1- \left({M_{b}^{2}-M_{c}^{2}\over M_{a}^{2}}\right)^{2}\Big)}\sum_{{\rm allowed}\,\, a}\Big|{\cal{S}}_{b+c \rightarrow a}\Big|^{2}
~.
\ee
In this expression, we restricted, by definition, the sum over the final states to single-particle states.

As reviewed in the main text, the absorption cross section can be derived from the optical theorem, through the computation of the imaginary part of the elastic $2\to 2$ scattering amplitude in the forward limit. In the above notation
\be\label{Opexpgen}
\begin{split}
\sigma_{abs}^{b+c\rightarrow \, {\rm anything}}&={{\rm Im}\, {\cal A}^{\rm forward}_{b+c \rightarrow b+c}\over  F^{(b,c)}_{\phi}}\\
%%%
&={1\over \phi^{(b,c)}_{flux}}{2\pi \delta\left(E_{b}{+}E_{c}{-}M_{a}\right)\over M_{a}^{3} \Big(1- \left({M_{b}^{2}-M_{c}^{2}\over M_{a}^{2}}\right)^{2}\Big)}\sum_{{\rm allowed}\,\, a}\Big|{\cal{S}}_{b+c \rightarrow a}\Big|^{2}+ {\rm inelastic}
\end{split}
~,
\ee
where $F_{\phi}$ is the standard M\o{}ller factor
\be
F^{(b,c)}_{\phi}=E_{b}E_{c}\phi_{flux}^{(b,c)}
~,
\ee
${\cal A}^{\rm forward}$ the forward limit of the 4-point amplitude $b+c \rightarrow b+c$ and `inelastic' the generic multi-particle contribution to the amplitude $b+c \rightarrow$ anything.
At leading order in perturbation theory the inelastic channels in \eqref{Opexpgen} do not contribute and the final result reduces to \eqref{baho}
\be\label{absLead}
\sigma_{abs}^{b+c \rightarrow\, {\rm anything}}
\simeq
\sigma^{b+c \rightarrow a}_{abs}
~.
\ee

The time-reversal symmetry $\TT$ allows us to relate the squared scattering matrix elements that appear in eqs.\ \eqref{difEmissionA}, \eqref{Opexpgen}. Specifically,
\be
\label{timerev}
\Big|{\cal{S}}_{a \rightarrow b+c}\Big|^{2}= \Big| \TT \cdot {\cal{S}}_{a \rightarrow b+c}\Big|^{2}=\Big|{\cal{S}}_{b+c \rightarrow a}\Big|^{2}
~,
\ee
which can be applied to Eqs.\ \eqref{ems} and \eqref{bahe} to obtain the relation
\be
\label{emfixstasim}
{d\Gamma_{em}^{a \rightarrow b+c}\over dE_{b}}={\Omega^{(d-2)}_{solid}\over 8 (2\pi)^{d-1} } {E_{b}^{d-3}\over E_{c} }M_{a}^{2}\left(1-{M_{b}^{2}\over E_{b}^{2}}\right)^{d-3\over2} \left(1- \left({M_{b}^{2}{-}M_{c}^{2}\over M_{a}^{2}}\right)^{2}\right) \phi^{(b,c)}_{flux}\sigma^{b+c \rightarrow a}_{abs}
~.
\ee

In the case of degenerate states in the $a, b, c$ sectors, with corresponding degeneracies $\rho(a),\rho(b),\rho(c)$, one can average over the $a$-sector states and sum over the $b, c$-sector states at fixed energies to find the following relation with the inclusive absorption cross section,
\bea
\frac{1}{\rho(a)} \sum_a \sum_{b,c}
{d\Gamma_{em}^{a\rightarrow b+c}\over dE_{b}}&=&
{\Omega^{(d-2)}_{solid}\over 8 (2\pi)^{d-1} } {E_{b}^{d-3}\over E_{c} }M_{a}^{2} \left(1-{M_{b}^{2}\over E_{b}^{2}}\right)^{d-3\over2} \left(1- \left({M_{b}^{2}{-}M_{c}^{2}\over M_{a}^{2}}\right)^{2}\right)
\nonumber\\
&&\times\, \phi^{(b,c)}_{flux}
\frac{1}{\rho(a)}  \sum_a \sum_{b,c}
\sigma^{b+c {\rightarrow} a}_{abs}
~.
\eea
When $\sigma^{b+c {\rightarrow} a}_{abs}$ is independent of the details of the degenerate states in the sectors $a,b,c$ (as we show for excited string states in the main text), this formula simplifies further to
\bea
\frac{1}{\rho(a)} \sum_a \sum_{b,c}
{d\Gamma_{em}^{a\rightarrow b+c}\over dE_{b}}&=&
{\Omega^{(d-2)}_{solid}\over 8 (2\pi)^{d-1} } {E_{b}^{d-3}\over E_{c} }M_{a}^{2} \left(1-{M_{b}^{2}\over E_{b}^{2}}\right)^{d-3\over2} \left(1- \left({M_{b}^{2}{-}M_{c}^{2}\over M_{a}^{2}}\right)^{2}\right)
\nonumber\\
&&\times\, \phi^{(b,c)}_{flux}
\frac{\rho(b)\rho(c)}{\rho(a)}
\sigma^{b+c {\rightarrow} a}_{abs}
~.
\eea

\section{Probabilities for non-orthogonal states\label{ortho}}

In this appendix, we work a simple quantum mechanical example of calculation of probabilities, as a simple reminder of the issues that appear when we use non-orthogonal bases.

We consider an initial state $|\psi_0\rangle$ in the Hilbert space, that we assume normalized, $\langle \psi_0|\psi_0\rangle=1$. We also consider a final state that belongs to a finite dimensional subspace $V_n$ that is spanned by an orthonormal basis $\psi_i$ with $\langle \psi_i|\psi_j\rangle=\delta_{ij}$.
A generic (normalized) vector in $V_n$ can be written as
\be
|\psi (\vec a)\rangle\equiv \sum_{i=1}^n a_i|\psi_i\rangle\sp \sum_{i=1}^n|a_i|^2=1
\ee
Therefore, the manifold of normalized states of $V_n$ is isometric to $U(n)$.

The amplitude for $|\psi_0\rangle \to|\psi(\vec a)\rangle $ is
\be
A(\vec a)=\langle\psi_0|\psi(\vec a)\rangle=\sum_{i=1}^n a_i A_{0i}\sp A_{0i}\equiv \langle\psi_0|\psi_i\rangle
\ee
The probability of finding any state of $V_n$ in $|\psi_0\rangle$ is given by the sum of probabilities $P(\vec a)=|A(\vec a)|^2$ of ending
in any vector of $V_n$:
\be
P_{0\to V_n}={1\over U_n}\int_{U(n)}d U_{n}  |A(\vec a)|^2={1\over U_n}\sum_{i,j=1}^{n}A^*_{0i}A_{0j}\int_{U(n)}d U_{n}  ~a_i^*a_j=
\label{g3}\ee
$$
=\sum_{i,j=1}^{n}A^*_{0i}A_{0j}\delta^{ij}=\sum_{i=1}^n|A_{0i}|^2
$$
where above, $dU_n$ is the Haar measure on U(n), and $U_n=\int_{U(n)}d U_{n}$ is the invariant volume of U(n). The end result is the standard sum of squared amplitudes formulae that is valid as we see in
an orthonormal basis.

We now translate the calculation in a non-orthogonal basis of final states.
To do this we start from the orthonormal basis above and we rotate it to generic basis by an GL(C,n) rotation $M_{ij}$,
\be
|\psi_i\rangle=\sum_{j=1}^n M_{ij}~|\bar \psi_j\rangle\sp det M\not=0
\ee
Now the inner products of the new basis have a nontrivial metric
\be
G_{ij}\equiv \langle \bar \psi_i|\bar \psi_j\rangle=\sum_{k,l=1}^nM^*_{ik}M_{jl}\langle  \psi_i| \psi_j\rangle=\sum_{k=1}^nM^*_{ik}M_{jk}=(M\cdot M^{\dagger})_{ji}
\ee
We also obtain
\be
\bar A_{0i}\equiv \langle \psi_0|\bar \psi_i\rangle =M_{ij}A_{0j}\ar A_{0i}=M^{-1}_{ij}\bar A_{0j}
\ee
so that the probability can be written as
\be
P_{0\to V_n}=\sum_{i=1}^n|A_{0i}|^2=\sum_{i=1}^n   A^*_{0i}A_{0i}=\sum_{i=1}^n\sum_{k,l=1}^n   (M^*)^{-1}_{il}A^*_{0l}   M^{-1}_{ik}\bar A_{0k}=
\ee
$$
=\sum_{k,l=1}^n (M\cdot M^{\dagger})^{-1}_{lk}A^*_{0l}\bar A_{0k}=\sum_{k,l=1}^n G^{kl}A^*_{0l}\bar A_{0k}
$$
where $G^{ij}$ is the inverse metric of $G_{ij}$.

\section{Systematics of amplitudes in the forward limit}\label{SForLimAmp}

In this appendix we discuss the forward limit of the 4-point amplitude in detail. We consider first a specific example of scattering between two tachyons and two leading Regge states and then extend to the most general case of four generic highly excited string states.

The scattering amplitude of two tachyons, represented by the vertex operators
\be
V_{T}^{in}(p_{2})=e^{ip_{2}{\cdot}X}\,,\quad V_{T'}^{out}(p_{3})=e^{ip_{3}{\cdot}X}
\ee
and two excited states of the leading Regge trajectory, represented by the vertex operators
\be
V_{H}^{in}(p_{1})={1\over \sqrt{2!}}H^{\mu^{1}_{1}\mu^{2}_{1}}_{(1)(1)}i\partial X_{\mu^{1}_{1}}i\partial X_{\mu^{2}_{1}} e^{ip_{1}{\cdot}X}\,,\quad V_{H'}^{out}(p_{4})={1\over \sqrt{2!}}H'^{\,\nu^{1}_{1}\nu^{2}_{1}}_{(1)(1)}i\partial X_{\nu^{1}_{1}}i\partial X_{\nu^{2}_{1}} e^{ip_{4}{\cdot}X}\,,
\ee
%where we use an explicit notation for the identification of the polarizations with the chosen excited harmonics
%\be
%H^{\mu^{1}_{1}\mu^{2}_{1}}_{(1)(1)}\Leftrightarrow \alpha_{-a}^{\mu^{1}_{a}}\alpha_{-a}^{\mu^{2}_{a}}
%\ee
%and by construction all the indices connected to the same excitation are symmetric, with the general representation
%\be
%{1\over \sqrt{a^{g_{a}}g_{a}! ... d^{g_{d}}g_{d}! ... }}H^{\overbrace{\overbrace{\mu^{1}_{a}\mu^{2}_{a}...\mu^{g_{a}}_{a}}^{Symm}...\overbrace{\mu^{1}_{d}\mu^{2}_{d}...\mu^{g_{d}}_{d}}^{Symm}...}^{generic}}_{\underbrace{(a)(a)...(a)}_{g_{a}}...\underbrace{(d)(d)...(d)}_{g_{d}}...}\Leftrightarrow \alpha_{-a}^{\mu^{1}_{a}}\alpha_{-a}^{\mu^{2}_{a}}...\alpha_{-a}^{\mu^{g_{a}}_{a}}...\alpha_{-d}^{\mu^{1}_{d}}\alpha_{-d}^{\mu^{2}_{d}}...\alpha_{-d}^{\mu^{g_{d}}_{d}}...
%\ee
can be computed according to \eqref{open1ak}
\be
{\cal A}_{H_2 + T \to H_{2}' + T}={1\over 2}H^{\mu^{1}_{1}\mu^{2}_{1}}_{(1)(1)}H'^{\,\nu^{1}_{1}\nu^{2}_{1}}_{(1)(1)} A_{Ven}(s,t)\, A_{\mu^{1}_{1}\mu^{2}_{1}\nu^{1}_{1}\nu^{2}_{1}}(\partial_{\beta_{s}},\partial_{\beta_{t}})\Phi_{\beta_{s}\beta_{t}}(s,t)\Big|_{\beta_{s,t}=0}
\ee
with $\Phi_{\beta_{s}\beta_{t}}(s,t)$ defined in \eqref{ffunc}. The differential operator acting on $\Phi$ is given by
\be\label{StaQua}
\begin{split}
A_{\mu^{1}_{1}\mu^{2}_{1}\nu^{1}_{1}\nu^{2}_{1}}& =(p_{2\,\mu_{1}^{2}}{+}p_{3\,\mu_{1}^{2}}\partial_{\beta_{s}})(p_{2\,\mu_{1}^{1}}{+}p_{3\,\mu_{1}^{1}}\partial_{\beta_{s}})(p_{1\,\nu_{1}^{1}}{+}p_{2\,\nu_{1}^{1}}\partial_{\beta_{t}})(p_{1,\nu_{1}^{2}}{+}p_{2\,\nu_{1}^{2}}\partial_{\beta_{t}})\\
%%%
&+(p_{1,\nu_{1}^{2}}{+}p_{2\,\nu_{1}^{2}}\partial_{\beta_{t}})(p_{2\,\mu_{1}^{2}}{+}p_{3\,\mu_{1}^{2}}\partial_{\beta_{s}})\delta_{\mu_{1}^{1} \nu_{1}^{1}}\partial_{\beta_{s}}+(p_{1,\nu_{1}^{1}}{+}p_{2\,\nu_{1}^{1}}\partial_{\beta_{t}})(p_{2\,\mu_{1}^{2}}{+}p_{3\,\mu_{1}^{2}}\partial_{\beta_{s}})\delta_{\mu_{1}^{1} \nu_{1}^{2}}\partial_{\beta_{s}}\\
%%%
&+(p_{1,\nu_{1}^{2}}{+}p_{2\,\nu_{1}^{2}}\partial_{\beta_{t}})(p_{2\,\mu_{1}^{1}}{+}p_{3\,\mu_{1}^{1}}\partial_{\beta_{s}})\delta_{\mu_{1}^{2} \nu_{1}^{1}}\partial_{\beta_{s}}+ (p_{1,\nu_{1}^{1}}{+}p_{2\,\nu_{1}^{1}}\partial_{\beta_{t}})(p_{2\,\mu_{1}^{1}}{+}p_{3\,\mu_{1}^{1}}\partial_{\beta_{s}})\delta_{\mu_{1}^{2} \nu_{1}^{2}}\partial_{\beta_{s}}\\
%%%
&+\delta_{\mu_{1}^{1}\nu_{1}^{2}}\delta_{\mu_{1}^{2}\nu_{1}^{1}}\partial^{2}_{\beta_{s}}+\delta_{\mu_{1}^{1}\nu_{1}^{1}}\delta_{\mu_{1}^{2}\nu_{1}^{2}}\partial^{2}_{\beta_{s}}\\
%%%
&+2(p_{1,\nu_{1}^{1}}{+}p_{2\,\nu_{1}^{1}}\partial_{\beta_{t}})(p_{1,\nu_{1}^{2}}{+}p_{2\,\nu_{1}^{2}}\partial_{\beta_{t}})\delta_{\mu_{1}^{1}\mu_{1}^{2}}W^{(1)}_{1,1}(\partial_{\beta_{s}},\partial_{\beta_{t}})+\\
%%%
&+2(p_{2\,\mu_{1}^{1}}{+}p_{3\,\mu_{1}^{1}}\partial_{\beta_{s}})(p_{2\,\mu_{1}^{2}}{+}p_{3\,\mu_{1}^{2}}\partial_{\beta_{s}})\delta_{\nu_{1}^{1}\nu_{1}^{2}}W_{1,1}^{(4)}(\partial_{\beta_{s}},\partial_{\beta_{t}})+\\
%%%
&+4W_{1,1}^{(1)}(\partial_{\beta_{s}},\partial_{\beta_{t}})W_{1,1}^{(4)}(\partial_{\beta_{s}},\partial_{\beta_{t}})\delta_{\mu_{1}^{1}\mu_{1}^{2}}\delta_{\nu_{1}^{1}\nu_{1}^{2}}
\end{split}
\ee
with
\be
W^{(1)}_{1,1}(\partial_{\beta_{s}},\partial_{\beta_{t}})= {1\over 2}(1+\beta_{1}^{(1)})\beta_{1}^{(1)}\partial_{\beta_{s}}^{2}-\alpha_{1}^{(1)}\beta_{1}^{(1)}\partial_{\beta_{s}}\partial_{\beta_{t}}+{1\over 2}(1+\alpha_{1}^{(1)})\alpha_{1}^{(1)}\partial_{\beta_{t}}^{2}
\ee
\be
W^{(4)}_{1,1}(\partial_{\beta_{s}},\partial_{\beta_{t}})= {1\over 2}(1+\beta_{4}^{(1)})\beta_{4}^{(1)}\partial_{\beta_{t}}^{2}-\alpha_{4}^{(1)}\beta_{4}^{(1)}\partial_{\beta_{s}}\partial_{\beta_{t}}+{1\over 2}(1+\alpha_{4}^{(1)})\alpha_{4}^{(1)}\partial_{\beta_{s}}^{2}
\ee
and
\be
\alpha_{1}^{(1)}=-1-2\alpha'q_{1}{\cdot}p_{2}\,,\quad \beta_{1}^{(1)}=-1-2\alpha'q_{1}{\cdot}p_{4}
\ee
\be
\alpha_{4}^{(1)}=-1-2\alpha'q_{4}{\cdot}p_{1}\,,\quad \beta_{4}^{(1)}=-1-2\alpha'q_{4}{\cdot}p_{3}
\ee
To compute the forward limit, one has to use the following constraints
\be\label{FlimH}
p_{1}^{\mu}=-p_{4}^{\mu}\,,\quad p_{2}^{\mu}=-p_{3}^{\mu}\, , \quad 2\alpha'q_{1}{\cdot}p_{4}=2\alpha'q_{4}{\cdot}p_{1}={-}1\, ,
\ee
\be
 2\alpha'q_{1}{\cdot}p_{2}=-2\alpha'q_{1}{\cdot}p_{3}=f_{kin}(E,M_{N},M_{T})=-2\alpha'q_{3}{\cdot}p_{2}=2\alpha'q_{4}{\cdot}p_{3}\,,
 \ee
 where
 \be
 f_{kin}(E,M_{N},M_{T})\equiv {E-\sqrt{E^{2}{-}M_{T}^{2}}\over M_{N}}
 ~,
 \ee
 combined with the $t\rightarrow 0$ limit, which can be easily extracted by considering the non-zero contributions coming from the derivatives with respect to the parameter $\beta_{t}$ of the function
 \be
 \Phi_{\beta_{s}\beta_{t}}(s,t=0)=e^{\beta_{s}}\left(1-{\beta_{s}-\beta_{t}\over \alpha's{+}2}\right)
 ~.
 \ee
 In particular,
 \bea
&& \partial_{\beta_{t}}^{2}\Phi_{\beta_{s}\beta_{t}}(s,t{=}0)=0\,,\quad \partial_{\beta_{t}}\Phi_{\beta_{s}\beta_{t}}(s,t{=}0)={e^{\beta_{s}}\over \alpha's{+}2 }\,,\quad
\nonumber\\
&&\partial_{\beta_{s}}^{k}\Phi_{\beta_{s}\beta_{t}}(s,t{=}0)= e^{\beta_s} \left(1-\frac{k + \beta_s}{\alpha's{+}2 }+{\beta_{t}\over  \alpha's{+}2} \right)
 ~.
 \eea

We now study what contributes to Eq.\ \eqref{StaQua} in the forward limit. First, since in this limit $\beta_{1}^{(1)}=\alpha_{1}^{(4)}=0$, all the contributions to the operator $A$ in \eqref{StaQua} containing
 \be\label{Abboh1}
 W^{(1)}_{1,1}(\partial_{\beta_{s}},\partial_{\beta_{t}})\Big|_{forw}={1\over 2}f_{kin}(1+f_{kin})\partial_{\beta_{t}}^{2}
 ~,
 \ee
\be\label{Abboh2}
W^{(4)}_{1,1}(\partial_{\beta_{s}},\partial_{\beta_{t}})\Big|_{forw}={1\over 2}f_{kin}(1+f_{kin})\partial_{\beta_{t}}^{2}
\ee
vanish when they act on $\Phi$ (the latter is linear in $\beta_{t}$). The remaining differential operators in $A$ involve the first four lines of \eqref{StaQua}. In addition, the following two identities
\be
p_{2\mu}{+}p_{3\,\mu}\partial_{\beta_{s}}\,\Big|_{forw}=p_{2\mu}(1-\partial_{\beta_{s}})=p_{2\mu}\partial_{\beta_{t}}
~,
\ee
\be
p_{1\,\nu}{+}p_{2\,\nu}\partial_{\beta_{t}}\,\Big|_{forw}=p_{2\,\nu}\partial_{\beta_{t}}
~,
\ee
are valid in the forward limit \eqref{FlimH} when we use $1-\partial_{\beta_{s}}=\partial_{\beta_{t}}$, as shown in \eqref{phiID}.

For $H'^{\mu\nu}= H^{*\mu\nu}$ in the forward limit, terms involving $H'$ are annihilated when contracted with either $p_{1}$ and $p_{4}$. Therefore, the first three lines in \eqref{StaQua} vanish when acting on $\Phi$.

Incorporating all the above observations we obtain
\be
A_{\mu^{1}_{1}\mu^{2}_{1}\nu^{1}_{1}\nu^{2}_{1}}\Big|_{forw}=\delta_{\mu_{1}^{1}\nu_{1}^{2}}\delta_{\mu_{1}^{2}\nu_{1}^{1}}\partial^{2}_{\beta_{s}}+\delta_{\mu_{1}^{1}\nu_{1}^{1}}\delta_{\mu_{1}^{2}\nu_{1}^{2}}\partial^{2}_{\beta_{s}}
\ee
and the final result is given by
\be
\begin{split}
&{\cal A}_{H_2 + T \to H_2' +T} \Big|_{forw}=\\
%%%
&\hspace{2cm} A_{Ven}(s,t)\Big|_{forw}\,\,{1\over 2}H^{\mu^{1}_{1}\mu^{2}_{1}}_{(1)(1)}H^{*\,\nu^{1}_{1}\nu^{2}_{1}}_{(1)(1)} \, \left( \delta_{\mu_{1}^{1}\nu_{1}^{2}}\delta_{\mu_{1}^{2}\nu_{1}^{1}}+\delta_{\mu_{1}^{1}\nu_{1}^{1}}\delta_{\mu_{1}^{2}\nu_{1}^{2}}\right)\partial^{2}_{\beta_{s}}\Phi_{\beta_{s}\beta_{t}}(s,t)\Big|_{\beta_{s,t}=0}\\
%%%
&\hspace{4.4cm}= A_{Ven}(s,t)\Big|_{forw} \,\,{1\over 2}\Big(Tr(HH^{*})+Tr(H^{T}H^{*})\Big){\alpha's\over \alpha's+2}\\
%%%
&\hspace{4.4cm}= A_{Ven}(s,t)\Big|_{forw}\,\,{\alpha's\over \alpha's+2}~,
\end{split}
\ee
where we used the normalisation condition \eqref{NormCond}, which in the specific case becomes $Tr(HH^{*})+Tr(H^{T}H^{*})=2$.

Using formula \eqref{open2aa}, we obtain the following absorption cross section
\be
\sigma_{abs}^{H_2+T\rightarrow {\rm any}\,H'}(s)=\pi \alpha'^{\frac{d-2}{2}}g_{o}^{2}{\alpha's \over \alpha'F_{\phi}^{(HT)}}
\ee
as in \eqref{protobohh}.

Similar cancelations occur in more general examples. They can be implemented directly at the level of the generating function. The generating function of the 2 to 2 scattering ($1\rightarrow 4$ and $2\rightarrow 3$) of arbitrarily excited states, in the forward limit, is given by
\be
{\cal A}_{gen}(s,t)\Big|_{forw}={\cal A}_{Ven}(s,t)\Big|_{forw}\,e^{{\cal K}_{forw}\left(\{\zeta^{(\ell)}_{n}\};\partial_{\beta_{s}},\partial_{\beta_{t}}\right)}\Phi^{(forw)}_{\beta_{s},\beta_{t}}\left(s\right)\Big|_{\beta_{s,t}{=}0}
\ee
where
\be
\Phi^{(forw)}_{\beta_{s},\beta_{t}}\left(s\right)\equiv \Phi_{\beta_{s},\beta_{t}}\left(s,t{=}0\right)=e^{\beta_{s}}\left(1-{\beta_{s}-\beta_{t}\over \alpha's{+}2}\right)\,,
\ee
\be
\begin{split}
{\cal K}_{forw}\left(\{\zeta^{(\ell)}_{n}\}; \partial_{\beta_{s}},\partial_{\beta_{t}}\right):=&\sum_{\ell=1}^{4}\left(\sum_{n}\zeta_{n}^{(\ell)}{\cdot}V^{(\ell)}_{n}(\partial_{\beta_s},\d_{\beta_t}){+}\sum_{n,m} \zeta_{n}^{(\ell)}{\cdot}\zeta_{m}^{(m)}W_{n,m}^{(\ell)}(\partial_{\beta_s}, \d_{\beta_t})\right)\\
%%%
&\quad\quad+\sum_{n,m}\zeta^{(1)}_{n}{\cdot}\zeta^{(4)}_{m} I_{n,m}^{(1,4)}(\partial_{\beta_s}, \d_{\beta t})\Big|_{forw}\,,
\end{split}
\ee
and
\be\label{soFF1}
\alpha_{2}^{(n)}=\alpha_{4}^{(n)}=\beta_{1}^{(n)}=\beta_{3}^{(n)}=0\,,\quad \alpha_{1}^{(n)}=\beta_{4}^{(n)}\,,\quad \alpha_{3}^{(n)}=\beta_{2}^{(n)}\,,
\ee
with the parameters above defined in \eqref{coeffJac}.
Equations \eqref{soFF1} are valid because of the forward limit \eqref{FlimH}, and the DDF conditions \eqref{DDFqp}.

The above structures reduce as follows:
\be\label{V1exp}
\begin{split}
V_{n}^{(1)\mu}(\partial_{\beta_{s}},\partial_{\beta_{t}})\Big|_{forw}&=p_{2}^{\mu}\left(P_{n-1}^{\hspace{0 mm}^{\hspace{0 mm}^{(\alpha_{1}^{(n)},\beta_{1}^{(n)})}}}\hspace{-9mm}(\partial_{\beta_{t}}{-}\partial_{\beta_{s}})-\partial_{\beta_{s}}P_{n-1}^{\hspace{0 mm}^{\hspace{0 mm}^{(\alpha_{1}^{(n)}+1,\beta_{1}^{(n)})}}}\hspace{-9mm}(\partial_{\beta_{t}}{-}\partial_{\beta_{s}})\right)\\
%%%
&=p_{2}^{\mu}\,(-)^{n{-}1}\left({\partial_{\beta_{s}}^{n-1}\over (n{-}1)!}+{\partial_{\beta_{s}}^{n-2}\over (n{-}2)!} \right)\partial_{\beta_{t}} + p_{2}^{\mu}\,O(\partial_{\beta_{t}}^{2})
\end{split}
\ee
\be\label{V3exp}
\begin{split}
V_{n}^{(3)\mu}(\partial_{\beta_{s}},\partial_{\beta_{t}})\Big|_{forw}&=p_{4}^{\mu}\left(P_{n-1}^{\hspace{0 mm}^{\hspace{0 mm}^{(\alpha_{3}^{(n)},\beta_{3}^{(n)})}}}\hspace{-9mm}(\partial_{\beta_{t}}{-}\partial_{\beta_{s}})-\partial_{\beta_{s}}P_{n-1}^{\hspace{0 mm}^{\hspace{0 mm}^{(\alpha_{3}^{(n)}+1,\beta_{3}^{(n)})}}}\hspace{-9mm}(\partial_{\beta_{t}}{-}\partial_{\beta_{s}})\right)\\
%%%
&=p_{4}^{\mu}\,(-)^{n{-}1}\left({\partial_{\beta_{s}}^{n-1}\over (n{-}1)!}+{\partial_{\beta_{s}}^{n-2}\over (n{-}2)!} \right)\partial_{\beta_{t}} + p_{4}^{\mu}\,O(\partial_{\beta_{t}}^{2})
\end{split}
\ee
\be\label{V2exp}
\begin{split}
V_{n}^{(2)\mu}(\partial_{\beta_{s}},\partial_{\beta_{t}})\Big|_{forw}&= p^{\mu}_{4} P_{n-1}^{\hspace{0 mm}^{\hspace{0 mm}^{(\alpha_{2}^{(n)}{+}1,\beta_{2}^{(n)})}}}\hspace{-12mm}(\partial_{\beta_{s}}{-}\partial_{\beta_{t}})\partial_{\beta_{t}}   \\
%%%
&=p_{4}^{\mu}{(1{+}\alpha_{2}^{(n)})_{n}\over (n{+}\alpha_{2}^{(n)})(n{-}1)!}\partial_{\beta_{s}}^{n-1}\partial_{\beta_{t}} +p_{4}^{\mu}\,O(\partial_{\beta_{t}}^{2})
\end{split}
\ee
\be\label{V4exp}
\begin{split}
V_{n}^{(4)\mu}(\partial_{\beta_{s}},\partial_{\beta_{t}})\Big|_{forw}&=p^{\mu}_{2} P_{n-1}^{\hspace{0 mm}^{\hspace{0 mm}^{(\alpha_{4}^{(n)}{+}1,\beta_{4}^{(n)})}}}\hspace{-12mm}(\partial_{\beta_{s}}{-}\partial_{\beta_{t}})\partial_{\beta_{t}}\\
&=p_{2}^{\mu}{(1{+}\alpha_{4}^{(n)})_{n}\over (n{+}\alpha_{4}^{(n)})(n{-}1)!}\partial_{\beta_{s}}^{n-1}\partial_{\beta_{t}}+p_{2}^{\mu}\,O(\partial_{\beta_{t}}^{2})
\end{split}
\ee
where $P_{n}^{(\alpha,\beta)}$ are the well-known Jacobi polynomials, defined in \eqref{Jpppoly} and which have been expanded as follows
\be
P_{n-1}^{\hspace{0 mm}^{\hspace{0 mm}^{(\alpha_{\ell}^{(n)},\beta_{\ell}^{(n)})}}}\hspace{-8mm}(\partial_{\beta_{t}}{-}\partial_{\beta_{s}})=-{(-)^{n}(1{+}\beta_{\ell}^{(n)})_{n}\over (n{+}\beta_{\ell}^{(n)})(n{-}1)!}\partial_{\beta_{s}}^{n-1}+{ (-)^{n}(\alpha_{\ell}^{(n)}{+}n{-}1)(1{+}\beta_{\ell}^{(n)})_{n}\over (n{+}\beta_{\ell}^{(n)})(1{+}\beta_{\ell}^{(n)}) (n{-}2)!}\partial_{\beta_{s}}^{n-2}\partial_{\beta_{t}}+O(\partial_{\beta_{t}}^{3})
\ee
\be
P_{n-1}^{\hspace{0 mm}^{\hspace{0 mm}^{(\alpha_{\ell}^{(n)},\beta_{\ell}^{(n)})}}}\hspace{-8mm}(\partial_{\beta_{s}}{-}\partial_{\beta_{t}})={(1{+}\alpha_{\ell}^{(n)})_{n}\over (n{+}\alpha_{\ell}^{(n)})(n{-}1)!}\partial_{\beta_{s}}^{n-1}-{(\beta_{\ell}^{(n)}{+}n{-}1)(1{+}\alpha_{\ell}^{(n)})_{n}\over (n{+}\alpha_{\ell}^{(n)})(1{+}\alpha_{\ell}^{(n)}) (n{-}2)!}\partial_{\beta_{s}}^{n-2}\partial_{\beta_{t}}+O(\partial_{\beta_{t}}^{3})
~.
\ee
Similar to \eqref{Abboh1} and \eqref{Abboh2}, the $W$ polynomials reduce in the forward limit to two or higher derivatives with respect to $\beta_{t}$,
\be
 W^{(\ell)}_{n,m}(\partial_{\beta_{s}},\partial_{\beta_{t}})\Big|_{forw}=O(\partial_{\beta_{t}}^{2})
 \ee
 and therefore vanish when they act on $\Phi$.

The $I$ polynomials, defined in \eqref{Ibv} and \eqref{Ibl} are given in the forward limit by
 \be\label{ella1}
 I_{n,m}^{(1,4)}=\delta_{n,m} n \partial_{\beta_{s}}^{n}+ (1{-}\delta_{n,m})n(2\ell_{s}^{2}nq_{1}{\cdot}p_{2}{-}m{+}n)\partial_{\beta_{s}}^{max(n,m)-1}\partial_{\beta_{t}}+O(\partial_{\beta_{t}}^{2})
 ~,
 \ee
 \be\label{ella2}
  I_{n,m}^{(3,2)}=\delta_{n,m} n \partial_{\beta_{s}}^{n}+ (1{-}\delta_{n,m})n(2\ell_{s}^{2}nq_{3}{\cdot}p_{4}{-}m{+}n)\partial_{\beta_{s}}^{max(n,m)-1}\partial_{\beta_{t}}+O(\partial_{\beta_{t}}^{2})
~,
\ee
 \be\label{ella3}
I_{n,m}^{(2,1)}=n\, m\, \partial_{\beta_{s}}^{n+m-2}\partial_{\beta_{t}}+O(\partial_{\beta_{t}}^{2})=I_{n,m}^{(4,3)}
~,
\ee
\be\label{ella4}
I_{n,m}^{(2,4)}=n\, m\, \partial_{\beta_{s}}^{n+m-1}\partial_{\beta_{t}}+O(\partial_{\beta_{t}}^{2})= I_{n,m}^{(1,3)}
~.
\ee

It remains to consider the contributions of the $V$ and $I$ polynomials. As a first illustration, we can consider the amplitude with two tachyons and two arbitrarily excited states. This amplitude contains two polarizations tensors, $H_{\mu_{1}...\mu_{Q}}$ and its complex conjugate. Therefore, $2Q$ polarization indices are contracting with momenta or among themselves. As shown above the $W$ operators vanish, and we have to consider the $V$ and $I$ operators. It is important to remember that the $I$ operator contracts one index of $H$ with another of $H^{*}$ while the $V$ operators contract a single index of $H$ or $H^{*}$.

In the expansion of the exponential we can have arbitrary powers of $I$ and $V$ operators. Consider now a term that contains $k$ powers of $I$. This involves the contraction of $k$ indices from $H$ and $k$ indices from $H^{*}$. The remaining $Q-k$ indices of $H$ contract either with indices from $H^*$ or with indices from $V$. In this fashion, an even number (from 0 to $2(Q-k)$) of $V$'s appear. From \eqref{V1exp}-\eqref{V4exp} each power of $V$ appears with a single derivative of $\beta_{t}$. As $\Phi$ is linear in $\beta_{t}$, the only non zero contribution comes from terms without any $V$ operators\footnote{It should be remembered that the transversality conditions on the polarization tensors are important for $V$ to be linear in $\beta_{t}$ derivatives.}. That leaves only contributions from terms that contain $I^Q$.

Now from \eqref{ella1}-\eqref{ella4} it is clear that when an operator $I$ connects the polarization indices corresponding to two different harmonics, it is proportional to $\partial_{\beta_{t}}$. As such contractions always come in pairs, such cases give zero when acting on $\Phi$. We conclude that the only non zero terms originating from $I^{Q}$, contract polarization indices corresponding to oscillators with the same harmonic. Therefore, the non zero contributions are in one to one correspondence with the contractions of the two polarization tensors that appear in the normalization condition \eqref{NormCond}. Consequently, the final answer is \eqref{opabsfinL}.

In the previous argument we assumed two particles to be scalars. Now, we generalize to arbitrarily excited states. The amplitude contains two polarization tensors $H_{1}$ and $H_{2}$ and in forward limit, their conjugates  $H_{1}^{*}$ and $H_{2}^{*}$. As before, $W$ operators give zero and $V$ operators appear in even numbers and also give zero. We are left with products of $I$ operators, which provide contractions of two types:
\begin{itemize}
\item Contractions between $H_{1}$ and $H_{2}^{*}$ and vice-versa. Such contractions appear in pairs and each contraction provides a derivative of $\beta_{t}$ as seen in \eqref{ella1}-\eqref{ella4}. Therefore, we always get at least terms with two derivatives of $\beta_{t}$ acting on $\Phi$, which vanish.

\item
 Contractions between $H_{1}$ and $H_{1}^{*}$ or between $H_{2}$ and $H_{2}^{*}$. According to the above discussion, these are the only contractions that could give a non-zero answer. For each one of them, what we said earlier applies, and the final amplitude is the product of such contributions, as in \eqref{finABSopHH}.
\end{itemize}

\end{appendix}
\newpage

\bibliography{HES_AbsEmiss_Final_Revised}

\end{document}